\documentclass[online,showlabels,coverfontpercent=100,final]{adsphd}

\usepackage{cite}
\usepackage[disable]{todonotes} 

\usepackage[english,dutch]{babel}
\usepackage{cancel}

\usepackage{nomencl}   
\usepackage[style=long,number=none]{glossary} 

\usepackage[utf8]{inputenc} 

\title{Phenomenological aspects of supergravity theories in de Sitter vacua}	

\author{Rob}{Knoops}

\supervisor{Prof.~dr.~I.~Antoniadis}{}
\supervisor{Prof.~dr.~A.~Van~Proeyen}{}
\president{Prof.~dr.~C.~Maes}
\jury{Prof.~dr.~N.~Bobev \\
 Prof.~dr.~M.~Fannes \\
Prof.~dr.~T.~Hertog\\
Prof.~dr.~T.~Van Riet 
    }
\externaljurymember{Prof.~dr.~J.-P.~Derendinger}{Bern University}

\phddegree{Science (Phd): Physics} 
\faculty{Faculty of Science}
\department{Department of Physics and Astronomy}

\researchgroup{Institute for Theoretical Physics}
\website{https://fys.kuleuven.be/itf} 
\email{} 
\address{Celestijnenlaan 200B}
 \addresscity{B-3001 Leuven} 
\date{August 2016}
\copyyear{2016}

 \setcustomcoverpage{mycoverpage.tex} 

\renewcommand{\nomname}{List of Symbols}
\makeatletter
\let\@printnomenclatureorig\@printnomenclature
\def\@printnomenclature[#1]{%
  \cleardoublepage%
  \chaptermark{\nomname}
  \@printnomenclatureorig[#1]
}
\makeatother
\makenomenclature

\newcommand{\glossname}{Abbreviations}
\makeglossary

\let\printglossaryorig\printglossary
\renewcommand{\printglossary}{%
  \renewcommand{\glossaryname}{\glossname}
  \cleardoublepage%
  \printglossaryorig\chaptermark{\glossname}}

\bibliography{allpapers}

\bibstyle{acm}

\usepackage{amssymb,amsthm}
\usepackage{amsmath}

\usepackage{todonotes}
  \newcommand{\Tr}{\text{Tr}}
  \newcommand{\im}{\text{Im}}
  \newcommand{\re}{\text{Re}}
\newcommand{\cA}{{\cal A}} \newcommand{\cB}{{\cal B}} \newcommand{\cS}{{\cal S}} 
\begin{document}

\makefrontcoverXII

\maketitle

\frontmatter

\includepreface{preface}
\includeabstract{abstract}

\printglossary

\tableofcontents
\listoftodos

\mainmatter

  \graphicspath{{\chaptersdir/introduction/image/}}%
  \chapter{Introduction}\label{ch:introduction}

\section{Particle physics}

What is the smallest thing you can imagine? A grain of salt, a hair, or perhaps it is an atom?
Ever since ancient history human kind has been fascinated with questions like ''What is matter made of?'', and ''which are nature's most fundamental building blocks?''.
Everything around you, from the book (or thesis) in front of you to the chair you are sitting on, is made of tiny atoms and molecules.
These atoms however, consist of even smaller constituents, called protons, neutrons and electrons. 
While it is assumed that the electron is a fundamental particle, protons and neutrons on the other hand consist of even smaller particles called quarks.
As particle physicists it is our goal to try to understand these most fundamental building blocks of nature and the interactions that govern them.

One approach is that of the theorist who applies mathematical rigor 
based on physical motivations and appropriate questions
to establish a theory of nature at the most fundamental level.
On the other hand there is the experimenter, who puts these theories to the test.
The content of this thesis is of the former approach. 
Our goal is to establish a mathematically consistent model based on simple symmetry principles, 
and to identify possible experimental signatures that could distinguish it from other models.

\section{The Standard Model}

\subsection{Its successes\ldots}

One of the most impressive theories describing the fundamental particles and their interactions is the Standard Model. 
The Standard Model unifies three of the fundamental forces of nature, namely the electromagnetic interaction, 
the weak force, which is responsible for the nuclear decay that is for example the source of energy in a nuclear power plant, 
and the strong interaction, which gets its name because it is so strong that it can keep a nucleus consisting of (neutrons and) positively charged protons together, 
even though they repel each other due to the electromagnetic interaction.  
Perhaps one of the most baffling accomplishments of the Standard Model is its incredibly accurate predictions that have been confirmed by numerous experiments. 
For example, the magnetic moment (or g-value) of the electron has been confirmed up to 13 significant digits.
Moreover, the Standard Model successfully predicted the existence of the $W$ and $Z$ bosons, the gluons, which are responsible for the weak and strong force respectively,
and the top and charm quarks as well as the Higgs Boson, even before these particles were observed.
Even their predicted experimental properties, as for example the masses of the W and Z bosons, were confirmed with an incredible precision. 

\subsection{\ldots and its flaws}

Despite its incredible experimental successes, physicists are convinced that the Standard Model is not the end of the story: 

\begin{itemize}
\item First, although it neatly describes the electromagnetic, the weak and the strong interactions, 
the Standard Model does not describe a fourth fundamental force of nature, namely gravity.
Although gravity is quite well understood at a long distance scale by Einstein's theory of General Relativity, 
physicists can only guess how gravity 'works' when look at it very closely, at the smallest scales.
In order words, although physicists can very well describe 'how fast the apple falls on the floor' or 'how the speed and orbit of a planet around the sun are related', 
we fail to understand how gravity is described at the quantum level.

\item Second, certain elementary particles, called neutrinos, are massless in the Standard Model.
Since neutrinos interact only very weakly (and only through the weak force) with other matter and therefore tend to just fly through most particle detectors without leaving any trace, 
they are extremely difficult to detect and an accurate mass measurement of these neutrinos has never been accomplished. 
However, neutrinos come in three lepton flavors, namely electron neutrinos, muon neutrinos and tau neutrinos, 
and it is observed a neutrino can change flavor after traveling some distance. 
These 'flavor oscillations' can be used to calculate the mass differences between these neutrinos of different lepton flavor.
Various neutrino oscillation experiments have indeed confirmed a very small mass difference between the three neutrinos, 
which is inconsistent with the Standard Model, where all neutrinos are assumed to be massless.

\item The particle content of the Standard Model, which in this context often referred to as baryonic matter (even though the Standard Model technically does not only contain baryons),
makes up only about 5 percent of the known matter of the universe. The rest is dark matter (about 27 percent) and dark energy (about 68 percent).
Although dark matter can not be directly seen with telescopes since it does not (or very little) interact with ordinary matter, 
its existence can be confirmed from its gravitational effects on visible matter~\cite{dark_matter}. 
While mass of large astronomical objects can be calculated from visible effects, like stars, dust or gas, their gravitational interactions suggest that the mass of these objects is much larger.
Dark energy on the other hand, is hypothesized to accommodate for recent observations that the universe seems to be expanding at an accelerating rate.

\item Perhaps one of the most compelling arguments in favor of the existence of a theory beyond the standard model, at least from a theoretical point of view, is the hierarchy problem~\cite{HP1,HP2,HP3,HP4,HP5}. 
The hierarchy problem is the large discrepancy between the weak force and gravity, the former being about $10^{32}$ times stronger than gravity.
The discovery of the Higgs boson at CERN in 2012, which led to a Nobel Prize for Francois Englert and Peter Higgs\footnote{
Actually, besides Francois Englert another Belgian physicist, namely Robert Brout, made significant contributions to the Brout-Englert-Higgs mechanism 
and if he were still alive would most probably have participated in the Nobel Prize.},
also meant the first experimental confirmation of the existence of a scalar particle. 
In the Standard Model, the measured Higgs boson mass of roughly 125 GeV 
\glossary{name=GeV,description=Giga ElectronVolt}
is the sum of the 'bare' (classical) mass squared $m^2_0$ and its quantum corrections $\delta m^2$, such that $m^2_{\text{measured}} = m^2_0 + \delta m^2$. 
Here, the bare mass $m_0$ is an input parameter for the theory which should be determined by experiment, while its quantum corrections $\delta m^2$ can be calculated.
The problem however is, that these quantum corrections turn out to be huge. 
If one assumes that the Standard Model is a good theory of nature all the way up to an energy where gravitational effects become important, say the Planck scale $M_P \sim 10^{18}$ GeV,
this implies that $m_0$ should be chosen to be very big and very precise such that one can recover the measured value of the Higgs mass of about 125 GeV.
This a naturalness problem, in the sense that it seems unnatural\footnote{See also section~\ref{sec:Naturalness} on naturalness.} that $m_0$ should be fine-tuned so precisely to result in a low $m_{\text{measured}}$, 
and is regarded as a clear indication that the Standard Model is not a fundamental theory of nature.

\item If the recent measured values for the top quark mass of the Higgs masses are naively extrapolated to high scales, without introducing any new physics beyond the Standard Model,
it turns out that the electroweak vacuum is metastable, or even unstable~\cite{Instableweakvacuum}, depending on the result of future greater accuracy measurements for example for the mass of the top quark.
Even if experimental measurements turn out in favor of a metastable electroweak vacuum, this could still pose several questions. 
For example, if there exists another vacuum and our vacuum is not the true one, why is the vacuum energy in our vacuum so low? 
Or perhaps more importantly, why has not most of the universe already fallen into this lower energy state due to the large Higgs fluctuations in the early universe~\cite{Higgs_fluctuations_EW}?
These questions could be solved by dropping the assumption that the Standard Model remains valid at high energies, and point in the direction of the existence of new physics before the Planck scale.
\end{itemize}

For more information and a more complete list of problems with the Standard Model, including also for example the strong CP problem and the matter-antimatter asymmetry, see for example~\cite{Binetruy_Book}.

\section{Beyond the Standard Model}

There are many theories that extend the Standard Model and potentially solve at least a subset of the problems outlined before.
Perhaps one of the most appealing candidates is supersymmetry. \glossary{name=SUSY,description=supersymmetry}
Supersymmetry extends the symmetries of space and time (the Poincar\'e algebra) by adding extra transformations that relate the two kinds of fundamental particles, 
namely fermions which have a half-integer spin and bosons which have an integer spin.
Such theories typically predict the existence of supersymmetric particles, or sparticles in short.
Each particle in the Standard Model, would have a supersymmetric counterpart (superpartner) whose spin differs with 1/2 and whose mass is typically much larger than its Standard Model counter part. 
Although supersymmetric particles have yet to be discovered, certain theories like the Minimally Supersymmetric Standard Model (MSSM) 
\glossary{name=MSSM,description=Minimally Supersymmetric Standard Model}
include promising candidates that might be discovered within the next run of the LHC (Large Hadron Collider, see below) \glossary{name=LHC,description=Large Hadron Collider}
or a future collider.

Before we continue to introduce certain technical aspects involving supersymmetric theories, let us readdress the problems with the Standard Model outlined in the previous section in view of supersymmetry\footnote{
For a recent review and current experimental status of these points, see for example~\cite{Ellis_status_of_SUSY}.
}:
\begin{itemize}
 \item Supersymmetry provides a potential solution to the hierarchy problem. 
 For every fermion in a supersymmetric theory, there is a corresponding boson which is related by a supersymmetry transformation.
 Moreover, the invariance of the Lagrangian under supersymmetry transformations implies that for every fermion contribution to $\delta m^2$, 
 its bosonic superpartner gives exactly the same quantum correction to the mass of the Higgs boson, but with an opposite sign, which drastically reduces the value of $\delta m^2$ and 
 thus solves the hierarchy problem~\cite{Hierarchy1,Hierarchy2,Hierarchy4,Hierarchy5}.
This cancellation however, is only exact when supersymmetry is unbroken and the mass of each sparticle equals that of their Standard Model counter part.
 Recent experimental bounds are pushing lower limits on the sparticle masses upward, which reintroduces the question of naturalness (see below). 
 Although some new ideas are emerging~\cite{Axion_Hierarchy1,Axion_Hierarchy2}, none of them is yet as compelling as supersymmetry.
For a recent review on naturalness in supersymmetric theories, see for example~\cite{Feng:2013pwa,Baer:2014ica}.
 
 \item In order to prevent the proton in the theory from decaying, a discrete symmetry called R-parity is imposed on the MSSM.
 This R-parity implies that the Lightest Supersymmetric Particle (LSP) is stable. \glossary{name=LSP,description=Lightest Supersymmetric Particle}
 If additionally this LSP 
 also interacts very weakly with all the Standard Model particles, it is a natural dark matter candidate in the form of a WIMP (weakly interacting massive particle)~\cite{dark_matter_SUSY1,dark_matter_SUSY2}.
 \glossary{name=WIMP,description=Weakly Interacting Massive Particle}
 
 \item Perhaps one of the most convincing arguments in favor of supersymmetry is the apparent gauge coupling unification at a high energy scale. 
The gauge couplings of the three fundamental interactions of the Standard Model depend on the energy scale that they are probed, 
and their dependence on the energy scale can be described by a set of differential equations called the RGEs (Renormalization Group Equations). 
\glossary{name=RGE,description=Renormalization Group Equation}
When corrections to these RGEs from the particle content of the MSSM are included, 
they seem to intersect quite accurately (within experimental bounds) at a certain energy scale. 
This energy scale can then be interpreted as the energy scale of some Grand Unified Theory (GUT).
\glossary{name=GUT,description=Grand Unified Theory}

\item Although other solutions to the electroweak vacuum instability exist (see for example~\cite{InstableEW_dim61,InstableEW_dim62} based on dimension 6 operators),
supersymmetry also provides a solution to the stability of the electroweak vacuum~\cite{InstableEW_Ellis}.

\item Supersymmetry appears naturally in superstring theory, which is a possible candidate for a theory that quantizes gravity.

\item According to Coleman-Mandula~\cite{coleman-mandula} 
there are no nontrivial extensions of the Poincar\'e algebra, which are the symmetries of spacetime, that are consistent with a non-vanishing scattering amplitude.
However,the Haag–Lopuszanski–Sohnius theorem \cite{HaagLopuszanskiSohnius}
provides a loophole in this argument by allowing graded Lie algebras. 
In this case,  the possible symmetries of a consistent 4-dimensional quantum field theory do not only consist of internal symmetries and Poincaré symmetry, 
but they can also include supersymmetry as a nontrivial extension of the Poincaré algebra.
\end{itemize}

It is important to realize that supersymmetry is a mathematical framework, rather than a theory. 
A quantum field theory, like the Standard Model, is often defined by its Lagrangian. 
In a supersymmetric theory, this Lagrangian is not only invariant under the symmetries of spacetime and possible internal symmetries, but also under supersymmetry transformations.
Obviously, there are many different possible supersymmetric Lagrangians, and thus supersymmetric theories. 
The simplest such model that is consistent with the particle content and internal symmetries of the Standard Model is the MSSM.
However, even is we restrict ourselves to the MSSM,
the exact properties and experimental signatures of these supersymmetric theories are very dependent on the underlying supersymmetry breaking mechanism.
Acquiring a better understanding of supersymmetry breaking mechanisms and the potential underlying theories is therefore of great importance.

If supersymmetry were realized as an exact symmetry of nature, 
this would imply that the masses (as well as the couplings) of all particles would be equal to that of their superpartners. 
If this were the case, sparticles would have showed up in experiments already a long time ago.
One is therefore forced to conclude that supersymmetry can only be realized in nature as a broken symmetry, 
i.e.~there are terms present in the Lagrangian which violate the invariance under supersymmetry transformations. 
From a practical point of view, it can be useful to simply parametrize our ignorance of any underlying mechanism that breaks supersymmetry
by adding extra terms to MSSM Lagrangian that explicitly break supersymmetry.
The extra terms are called soft supersymmetry breaking terms,
where 'soft' means these terms are of a positive mass dimension and therefore it can be argued (see for example~\cite{SteveMartin}) that they
maintain a hierarchy between the electroweak and the Planck scale.
This way, there are no large quantum corrections ($\delta m$) to the Higgs mass and
 the theory to still provides a solution to the hierarchy problem, even in the presence of these extra terms that break supersymmetry.

These soft supersymmetry breaking terms however, are not very appealing from a theoretical point of view. 
While in quantum field theories one usually prefers to write down the most general Lagrangian based on certain symmetry principles, 
these terms are not only 'added by hand', they even violate the symmetries (supersymmetry) that we started with in the first place.
However, it might be possible that while supersymmetry is an exact symmetry of nature, it is broken spontaneously.
In other words, the Lagrangian of the underlying model is invariant under supersymmetry transformations, but the vacuum is not.
If one of the scalar fields in the MSSM were to be responsible for the spontaneous breaking of supersymmetry by acquiring a VEV (Vacuum Expectation Value),
\glossary{name=VEV,description=Vacuum Expectation Value}
this would additionally break at least one of the Standard Model gauge symmetries. 
It is therefore necessary to introduce additional fields to the model, which are responsible for spontaneous supersymmetry breaking.
These fields are in a so-called 'hidden sector', since it is assumed that these fields do not interact (or interact very weakly) with the Standard Model.
This leaves us with the question how supersymmetry is communicated from the hidden sector to the MSSM.

One such idea\footnote{ 
Other ideas include Gauge Mediation \cite{gauge_mediation1,gauge_mediation2}, where supersymmetry is mediated from a 'hidden sector' to the MSSM by gauge interactions, 
 or dynamical supersymmetry breaking~\cite{witten:1981nf} (or see \cite{Dynamical_SUSY_Breaking1,Dynamical_SUSY_Breaking2,Dynamical_SUSY_Breaking3} for more recent works)
 where fermion condensates of a strongly coupled gauge group are responsible for the breaking of supersymmetry. 
}   
is to promote supersymmetry to a local (gauge) symmetry. 
Theories with local supersymmetry necessarily contain gravity and are therefore called supergravity theories. \glossary{name=SUGRA,description=supergravity}
The idea is that besides the Standard Model particles and their superpartners, 
there also exists a 'hidden sector' of fields that does not interact with the Standard Model sector.
supersymmetry is then broken spontaneously in the hidden sector when a (hidden sector) scalar field acquires a VEV due to a nontrivial scalar potential, 
similar to the Brout-Englert-Higgs mechanism, which breaks the Standard Model electroweak gauge symmetry.
Next, gravitational effects are responsible for the communication of the supersymmetry breaking from the hidden sector to the visible (MSSM) sector. 
This can result in the necessary soft supersymmetry breaking terms of the MSSM, thus providing a framework for a possible origin for these terms.

On the other hand, certain supergravity theories arise as a low energy limit (where this is where the limit where the string length goes to zero) of string theory.
String theory, although it is still poorly understood, provides a quantum theory of gravity and is a candidate for a theory that unifies all interactions in a consistent quantum framework. 
It is therefore interesting whether such a supergravity theory can correspond to a certain string theory at a very high energy.

We conclude that trying to recover the MSSM from a supergravity theory can therefore be useful for two reasons: 
On the one hand, by breaking supersymmetry spontaneously one can obtain the desirable soft supersymmetry breaking terms, which in turn provides information on the low energy spectrum of this theory.
On the other hand, if one manages to identify this supergravity theory as the low energy limit of a certain string theory, 
this could give some useful insight in possible UV-completions of this theory. \glossary{name=UV, description= Ultra-Violet}

\section{de Sitter vacua}

In cosmology, the cosmological constant, usually denoted as $\Lambda$, is the value of the energy density of the vacuum in space.  
In the context of his theory of General Relativity, 
the cosmological constant was introduced by Albert Einstein in order to make to the theory consistent with the world view at that time. 
Since it was believed that we live in a static universe, an additional (constant) contribution $\Lambda$ was introduced to allow for static solutions of the Einstein equations.
This idea was abandoned however, when Georges Lema\^itre in 1927~\cite{lemaitre} and Edwin Hubble in 1929~\cite{hubble} discovered that we live in an expanding universe.
Since the 1990s, more and more experimental evidence confirmed that the universe is not only expanding, 
but it appears to be expanding at an accelerating rate\footnote{For a recent review, see for example~\cite{Frieman:2008sn}. For the latest measurements, see~\cite{Ade:2015xua}.}. 
This lead to the reintroduction of the cosmological constant.
It turns out that this cosmological constant is very small and positive ($\Lambda \approx (10^{-3} \text{eV})^4$), 
where a universe with a positive cosmological constant is called a de Sitter (dS)
\glossary{name=dS,description=de Sitter	}
\glossary{name=AdS,description=Anti-de Sitter} universe\footnote{
A vanishing cosmological constant would correspond to flat space (Minkowski), while a negative cosmological constant corresponds to an Anti-de Sitter (AdS) universe.
}.

In a quantum field theory without gravity, like the Standard Model, the absolute value of the minimum of the potential has no significance. 
A constant term can be added to the Lagrangian without changing the laws of physics that it describes.
However, this changes drastically when gravity is taken into account. 
In a theory containing gravity, like supergravity or string theory, 
the value of the minimum of the potential for the scalar fields is a contribution to the cosmological constant,
as it results in a similar constant contribution to the resulting Einstein equations.
It is still one of the biggest challenges of string theory at this date to accommodate for a positive cosmological constant.
Although interesting progress has been made in the last decade \cite{KKLT1,KKLT2,KKLT3,KKLT4,KKLT5,KKLT6,KKLT7,KKLT8,KKLT9}, 
the hunt for de Sitter spaces in string theory remains a challenging topic. 
At is therefore interesting to study supergravity theories which allow for a positive (and small) value of the minimum of the potential.

In this thesis we aim to obtain a supergravity theory with a tunably small and positive cosmological constant based on a minimal particle content and simple symmetry principles. 
We then relate such a theory to a string theory result that was obtained in the literature, 
and we continue to calculate the MSSM soft supersymmetry breaking terms and low energy spectrum 
that allows us to distinguish such a model from other models. 
In doing so, we find that anomaly cancellation conditions, which will be introduced in the next section, impose very stringent constraints on the consistency of such a theory.

\section{Quantum anomalies}

A quantum field theory, as was already mentioned above, is defined by its Lagrangian. 
Such a Lagrangian is usually constructed from symmetry principles: 
The Lagrangian, or rather its spacetime integral called the action, should be invariant under the symmetry transformations.
Sometimes however, it can occur that although the classical action is invariant under a symmetry, this symmetry is violated when quantum corrections are included.
In this case the symmetry is said to be anomalous.

An anomalous global symmetry of the Lagrangian usually does not pose a problem. 
It merely indicates that the classical selection rules related to this symmetry are not obeyed in the quantum theory. 
If these symmetries are local (gauge) symmetries on the other hand, 
this would be disastrous since the gauge symmetry is required in order to cancel the unphysical degrees of freedom with a negative norm.
In this case, the theory is said to be anomalous and is, in principle, useless. 
It is therefore of tremendous importance to check that a theory is anomaly-free. 
Typically, the anomalous contributions from various fermion triangle diagrams corresponding to all fermions in the theory can cancel each other,
provided that the anomaly cancellation conditions are satisfied.
Or, occasionally, counterterms can be introduced in order to cancel certain anomalies.

\section{Naturalness} \label{sec:Naturalness}

A good theory should provide a \emph{natural} explanation for the experimental data.
What is exactly meant with 'naturalness' is something many physicists disagree about.
However, in particle physics one could define naturalness as the property that 
the physical constants and parameters in the theory should not differ by too many orders of magnitude.
One example of a so-called naturalness problem is the gauge hierarchy problem, which concerns the question why 
the weak scale ($m_{\text{weak}} \sim 0.1 - 1$ TeV) is so far below the (reduced) Planck scale $m_p \approx 2.4 \times 10^{15}$ TeV.
Answering this question has been one of the guiding principles motivating proposals for new particles and interactions during the last decades.

The Higgs hierarchy problem, which as already introduced above, is that the quantum corrections to the Higgs mass ($\delta m^2$) are quadratically sensitive to the scale of new physics.
This scale can be the mass of a new (and yet undiscovered) particle, 
the UV cut-off scale after integrating out any new physics,
or even the Planck scale if gravity is included in a quantum theory.
The latter gives an alternative facet of the question why gravity is so weak compared to other forces.

Another example of a naturalness problem involves the cosmological constant.
It will be explained later in more detail how the cosmological constant in related to the vacuum energy in the theory.
Although particle physicists do not know how to calculate the vacuum energy density $\rho_{\text{vac}}$ exactly, 
theory allows one to estimate its value.
Depending on how this estimate is made\footnote{
For a more precise statement of the cosmological constant problem and on how this estimate is made, the reader is refered to the literature.
See for example~\cite{Carroll:1991mt}.}, its value differs between 40 to 120 orders of magnitude with its measured value.
This discrepancy is sometimes called 'The worst prediction in the history of physics'~\cite{Hobson:2006se}.
The cosmological constant problem can then be stated as follows: 
There are several independent contributions to the vacuum energy density (for example from the potential energy of each field in the theory, 
or from a bare cosmological constant in the Lagrangian).
Each of these contributions are estimated to be much larger than the observed value of the cosmological constant, 
and yet they combine to result in an uncanny small (measured) value.
This is widely seen as an indication that some unknown physics plays a decisive role.

The strong CP problem (CP standing for charge parity) is an example of a fine-tuning problem in physics. 
\glossary{name=CP,description=Charge Parity}
\glossary{name=QCD,description=Quantum Chromo Dynamics}
This problem has its origin in the fact that the symmetries of the Lagrangian describing the strong interactions 
a priori allow for a term that violates CP. However, since there is no experimentally known CP violation in the strong interactions, this term should be severely supressed (or even vanish).
The strong CP problem then questions why this CP-violating term vanishes (or is so small), while from a theoretical point of view there is no reason why this term should be absent.

It was already mentioned above that supersymmetry provides a solution to the Higgs hierarchy problem. 
Since bosons and fermions are related by supersymmetry transformations in the Lagrangian, their quantum corrections to the Higgs mass cancel out, which would in principle solve the Higgs hierarchy problem.
However, since the superpartners should be much heavier than their Standard Model counterparts to escape experimental bounds, supersymmetry must be a broken symmetry of nature (if it is a symmetry of nature at all).
This reintroduces corrections to the Higgs mass that depend on the sparticle masses. 
If these masses are too large, this results in a new fine-tuning problem, 
which questions the theoretical motivatons behind supersymmetry as a solution to the hierarchy problem.
While the current experiments are pushing the lower bounds of the sparticle masses upwards, this is indeed a problem that should be addressed.

Although numerous different definitions of naturalness exist in this context, it is commonly agreed that sparticle masses around the TeV-scale can still be considered as 'natural'.
A particular idea is that of 'split supersymmetry'~\cite{ArkaniHamed:2004fb}. 
In a split supersymmetry scenario, certain superpartners (the gauginos and higgsinos) are relatively light, while for example the squarks and gravitino are allowed to be heavier ($\sim 10$ TeV).
Although by most definitions of 'naturalness', such a scenario is considered to be 'more natural', split supersymmetry is not considered to be a solution to the Higgs hierarchy problem.

Supersymmetry, in particular supergravity, also has an impact on the cosmological constant problem. 
It was already explained above, and it will be quantified in later chapters, that the value at the minimum of the potential governing the scalar fields in the theory can be interpreted as the cosmological constant.
The cosmological constant problem can in this case be interpreted as the question why the value of the minimum of the potential is so incredibly small.
The scalar potential in supergravity has two contributions: 
A so-called F-term contribution which can be negative due to the presence of a contribution coming from the elimination of a gravity auxiliary field, 
and a D-term contribution which is positive (nonnegative).
A very small (and positive) cosmological constant can then be obtained by a nearly exact cancellation between the F-term and the D-term contributions of the scalar potential.
This results in a scenario where supersymmetry is broken exactly in the right way, such that the F-term and the D-term contributions cancel to a incredible accuracy.
However, this comes at the cost of a new and unexplained fine-tuning.

The fine-tuning problem of the cosmological constant has then been reduced to a new fine-tuning problem: What is the mechanism behind this miraculous cancellation between the F-terms and the D-terms?

The approach in this thesis is indeed of the former, where the minimum of the potential is tuned by a cancellation between an F-term contribution and a D-term contribution to the scalar potential.
The resulting sparticle spectrum will turn out to be similar to the spectrum of split supersymmetry.

Another approach has its origin in the fact that there are an extraordinary large amount of different vacua in string theory.
In each of these vacua, the low energy physics (and in particular the cosmological constant) is different.
Finding realistic vacua in the string theory landscape that correspond to the laws of physics as we known them  is a very contemporary problem.
Under the assumption that there exists at least one particular of those vacua, one could explain the smallness of the cosmological constant through an anthropic principle 
(which is liked as much as disliked among physicists)~\cite{Schellekens:2015cia}.

\section{Experiments}

Although this work is of a theoretical nature, most of it was carried out at CERN, \glossary{name=CERN,description=European Council for Nuclear Research} 
the European Council for Nuclear Research near Geneva, Switzerland, which is one of the biggest particle physics laboratories in the world. 
CERN hosts the Large Hadron Collider, which is undeniably one of the most impressive machines that was ever made: 
The LHC is constructed in a circular tunnel with a circumference of 27~km and about 100~m below the ground on the border between France and Switzerland.
Its goal is to accelerate protons to nearly the speed of light in two separate tubes, in opposite directions in this tunnel. 
When the two beams of protons reach an energy of about 7~TeV each, \glossary{name=TeV,description=Tera ElectronVolt}
they are collided inside four particle detectors.
It is then possible to look for possible signatures in the particle detectors that could indicate the existence of any new physics beyond the known Standard Model,
say, perhaps a supersymmetric particle..

\section{Overview of the thesis}

\subsection{Personal research work}

Our first goal is to obtain a supergravity model with an infinitesimally small, but positive, value of the minimum of the scalar potential.
It was argued above that in the presence of gravity this can be interpreted as the cosmological constant, which is indeed small and positive.
After such a model is obtained based on simple symmetry principles and a minimal particle content, our next goal is two-fold:
On the one hand we aim to find possible UV completions from which this model can be obtained.
In particular, we explain how such a potential can arise in the context of string theory.
On the other hand, this model is used as a hidden sector responsible for supersymmetry breaking,
which is then communicated to the visible sector (MSSM) through gravity effects.
We calculate the resulting soft supersymmetry breaking terms and the resulting low energy phenomenology and particle spectrum.
In particular, we explain how this model distinguishes itself from other scenarios.
Finally, we show how global symmetries of the MSSM can be identified with a gauge symmetry in our model
 and we work out the corresponding implications on the phenomenology of this model.

Our main result includes a model in work with I.~Antoniadis~\cite{RK1}, which was originally proposed in~\cite{Zwirner1,Zwirner2},
based on a single chiral multiplet $S$ with a gauged shift symmetry.
The Lagrangian of a supergravity theory is determined by three functions, namely a K\"ahler potential, a superpotential and gauge kinetic functions.
The shift symmetry of the dilaton implies that the K\"ahler potential of the theory should be a function of the real part of $S$, 
which is taken to be a logarithm, where in the context of string theory $S$ can then be identified with the heterotic string dilaton or a compactification modulus.
The most general superpotential consistent with the shift symmetry is an exponential of $S$, and the gauge kinetic function can be at most linear in $S$. 
In the case of an exponential superpotential, the shift symmetry becomes a gauged R-symmetry which results in a (positive) D-term contribution to the scalar potential. 
This way, the F-term contribution to the scalar potential 
(which can be negative because of a contribution from a gravity auxiliary field) can be fine-tuned with respect to the positive D-term contribution 
in order to allow for a very small and positive value of the minimum of the potential.
The resulting model allows for a tunable small (and positive) cosmological constant, 
while keeping the gravitino mass parameter and thus the supersymmetry breaking scale as an independent parameter.

The resulting model consists of one complex scalar field whose imaginary part becomes the longitudinal component of the $U(1)$ gauge boson, 
which in turn becomes massive  due to a St\"uckelberg mechanism, 
 a gravitino which becomes massive due to a super-Brout-Englert-Higgs mechanism by eating the Goldstino, 
 which in turn is a linear combination of the chiral fermion (the superpartner of the complex scalar $s$)
and the R-gaugino (the superpartner of the $U(1)$ gauge boson), and one physical fermion which is orthogonal to the Goldstino.
It is however known that for theories with a shift symmetry there exists an alternative formulation in terms of a 2-index antisymmetric tensor which is dual to Im$S$ (under Poincar\'e duality of its field strength).
In this work we provide a dual formulation of the model introduced above, which is to our knowledge the first dual formulation of a model with a field dependent superpotential. 
The importance of this result lies in the fact that the  2-form  field  appears naturally  in  the  string  basis.
Another interesting question is whether this model can be obtained in the context of string theory.
Indeed, it turns out~\cite{RK1} that the scalar potential can be identified with the one in~\cite{Antoniadis:2008uk} derived from D-branes in non-critical strings. 

It is important to check whether the new gauge symmetry that was introduced, namely the gauged shift symmetry (which is an R-symmetry when the superpotential is an exponential of $S$),
is not violated at the quantum level.
The work by Elvang, K\"ors and Freedman~\cite{F1,F2}, 
on quantum anomalies in supergravity theories with gauged R-symmetries, 
was extended, in work with I.~Antoniadis and D.~Ghilencea~\cite{RK2}, and their final result can be interpreted in from a field-theoretic point of view.   
We show how in the above model, the cubic anomalies, corresponding to the gauged shift symmetry, 
generated by the corresponding triangle diagrams are canceled by a Green-Schwarz mechanism. 
It turns out that the requirement of gravitino mass to be in the TeV range, in combination with the anomaly cancellation conditions,
severely restricts the allowed parameter range.
There exist certain choices for the parameters however, 
for which the model is consistent and can be successfully used as a hidden sector supersymmetry breaking sector coupled to the MSSM. 

The anomaly-consistent model was analyzed in full detail in work with I.~Antoniadis~\cite{RK3}.
Here, it was shown that when this model is coupled to the MSSM, 
these fields suffer from a negative scalar mass squared. 
This problem can however be avoided by introducing a non-canonical K\"ahler potential for the Standard Model fields, or by including an extra Pol\'onyi-like field. 
The model is then coupled to the MSSM, 
leading to calculable soft supersymmetry masses and a distinct low energy phenomenology that allows to differentiate it from other models of supersymmetry breaking.

Finally, in work with I.~Antoniadis~\cite{RK4}, 
the gauged shift symmetry is identified with a global symmetry of the Standard Model (or rather, the MSSM) and the corresponding phenomenology is worked out in full detail. 
One particularly interesting possibility is to identify the gauged shift symmetry with a combination of baryon and lepton number that contains the known matter parity (or R-parity) of the MSSM,
and guarantees the absence of dimension-four and five operators that violate baryon and lepton number, 
implying proton stability and smallness of neutrino masses at phenomenologically acceptable levels.

\subsection{Overview of the following chapters}

In chapter \ref{ch:supersymmetry} we introduce supersymmetry along with the necessary conventions and notation which will thoroughly be used in the following chapters. 
In particular, we will focus on the necessary ingredients to build a supersymmetric theory: 
The particle content in the form of supermultiplets on the one hand, and the K\"ahler potential, the superpotential and gauge kinetic functions which define the supersymmetric Lagrangian on the other hand. 
Next, we introduce the MSSM. 
We focus on its particle content, its superpotential and the soft supersymmetry breaking terms. 
We close this chapter by introducing R-parity, which in the MSSM is sufficient to guarantee the stability of the proton. 
We show its relation with the global symmetry $B-L$ (baryon minus lepton number) \glossary{name=B-L, description=Baryon minus Lepton number}
of the MSSM and show how certain (baryon and lepton number violating) dimension 5 operators are still allowed.
  
In chapter \ref{ch:supergravity} we introduce supergravity theories, which are theories with local supersymmetry. 
We readdress the K\"ahler potential, the superpotential and gauge kinetic functions in this context, and we introduce
the scalar potential. We comment on the importance of R-symmetries and K\"ahler transformations in supergravity theories. 
Next, we discuss spontaneous supersymmetry breaking, where we focus on the super-Brout-Englert-Higgs mechanism. \glossary{name=super-BEH,description=super-Brout-Englert-Higgs} 
We elaborate on hidden sector models where supersymmetry is mediated to the visible sector (MSSM) 
due to gravity effects and we relate this discussion to the soft supersymmetry breaking terms of the MSSM.

The presentation of our own work starts in chapter \ref{ch:dS}, where a model based on a gauged shift symmetry is introduced. 
We discuss in full detail for which values of the parameters a (meta)stable de Sitter vacuum exists 
and how these parameters can be tuned to obtain a small but positive cosmological constant. 
For particular values of the parameters, the scalar potential can be obtained in the context of string theory. 
Finally, we provide a dual description of this model in terms of a linear multiplet.
This is the first time a such a duality has been performed in the presence of a nontrivial superpotential.
This chapter is mainly based on work with I.~Antoniadis~\cite{RK1}.

In chapter~\ref{ch:anomalies}, we briefly review the work of Elvang, K\"ors and Freedman~\cite{F1,F2} on anomalies in supergravity theories with Fayet-Iliopoulos terms.
\glossary{name=FI,description=Fayet-Iliopoulos}
In our results, based on work with I.~Antoniadis and D.~Ghilencea~\cite{RK2}, 
their anomaly cancellation conditions are rewritten in a form that is more intuitive from a ``naive`` field theory point of view.
These results are then applied to the above model. It turns out that the anomaly cancellation conditions force the gravitino mass parameter to exceed the Planck scale for a certain value of the parameters in the model.
Finally, We introduce the necessary notation concerning Green-Schwarz counterterms and how the gaugino mass parameters can be generated at one loop, 
either due to the Green-Schwarz counterterms, or due to an effect called 'anomaly mediation'.

In chapter~\ref{ch:softterms}, we calculate the soft supersymmetry breaking terms for the parameters where this model is anomaly-free. 
It is shown that in the minimal case the model suffers from negative soft scalar masses squared. 
These can however be cured by introducing another (Pol\'onyi-like) field or by allowing non-canonical kinetic terms for the MSSM superfields,
while the desirable properties of the model such as a tunably small and positive cosmological constant and a TeV gravitino mass remain intact.
We discuss the low energy particle spectrum of this model and show how it can be distinguished from other minimal scenarios.

In chapter~\ref{ch:chargedMSSM}, we demonstrate how the shift symmetry of the model can be identified with a known global symmetry of the MSSM. 
We work out the particularly interesting case where this symmetry is a combination of baryon and lepton number, 
since in this case it reduces to the known matter parity (or R-parity) of the MSSM.
Following, the phenomenology of this model is worked out in full detail, 
and it is compared with the case of the previous chapter where the Standard Model fields are inert under the shift symmetry.
Finally, we demonstrate how a third possibility to avoid the tachyonic masses based on a D-term contribution 
proportional to the charge of the MSSM fields under the shift symmetry does not lead to a viable electroweak vacuum.

Our conclusions are listed in chapter~\ref{ch:conclusions}.

The masses of the hidden sector scalars, fermions and gauge field are calculated in Appendix~\ref{Appendix:masses}.
In Appendix~\ref{ch:appendix_linear_chiral} we present the full details of the calculation involving the linear-chiral duality in section~\ref{sec:linearchiral}.
In Appendix~\ref{ch:Appendix_anomalies} 
we present the anomaly cancellation conditions for models with a gauged R-symmetry from a field-theoretic point of view. 
Finally, in Appendix~\ref{ch:Appendix_p1} 
we comment on the certain values of the parameters that were ignored in the rest of the work.

\cleardoublepage
  \graphicspath{{\chaptersdir/supersymmetry/image/}}%
  \chapter{Supersymmetry}\label{ch:supersymmetry}

In this chapter we introduce supersymmetry. Supersymmetry is, as was already explained in the introduction, perhaps one of the most promising theories that extends the Standard Model.
From a technical point of view it provides the only 'loophole' in the Coleman-Mandula theorem \cite{coleman-mandula}, 
which under very reasonable assumptions prohibits the symmetries of spacetime (the Poincar\'e algebra) and internal symmetries from being extended in a nontrivial way.
The Haag-Lopuszanski-Sohnius theorem \cite{HaagLopuszanskiSohnius} evades this problem by allowing for so-called graded Lie algebras. 
Indeed, they demonstrate that supersymmetry allows the symmetries of spacetime  and internal symmetries to be combined consistently\footnote{
In fact, for theories including only massless fields, the Coleman-Mandula theorem allows for an extension involving the conformal group, 
and the Haag-Lopuszanski-Sohnius theorem allows the superconformal group.
}:
The Poincar\'e algebra is extended with supersymmetry generators $Q_\alpha$, which obey the (anti-)commutation relations
\begin{align}
 \left\{Q_\alpha, Q^\beta \right\} &= - \frac{1}{2} \left( \gamma_\mu \right)_\alpha^{\  \beta} P^\mu, \notag \\
 \left[M_{\mu \nu}, Q_\alpha \right] &= - \frac{1}{2} \left(\gamma_{\mu \nu} \right)_\alpha^{\  \beta} Q_\beta, \notag \\
 \left[P_\mu, Q_\alpha \right] &=0. \label{SUSY_algebra}
\end{align}
Here (and below), Greek indices in the middle of the alphabet ($\mu,\nu$) indicate spacetime indices, where the metric is assumed to be 'mostly plus', 
i.e. $\eta_{\mu \nu}= \text{diag}(-+++)$, 
and we assume four spacetime dimensions.
Greek indices in the beginning of the alphabet ($\alpha,\beta$) are two-index spinor indices which will be dropped whenever possible\footnote{
Below, Greek indices $\alpha,\beta$ will instead label the various chiral supermultiplets in a theory, see section \ref{sec:ingredients}.}, 
$P^\mu$ is the generator for spacetime translations, and $M_{\mu \nu}$ is the generator for Lorentz transformations.
The squared brackets stand for commutation relations, while curly brackets stand for anti-commutators. The $\gamma-$matrices are defined in~\cite{TheBook}.
Although, several copies of the supersymmetry generators (extended supersymmetry) are allowed, i.e. $Q_{i\alpha}, \ i=1\dots \mathcal N$, 
we limit ourselves for phenomenological reasons to so-called simple supersymmetry where there is only one copy of the supersymmetry generators ($\mathcal N = 1$). 

Although supersymmetry is a very interesting and technical subject, we will introduce merely its aspects that will turn out important for our work. 
For a more detailed introduction to supersymmetry, the reader is referred to standard textbooks, such as~\cite{TheBook} whose notation will be used throughout this thesis.

In section \ref{sec:ingredients} we introduce our notation and the basic ingredients that are needed to construct a supersymmetric theory. 
In particular, we introduce the various multiplets and discuss how a supersymmetric Lagrangian is constructed by specifying a K\"ahler potential, a superpotential and gauge kinetic functions. 

In section \ref{sec:MSSM} we introduce the MSSM, with in particular its particle content and the soft supersymmetry breaking terms that parametrize the breaking of supersymmetry.
We introduce global R-symmetries in supersymmetric theories, and a discrete R-parity that is usually imposed on the MSSM to ensure proton stability.
It is shown how R-parity can be reformulated in terms of another discrete symmetry called matter parity, and how it is related to certain global symmetries of the MSSM, such as Baryon minus Lepton number ($B-L$).

\section{The basic ingredients of a $\mathcal N=1$ supersymmetric theory} \label{sec:ingredients}

\subsection{Multiplets}

Before we can present the basic ingredients of a supersymmetric theory, we first specify its particle content. 
Since supersymmetry relates bosons and fermions, they appear together in so-called supermultiplets\footnote{
In the literature, a superfield method is very often used to combine bosons and fermions in a single superfield by the use of anti-commuting Grassmann variables.
Since some concepts, as for example R-symmetry can be interpreted more conveniently in this formalism, 
a superfield method will be used in certain appendices. We will however postpone the necessary notation involving the superfield method to appendices 
and refer the reader to standard textbooks, as for example~\cite{wessbagger}.}.
A supermultiplet is a set of boson and fermion fields which transform among themselves under supersymmetry transformations.
In this thesis however, we will not introduce the supersymmetry transformations, and we merely list the various multiplets.
For the SUSY variations and an introduction to supermultiplets the reader is referred to standard textbooks, such as \cite{TheBook}.

A chiral multiplet consists of a complex scalar $Z(x)$, a Majorana spinor $\chi(x)$ and a complex scalar auxiliary field $F(x)$. 
The complex scalar field $F$ is auxiliary in the sense that the Lagrangian does not contain any kinetic terms for $F$ such that it can therefore be eliminated by its equations of motion. 
The field $F$ does not correspond to a physical field. 
Its elimination by its equations of motion results in a potential for the physical scalars in the theory, called the scalar potential, which will be very relevant in later chapters.
The complex conjugate of a chiral multiplet is an anti-chiral multiplet.
We will denote a (anti-)chiral multiplet as
\begin{align} \text{Chiral multiplet:}& \ \ \ \ \left(Z,\chi,F \right), \notag \\  \text{Anti-chiral multiplet:}& \ \ \ \ \left( \bar Z, \bar \chi,\bar F \right). \label{chiral_multiplet}  \end{align}

A gauge (or vector) multiplet consists of a spin-1 gauge boson $A_\mu(x)$, a spin-$1/2$ fermion, called a gaugino, described by a Majorana spinor $\lambda(x)$ (or the corresponding Weyl field $P_L \lambda$), and a real scalar auxiliary field $D(x)$.
As above, the kinetic terms of the auxiliary field $D$ are absent in the Lagrangian and it can be eliminated by its equations of motion, which results in another contribution to the scalar potential.
A gauge multiplet is denoted as
\begin{align} \text{Gauge multiplet:} \ \ \ \ \left(A_\mu, \lambda, D \right). \label{gauge_multiplet}\end{align}

A third (and lesser known) multiplet is a linear multiplet. 
These are usually omitted in the literature since due to a linear-chiral duality (see section \ref{sec:linearchiral}) a theory containing a linear multiplet can be rewritten in terms of a chiral multiplet. 
A linear multiplet consists of a real scalar field $l(x)$, a Majorana fermion $\chi(x)$ and divergenceless\footnote{
In a superspace formalism, in the conventions of \cite{wessbagger}, a linear multiplet can be obtained by imposing the constraints 
\begin{align} D^2 L = \bar D^2 L = 0, \notag \end{align}
 on a general superfield, where $\bar D^2$ and $D^2$ are the chiral and anti-chiral projection operators.} 
field strength $v_\mu (x)$, satisfying 
\begin{align} \partial^\mu v_\mu = 0. \label{linear:vmu_constraint} \end{align} 
It follows that  $v_\mu$ can be rewritten in terms of an antisymmetric tensor $b_{\mu \nu}$ by a Poincar\'e duality
\begin{align} v^\mu = \frac{1}{\sqrt 6} \epsilon^{\mu \nu \rho \sigma} \partial_\nu b_{\rho \sigma}, \label{linear:vmu} \end{align} 
which has a gauge symmetry 
\begin{align} b_{\mu \nu } \longrightarrow b_{\mu \nu} + \partial_\mu b_\nu - \partial_\nu b_\mu. \label{linear:bgauge} \end{align}
A linear multiplet is denoted as 
\begin{align}  \text{Linear multiplet:} \ \ \ \ \left( l, \chi, v_\mu \right), \ \ \ \ \text{with} \ \ \partial^\mu v_\mu = 0. \label{linear_multiplet} \end{align}

\subsection{How to construct a supersymmetric Lagrangian} 

For a certain set of chiral multiplets (labeled by a Greek index $\alpha, \beta, \dots$), 
and their respective gauge transformations with the corresponding gauge multiplets (labeled by capital Roman indices $A,B,\dots$), 
a supersymmetric Lagrangian can be constructed in terms of the following functions of the scalar component fields of the chiral multiplets:
\begin{itemize}
\item A K\"ahler potential $\mathcal K(Z,\bar Z)$, which is a gauge invariant function of the scalar components of the chiral multiplets and their conjugates, 
gives rise to the kinetic terms for the components of the chiral multiplets. 
\item A gauge invariant superpotential $W(Z)$, which is a holomorphic function of the scalars and therefore only depends on $Z$, and not on $\bar Z$. The superpotential contributes to the various interaction terms as well as the scalar potential.
\item Gauge kinetic functions $f_{AB}(Z)$, whose (among other contributions) real component $\text{Re}f_{AB}$ multiplies the kinetic terms of the gauge bosons, the gauginos and the $D$-auxiliary fields. 
\end{itemize}

The Lagrangian is given by
\begin{align} \mathcal L &= \mathcal L_{\text{kin,chir}} + \mathcal L_{\text{kin,gauge}} + \mathcal L_{\text{pot,chir}}, \label{SUSY_Lagrangian} \end{align}
where $\mathcal L_{\text{kin,chir}}$ and $\mathcal L_{\text{kin,gauge}}$ include the kinetic terms for the component fields of the chiral multiplets and the gauge multiplets respectively,
and $\mathcal L_{\text{pot,chir}}$ includes the interaction terms.
For the full Lagrangian, together with the relevant conventions and notation, the reader is referred to the literature (see for example eq.~(14.57) in~\cite{TheBook}).
Below, we focus only on the resulting scalar potential.

As was already anticipated in the previous section, 
the auxiliary fields $F^\alpha$ and $D^A$ in the supersymmetric Lagrangian (\ref{SUSY_Lagrangian}) do not appear with kinetic terms and can therefore be eliminated by their equations of motion.
After the elimination of the auxiliary fields, one obtains (among other terms) a potential for the scalar fields in the theory. The scalar potential is given by
\begin{align}  \mathcal V = g^{\alpha \bar \beta} W_\alpha \bar W_{\bar \beta} + \frac{1}{2} (\re f)^{-1 \ AB} \mathcal P_A \mathcal P_B, \label{SUSY:scalarpot} \end{align}
where $W_\alpha$ is short for $\frac{\partial W}{\partial Z^\alpha}$ and the K\"ahler metric is given by
\begin{align} g_{\alpha \bar \beta} = \partial_\alpha \partial_{\bar \beta} \mathcal K(Z, \bar Z).\end{align}
The moment maps are 
\begin{align}\mathcal P^A = i k_A^\alpha \mathcal K_\alpha + p_A, \label{SUSY:moment maps} \end{align}
where $k_A^\alpha$ are the Killing vectors,
which appear in the gauge transformations (with gauge parameters $\theta^A$) of the scalar fields
\begin{align} \delta Z^\alpha = \theta^A k_A^\alpha (Z), \label{Killing_vector} \end{align}
and the $p_A$ are the so-called Fayet-Iliopoulos (FI) constants~\cite{FI}. 
These constants turn out very useful to break supersymmetry and have their origin from the fact that the integral $S_{\text{FI}} = - \int d^4 x \, p_A D^A$ 
is invariant under supersymmetry transformations as well as gauge transformations, provided that the gauge symmetry $A$ is abelian,
and can therefore be freely added to the theory for any constant $p_A$.
This is however in contrast with theories with local symmetry in the next chapter, 
where the value of the Fayet-Iliopoulos constant parameter is severely constrained. 

Notice that the scalar potential~(\ref{SUSY:scalarpot}) satisfies $\mathcal V \geq 0$. 
An important question arises whether the symmetries of the action are broken in the vacuum. 
If the broken symmetry is a global symmetry, then according to the Goldstone theorem the theory contains a massless scalar particle.
In case the broken symmetry is a gauge symmetry, the Goldstone boson will disappear from the spectrum, and, by the Higgs mechanism, a spin-1 gauge boson becomes massive.

If supersymmetry would be a symmetry of nature, it can only appear as a broken symmetry (see next section).
Although we will not present a proof in this work, 
it can be shown that supersymmetry is broken if and only if the minimum  of the potential is non-zero, $\mathcal V_{\text{min}} >0$ \cite{secretsymmetry}. 
In this case, there is a massless fermion called a Goldstino in the spectrum.

In the next chapter, we show that in a theory with local supersymmetry there is an additional negative F-term contribution to the scalar potential which has its origin from a gravity auxiliary field. 
This way, the scalar potential can have negative values. 
Moreover, this allows one to construct theories with a vanishing minimum of the scalar potential, while supersymmetry is spontaneously broken even though the vacuum energy vanishes.
Since supersymmetry is a local symmetry,  the massless Goldstino disappears from the spectrum by the super-Brout-Englert-Higgs mechanism, and the spin-$3/2$ gravitino becomes massive 
(see section~\ref{sec:super-BEH}).

\section{MSSM} \label{sec:MSSM}

Supersymmetry provides a mathematical framework to construct theories that have desirable properties,
as for example solving the hierarchy problem or including a dark matter candidate in the sparticle spectrum.
The supersymmetric theory with the minimal particle content that includes the Standard Model is the MSSM \cite{MSSM1,MSSM2}\footnote{For more recent reviews, see for example \cite{MSSMreview1,MSSMreview2}.}.
In this section we review the basic structure of the MSSM, its particle content and the superpotential. 
Following, we introduce R-symmetries, which are internal symmetries that do not commute with supersymmetry.
Finally, we discuss a discrete symmetry called R-parity, which is usually imposed on the MSSM to exclude terms in the Lagrangian that may lead to proton decay.
We show how one can equivalently impose matter parity to prohibit the same terms from the Lagrangian, and we discuss its relation with known global symmetries of the MSSM.

\subsection{Structure of the MSSM}

The MSSM is the minimal supersymmetric model that includes the Standard Model.
An obvious question arises to identify the supermultiplets in which these Standard Model fields reside.
Since all components of a supermultiplet carry the same quantum numbers, the Standard Model fermions can not reside in the same multiplet with any of the Standard Model gauge bosons.
Instead, the Standard Model fermions are the fermionic components of a chiral multiplet, 
together with their scalar superpartners, called sfermions\footnote{
For example, the superpartner of the electron is a selectron and the superpartner of a top quark is called a stop squark.},
 which have the same quantum numbers corresponding to the Standard Model gauge groups.
The Standard Model gauge bosons on the other hand are the spin-1 components in a gauge multiplet. Their fermionic superpartners are called gauginos\footnote{
In particular, the fermionic superpartners of the Standard Model gauge bosons are called photino, bino, wino and gluiono.}.
However, in contrast with the Standard Model which contains a single Higgs doublet, 
the MSSM contains two Higgs chiral supermultiplets\footnote{A single chiral multiplet containing the Higgs boson would lead to quantum anomalies.} called $H_u$ and $H_d$.

The chiral multiplets together with their Standard Model quantum numbers are summarized in table \ref{table:MSSM}.
Here, $Q$ stands for the $SU(2)_L$-doublet chiral multiplet containing the left-handed up quark  (or since family indices are suppressed in $Q_i,\ \bar u_i,\ \bar d_i, \ \bar e_i, \ L_i$, it could represent the multiplet containing the charm or the top quark),
$\bar u$ ($\bar d$) stands for the $SU(2)_L$-singlet chiral multiplet containing $u^\dagger_R$ ($d^\dagger_R$). $L$ are the lepton doublets and $\bar e$ are the antilepton singlets.
\begin{table}[h]
  \begin{tabular}{l| lllllll}
	  & $Q$   & $\bar u$        & $\bar d$        & $L$    & $\bar e$ & $H_u$ & $H_d$ \\ \hline
   $U(1)_Y$ & 1/6 & -2/3     & 1/3      & -1/2 & 1 & 1/2   & -1/2  \\ 
  $SU(2)$   & 2   & 1        & 1        & 2    & 1 & 2     & 2     \\
  $SU(3)$   & 3   & $\bar 3$ & $\bar 3$ & 1    & 1 & 1     & 1     
  \end{tabular} 
  \caption{The MSSM particle content and their representations in the Standard Model gauge groups. The bar on $\bar u, \bar d, \bar e$ is part of the name and does not represent any sort of complex conjugation. Family indices $i$ are surpressed.} \label{table:MSSM}
  \end{table} 

It is important to note that the above superpartners are not necessarily the mass eigenstates of the theory. 
For example, there is a mixing between the electroweak gauginos and the Higgsinos when electroweak symmetry breaking (and supersymmetry breaking) effects are included.
Similarly, mass mixing can occur between squarks, sleptons and Higgs scalars that carry the same electric charge. Such masses and mixing of the superpartners contain very valuable information for the experimentalists.
We will however not discuss this here, and refer to standard textbooks~\cite{SteveMartin,Drees} instead.

The K\"ahler potential is assumed to be canonical, i.e. for the MSSM chiral superfields $\varphi_\alpha$, one has $\mathcal K(\varphi, \bar \varphi) = \sum_\alpha \varphi_\alpha \bar \varphi_\alpha$.
The MSSM superpotential is given by 
\begin{align} W_{MSSM}= y_u^{ij} \bar u_i Q_j \cdot H_u - y_d^{ij} \bar d_i Q_j \cdot H_d - y_e^{ij} \bar e_i L_j \cdot H_d + \mu H_u \cdot H_d, \label{MSSMsuperpot} \end{align}
where the family indices $i,j$ are explicitly written and all gauge indices are surpressed. The Yukawa matrices $y_u, y_d$ and $y_e$ are $3 \times 3$ matrices in family space.
The last term in eq.~(\ref{MSSMsuperpot}) is traditionally called the $\mu$-term and should be read out as $\mu (H_u)_\alpha (H_d)_\beta \epsilon^{\alpha \beta}$,
where $\epsilon^{\alpha \beta}$ is a two-index antisymmetric tensor used to tie together $SU(2)_L$ isospin indices.
Likewise, the first term is short for $\bar u^{ia} (y_u)_i^{ \ j} Q_{j\alpha a} (H_u)_\beta \epsilon^{ \alpha \beta}$ (and similarly for the second and third term), where $a$ is the color index.

The gauge kinetic functions are assumed to be constant and inversely proportional to the square of their respective gauge couplings, $f_A = 1/g_A^2$.
With the particle content of the MSSM in table~\ref{table:MSSM}, together with the K\"ahler potential, the superpotential and gauge kinetic functions above,
one can in principle now write down the MSSM Lagrangian $\mathcal L_{\text{SUSY}}$.
However, as we are not concerned with the exact form of $\mathcal L_{\text{SUSY}}$ in this thesis, the reader is referred to standard textbooks such as~\cite{SteveMartin} or \cite{Drees} for a more complete treatment of the MSSM Lagrangian.

\subsection{Supersymmetry breaking in the MSSM} \label{sec:softterms}

In the previous section we introduced the particle content of the MSSM.
It was already mentioned that the various components fields of a supermultiplet share the same representations of the (Standard Model) gauge groups. 
Moreover, the invariance of the Lagrangian $\mathcal L_{\text{SUSY}}$ under supersymmetry transformations also implies that the different components of a multiplet have the same mass.
If this were the case, this would imply that for example the selectron, the scalar superpartner of the electron has a mass of about $0.511$ MeV (of course,  similar statements are true for the squarks, gluinos, photinos, $\dots$).
Such a scalar particle is very easy to detect, and should have shown up in collider experiments if it existed. 
Since no superpartners were found yet, one can conclude that if supersymmetry is a symmetry is nature, it has to be a broken symmetry.

In practice, one usually 'manually' adds soft supersymmetry breaking terms~\cite{softterms} to the supersymmetric MSSM Lagrangian. 
The effective Lagrangian of the MSSM is given by
\begin{align}
 \mathcal L = \mathcal L_{\text{SUSY}} + \mathcal L_{\text{soft}}.
\end{align}
Here, $\mathcal L_{\text{SUSY}}$ is the Lagrangian of the MSSM defined above, which is invariant under supersymmetry transformations. The possible soft supersymmetry breaking interactions are\footnote{
Tadpole contributions $ t^i \phi_i $ to the soft supersymmetry breaking Lagrangian are not included in eq.~(\ref{softterms}) since they require the existence of a gauge singlet in the MSSM.}
\begin{align}
  \mathcal L_{\text{soft}} =& - \left( \frac{1}{2} M_A \lambda^A \lambda^A + \frac{1}{6} a^{ijk} \phi_i \phi_j \phi_k +\frac{1}{2}  b^{ij}\phi_i \phi_j  \right) + \text{c.c.}  \notag \\
&- (m^2)^i_{\ j} \phi^{j\dagger} \phi_i, \label{softterms}
\end{align}
where $M_A$ are the mass parameters for the gauginos $\lambda^A$ for each gauge group. The scalar fields of the MSSM are labeled by $i,j,k$, $a^{ijk}$ are the trilinear couplings, 
$b^{ij}$ is the 'B$\mu$-term' and $(m^2)^i_{\ j}$ are the scalar soft masses squared.
The terms in $\mathcal L_{\text{soft}}$ are soft in the sense that they have a positive mass dimension and their presence does not spoil supersymmetry as a potential solution to the hierarchy problem.

Without further justification, the addition of the soft supersymmetry breaking terms to the Lagrangian might seem anything but elegant and a rather arbitrary requirement. 
However, as was already mentioned before, supersymmetry can be spontaneously broken if one of the scalar fields acquires a VEV in an appropriate scalar potential.
While supersymmetry is spontaneously broken it can be shown that, given an appropriate scalar potential, the soft supersymmetry breaking terms (\ref{softterms}) appear after the scalar fields responsible for supersymmetry breaking are replaced with their VEVs.
This scalar field can however not be the superpartner of one of the Standard Model fermions, since this would additionally break some of the Standard Model gauge symmetries.
It is therefore required to introduce an additional 'hidden sector' sector, whose scalar fields are responsible for supersymmetry breaking. 
These fields are usually singlets under the Standard Model gauge groups and do not communicate with the Standard Model at low energies.

\subsection{R-symmetry} \label{sec:R-symmetry}

Many supersymmetric theories are invariant under a (global) chiral $U(1)$ symmetry called R-symmetry (which is consequently denoted by $U(1)_R$). 
An R-symmetry is characterized by the fact that (at the level of the algebra) its generator $T_R$ does not commute with the supersymmetry generators.
Instead, it acts on the supersymmetry generators $Q_\alpha$ as 
\begin{align}
 \left[T_R, Q_\alpha \right] &= -i \left(\gamma_* \right)_\alpha^{\ \beta} Q_\beta, \label{algebra:R-symmetry}
\end{align}
where $\gamma_*$ is given by $\gamma_* = i \gamma_0 \gamma_1 \gamma_2 \gamma_3$.
This has the consequence that, in contrast with other internal symmetries, 
like the Standard Model gauge groups under which all components of a chiral multiplet carry the same charge, 
an R-symmetry acts differently on each component of a chiral multiplet.
The scalar component $Z$, the fermionic component $P_L \chi$ and the auxiliary field $F$ carry R-charges $r,\ r_\chi=r-1,\ r_F=r-2$ respectively.
The transformations under  $U(1)_R$ are summarized as
\begin{align}  \delta_R Z &= i \rho r Z, \notag \\  \delta_R P_L \chi &= i \rho (r-1) P_L \chi \notag \\  \delta_R F &= i \rho (r-2) F. \label{R-transformations} \end{align}
In global supersymmetry there is an interesting connection between theories which have a global R-symmetry and supersymmetry breaking \cite{Nelson:1993nf}. 
It should also be noted that by eq.~(\ref{algebra:R-symmetry}) in global supersymmetry an R-symmetry can not be gauged. 
This is not the case however in theories with local supersymmetry.
Moreover, in local supersymmetry a gauged R-symmetries implies the presence of a Fayet-Iliopoulos D-term contribution to the scalar potential (see section~\ref{sec:sugra_ingredients}).

\subsection{R-Parity}

Although in the literature R-parity is usually introduced as a discrete R-symmetry, 
we point out in this section that R-parity can be formulated in such a way that it not an R-symmetry at all (in the definitions of the previous section).

The most general gauge-invariant and renormalizable superpotential for the MSSM would not only include the usual terms eq.~(\ref{MSSMsuperpot}), but also the following baryon- and lepton-number violating interactions
\begin{align} W_{\Delta L =1} &= \frac{1}{2} \lambda^{ijk} L_i L_j \bar e_k + \lambda'^{ijk } L_i Q_j \bar d_k + \mu'^i L_i H_u, \notag \\ W_{\Delta B =1} &= \frac{1}{2} \lambda''^{ijk}\bar u_i \bar d_j \bar d_k. \label{deltabl} \end{align}
The chiral superfields carry baryon number $B = 1/3$ for $Q_i$, $B=-1/3$ for $\bar u_i, \bar d_i$ and $B=0$ for all the others. 
Similarly, $L_i$ and $\bar e_i$ carry lepton number $+1$ and $-1$, respectively, while all other superfields have vanishing lepton number. 
Since the baryon and lepton number violating processes (\ref{deltabl}) are not seen experimentally, these terms should be absent (or sufficiently suppressed). 
This is usually done by imposing that a discrete R-parity is preserved. The R-parity of a field is given by
\begin{align} P_R = (-1)^{3(B-L) + 2s},\end{align}
with $s$ its spin. Note that the Standard Model particles and Higgs bosons carry $P_R = +1$, while the 'sparticles' (squarks, sleptons, gauginos and Higgsinos) have $P_R =-1$. 
Also, since every interaction vertex contains an even number of $P_R = -1$ particles, this implies that the lightest supersymmetric sparticle (LSP) with $P_R=-1$ must be absolutely stable. If this LSP interacts only weakly with ordinary matter it can be an excellent dark matter candidate. Note also that since the different fields of the same multiplet carry a different R-parity, this symmetry does not commute with supersymmetry. 

Although this assignment appears to be quite natural in a supersymmetric context, it should be stressed that one can equivalently forbid the terms (\ref{deltabl}) by imposing conservation of matter parity \cite{matterparity1,matterparity2,matterparity3,matterparity4}. The matter parity $P_M$ of a superfield (as opposed to R-parity, which is defined separately on each component field) is defined as
\begin{align} P_M = (-1)^{3(B-L)}. \label{matterparity} \end{align}
Note that since the matter parity of all fields within a given supermultiplet is the same, this symmetry does commute with supersymmetry. Since for the scalar components ($s=0$) the matter parity is the same as the R-parity, it is completely equivalent to impose either matter parity or R-parity as a symmetry on the theory. Moreover, R-parity and matter parity only differ by the fermion number, which is an exact parity symmetry by itself.

We conclude that imposing matter parity or R-parity is completely equivalent. 
While the R-parity interpretation can be useful to easily abstract its phenomenological consequences, 
from a model building point of view it is far more natural to impose matter parity (\ref{matterparity}), 
since (in contrast with R-parity) it commutes with supersymmetry. 
In fact, since R-parity is equivalent to the (non-R) matter parity, this shows that there is nothing intrinsically 'R' about R-parity.

In chapter~\ref{ch:dS}, we will introduce a model based on a gauged $U(1)$ (shift) symmetry. 
Since matter parity is nothing else but $3(B-L)$. 
It therefore seems an obvious choice in this context to see whether one can consistently identify it with the $U(1)$ symmetry we need in our toy model of supersymmetry breaking.
This can be done by giving a charge proportional to $B-L$ to the MSSM superfields.
As a result, the terms in eqs.~(\ref{deltabl}) are excluded from the superpotential, and thus the $U(1)$ symmetry takes over the role of R-parity. 
This will be discussed in full detail in chapter~\ref{ch:chargedMSSM}.

It should however be noted that in principle one can also have dimension 5 operators that violate baryon and/or lepton number (see for example \cite{dim5operators} and references therein)
\begin{align}  W_{\text{dim 5}} = & \kappa^{(0)}_{ij} H_u L_i H_u L_j + \kappa^{(1)}_{ijkl}Q_i Q_j Q_k L_l + \kappa^{(2)}_{ijkl} \bar u_i \bar u_j \bar d_k \bar e_l \notag \\  &+ \kappa^{(3)}_{ijk} Q_i Q_j Q_k H_d + \kappa^{(4)}_{ijk} Q_i \bar U_j \bar e_k H_d + \kappa^{(5)}_{i} L_i H_u H_u H_d. \label{dim 5} \end{align}
Here the various couplings $\kappa^{(n)}$ have inverse mass dimensions and can be generated by a high-energy microscopic theory, such as a supersymmetric grand unified theory or string theory.
While R-parity forbids the terms in the last line of eq.~(\ref{dim 5}), all terms in the first line are still allowed. 
Imposing a B-L symmetry additionally forbids the terms proportional to $\kappa^{(0)}_{ij}$. 
The terms proportional to $\kappa^{(1)}_{ijkl}$ and $\kappa^{(2)}_{ijkl}$ are still allowed.  
It should however be noted that a $3B-L$ symmetry (which has the same parity $(-1)^{3B-L} = (-1)^{3B-3L}$ on MSSM fields) forbids all the above dimension 5 operators.  
However, in contrast with a gauged $B-L$ which can be made anomaly-free upon the inclusion of three right-handed neutrinos to the MSSM,  a gauged $3B-L$ contains a cubic $U(1)_{3B-L}^3$, and mixed $U(1)_{3B-L} \times SU(2)$ and $U(1)_{3B-L} \times U(1)_Y^2$ anomalies which should be canceled by a Green-Schwarz mechanism (see chapter~\ref{ch:chargedMSSM}).

\cleardoublepage
  \graphicspath{{\chaptersdir/supergravity/image/}}%
  \chapter{Supergravity}\label{ch:supergravity}

In the previous section we introduced supersymmetry, along with the necessary notation, and its minimal model containing the Standard Model, the MSSM. 
While supersymmetry is realized as a global symmetry in the MSSM, we introduce in this chapter theories where supersymmetry is realized as local (gauge) supersymmetry.
If a bosonic symmetry is promoted to a local symmetry, the gauge theory requires the presence gauge bosons to preserve gauge invariance of the Lagrangian under this symmetry.
Local supersymmetry on the other hand is a fermionic symmetry and requires the presence spin-3/2 fermion called the gravitino. 
Since the gravitino appears together with a spin-2 graviton in a gravity multiplet, theories with local supersymmetry naturally contain gravity.

In this chapter we explain how a supergravity theory is constructed, 
where we particularly focus on gauged R-symmetries and Fayet-Iliopoulos terms, which play an important role in later chapters.
The scalar potential is introduced, together with the relevant conventions and notation.
We discuss spontaneous supersymmetry breaking, and in particular how the soft supersymmetry breaking terms arise in a gravity mediated supersymmetry breaking scenario.
We finish this chapter by explaining the relation between the cosmological constant and the minimum of the potential.
As before, we follow the notation of~\cite{TheBook}.

\section{How to build a supergravity theory} \label{sec:sugra_ingredients}

As was the case in a rigid supersymmetric theories (see section~\ref{sec:ingredients}), 
the particle content of a supergravity theory consists, besides the gravity multiplet, of chiral multiplets (see eq.~(\ref{chiral_multiplet})) and gauge multiplets (see eq.~(\ref{gauge_multiplet})).
One can construct a Lagrangian invariant under local supersymmetry transformations by defining a K\"ahler potential $\mathcal K$, a superpotential $W$ and gauge kinetic functions $f_{AB}$. 
The resulting Lagrangian however, is rather lengthy, and the reader is referred to standard textbooks (see for example eq.~(18.6) in \cite{TheBook}) for its full form.
Below however, we will introduce several relevant terms in the Lagrangian, such as the scalar potential, fermion mass terms and possible Green-Schwarz counterterms 
which can arise.

Although a supergravity theory is uniquely determined (up to Chern-Simons terms) by a K\"ahler potential $\mathcal K(z, \bar z)$, a superpotential $W(z)$, 
and the gauge kinetic functions $f_{AB}(z)$, 
there is a certain degeneracy in the definitions of the K\"ahler potential  and the superpotential.
Namely, a theory is invariant under a K\"ahler transformations
  \begin{align} \mathcal K(z ,\bar z) &\longrightarrow \mathcal K(z , \bar z) + J(z) + \bar J(\bar z),  \notag \\ W(z) &\longrightarrow e^{-\kappa^2 J(z)} W(z), \label{Kahler_tranformation} \end{align}
    where $J(z)$  $(\bar J(\bar z))$ is an (anti-)holomorphic function, and  $\kappa$ is the inverse of the reduced Planck mass, $m_p = \kappa^{-1} = 2.4 \times 10^{15} $ TeV,
  and the $z^\alpha$ are the scalar components of the chiral multiplets of the theory. 

  A supergravity theory can be described by a gauge invariant function (for theories with a non-vanishing superpotential)
  \begin{align} \mathcal G = \kappa^2 \mathcal K + \log(\kappa^6 W \bar W). \end{align}
  The gauge transformations (with gauge parameters $\theta^A$) of the chiral multiplet scalars are given by holomorphic Killing vectors eq.~(\ref{Killing_vector}). 
  In contrast with global supersymmetry, the K\"ahler potential and superpotential in a supergravity theory need not be invariant under a gauge transformation, 
  but can leave a K\"ahler transformation. The gauge transformation of the K\"ahler potential satisfies
  \begin{align} \delta \mathcal K &= \theta^A \left[r_A(z) + \bar r_A(\bar z)\right], \label{Ktransf} \end{align}
  for some holomorphic function $r_A(z)$, provided that the gauge transformation of the superpotential satisfies   $ \delta W = - \theta^A \kappa^2 r_A(z) W $.
  One then has from $\delta W = W_\alpha \delta z^\alpha$ that the superpotential satisfies
  \begin{align} W_\alpha k_A^\alpha = - \kappa^2 r_A W, \label{Wtransf} \end{align}
  where $W_\alpha = \partial_\alpha W$ and $\alpha$ labels the chiral multiplets.  
 
  The scalar potential is given by~\cite{Cremmer}
  \begin{align} \mathcal V &= V_F + V_D, \notag \\ V_F &= e^{\kappa^2 \mathcal K} \left( - 3 \kappa^2 W \bar W  + \nabla_\alpha W g^{\alpha \bar \beta} \bar \nabla_{\bar \beta} \bar W \right),  \notag \\ V_D &= \frac{1}{2} \left( \re f \right)^{-1 \ AB} \mathcal P_A \mathcal P_B,\end{align}
  where the K\"ahler covariant derivative of the superpotential is 
  \begin{align}\nabla_\alpha W = \partial_\alpha W(z) + \kappa^2 (\partial_\alpha \mathcal K) W(z). \label{nablaW} \end{align}
  The moment maps $\mathcal P_A$ are given by
  \begin{align} \mathcal P_A = i(k_A^\alpha \partial_\alpha \mathcal K - r_A). \label{momentmap} \end{align}
While the scalar potential of a globally supersymmetric theory (see eq.~(\ref{SUSY:scalarpot})) is always positive (or rather nonnegative), 
  the F-term contribution to the scalar potential of a supergravity theory contains a negative contribution proportional to $-3\kappa^2 W \bar W$.
 This negative contribution, which has its origin in the elimination of a gravity auxiliary field,
 makes it possible to construct spontaneously broken supergravity with a vanishing\footnote{
 In fact, we are interested in theories where the minimum of the potential is at a very small and positive value as to accommodate for the cosmological constant.} 
 minimum of the scalar potential.
    
  In this thesis we will be concerned with theories with a gauged R-symmetry\footnote{
  For an early paper on gauged R-symmetries, see for example~\cite{gauged_R-symmetry}.}, 
  for which $r_A(z)$ is given by an imaginary constant $r_A(z) = i \kappa^{-2} \xi $. 
  In this case, $\kappa^{-2} \xi$ is a Fayet-Iliopoulos (FI) \cite{FI} constant parameter (where $\xi$ is dimensionless). Note that in contrast with a globally supersymmetric theory, 
  the FI constant can not be chosen at will, but it is constrained by eq.~(\ref{Wtransf}). 
  Since a FI term gives a (positive) D-term contribution to the scalar potential, 
  these terms can be very interesting to study in the context of spontaneous supersymmetry breaking.

  The mass terms for the fermions in the Lagrangian are given by
  \begin{align}    \mathcal L_m =& \frac{1}{2} m_{3/2} \bar \psi_\mu P_R \gamma^{\mu \nu} \psi_\nu \notag \\    & \ - \frac{1}{2} m_{\alpha \beta} \bar \chi^\alpha \chi^\beta - m_{\alpha A} \bar \chi^\alpha \lambda^A - \frac{1}{2} m_{AB} \bar \lambda^A P_L \lambda^B + \text{h.c.},  \end{align}
where $\psi_\mu$ is the spin-3/2 gravitino\footnote{The generation of its physical mass will be discussed below in section \ref{sec:spontaneous susy breaking}} (which is the superpartner of the graviton) and its mass parameter is given by
\begin{align}  m_{3/2} = \kappa^2 e^{\kappa^2 \mathcal K/2} W. \label{gravitinomass} \end{align}
The spin-1/2 mass matrices of the chiral fermions $\chi^\alpha$ and the gauginos $\lambda^A$ are given by
\begin{align}  m_{\alpha \beta} &= e^{\kappa^2 \mathcal K/2} \left[\partial_\alpha + \left( \kappa^2 \partial_\alpha \mathcal K \right) \right] \nabla_\beta W   
- e^{\kappa^2 \mathcal K/2} \Gamma_{\alpha \beta}^\gamma \nabla_\gamma W, \label{theory:chiralmass}
\\  m_{\alpha A} &= i \sqrt{2} \left[ \partial_\alpha \mathcal P_\alpha - \frac{1}{4} f_{AB\alpha} (\re f)^{-1 \ AB} \mathcal P_C \right], \label{theory:gauginochiralmass}
\\  m_{AB} &= - \frac{1}{2} e^{\kappa^2 \mathcal K/2} f_{AB\alpha} g^{\alpha \bar \beta} \bar \nabla_{\bar \beta} \bar W, \label{theory:gauginomass} \end{align}
where $f_{AB\alpha} = \partial_\alpha f_{AB}$, and the connection components $\Gamma_{\alpha \beta}^\gamma$ are given by
\begin{align}
 \Gamma^\alpha_{\beta \gamma} &= g^{\alpha \bar \delta} \partial_\beta g_{\gamma \bar \delta}, \notag\\
 \Gamma^{\bar \alpha}_{ \bar \beta \bar \gamma} &= g^{\delta \bar \alpha} \partial_{\bar \beta} g_{ \delta \bar \gamma}. \label{connection components}
\end{align}
All other connection components vanish (i.e. $\Gamma^\alpha_{\beta \bar \gamma} = \Gamma^{\bar \alpha}_{\bar \beta \gamma} = 0$).
In the supergravity models in the following chapters it will often occur that the non-diagonal elements of the gaugino mass matrix $m_{AB}$ vanish. 
In this case, the diagonal elements will be denoted by a single index, $m_A$. Similarly for the gauge kinetic functions $f_A$.

\section{Spontaneous supersymmetry breaking} \label{sec:spontaneous susy breaking}

\subsection{The super-BEH mechanism} \label{sec:super-BEH}
It is a well-known fact that when a gauge symmetry is spontaneously broken, a Brout-Englert-Higgs (BEH) mechanism occurs; For example, in the Standard Model, the $SU(2)_L \times U(1)_Y$ gauge symmetry is spontaneously broken to a $U(1)$. 
This occurs because the potential of a scalar field (the Higgs boson) forces it to acquire a non-zero vacuum expectation value (VEV). 
For each of the three broken generators of the gauge groups, there exists a so-called Goldstone boson, which is subsequently 'eaten'\footnote{The gauge bosons 'eat' the Goldstone bosons in the sense that they absorb their degree of freedom. They indeed need an extra degree of freedom to become massive since a massless gauge boson has two (on-shell) degrees of freedom, while a massive one has three.} by the corresponding $W^+, W^-$ and $Z$ gauge bosons which in turn become massive.
The fermions in the theory then acquire masses when the Higgs field is replaced with its VEV in its interactions with the fermions.

A similar mechanism, called the super-BEH mechanism, occurs when local supersymmetry is spontaneously broken by the VEV of some scalar field(s).
An important difference with the above however, is that the supersymmetry generators (see eqs.~(\ref{SUSY_algebra})) are fermionic. 
Therefore, spontaneous supersymmetry breaking will lead to a Goldstone fermion (instead of a boson), called the Goldstino $P_L \nu$, which in general is a linear combination of the fermionic superpartners of all fields that contribute to the supersymmetry breaking
\begin{align}  P_L \nu = \chi^\alpha \delta_s \chi_\alpha + P_L \lambda^A \delta_s P_R \lambda_A, \label{theory:Goldstino} \end{align}
where the 'fermion shifts' (the scalar parts of the supersymmetry transformations of the fermions) are given by
\begin{align} \delta_s \chi_\alpha &= - \frac{1}{\sqrt 2} e^{\kappa^2 \mathcal K/2} \nabla_\alpha W, \notag \\ \delta_s P_R \lambda_A &= -\frac{i}{2} \mathcal P_A.  \label{theory:fermion_shifts}\end{align}
Due  to  the  super-BEH  effect,  the elimination of the  Goldstino  will  give a mass  to  the  gravitino given by eq.~(\ref{gravitinomass}), and the mass matrix for the fermions becomes
\begin{align}  m =   \begin{pmatrix}  m_{\alpha \beta} + m_{\alpha \beta}^{(\nu)}  & m_{\alpha B} + m_{\alpha B}^{(\nu)}    \\  m_{A \beta} + m_{A \beta}^{(\nu)}      &     m_{AB} + m_{AB}^{(\nu)}       \end{pmatrix}, \label{theory:fermion_matrices_BEH} \end{align}
where the corrections to the fermion mass terms due to the elimination of the Goldstino are given by
\begin{align} m_{\alpha \beta}^{(\nu)} &= - \frac{4 \kappa^2}{3 m_{3/2}} (\delta_s \chi_\alpha)(\delta_s \chi_\beta), \notag \\  m_{\alpha A}^{(\nu)} &=- \frac{4 \kappa^2}{3 m_{3/2}} (\delta_s \chi_\alpha)(\delta_s P_R \lambda_A ),\notag \\ m_{AB}^{(\nu)} &= - \frac{4 \kappa^2}{3 m_{3/2}} (\delta_s P_R \lambda_A)(\delta_s P_R \lambda_B ). \label{theory:fermion_matrices_BEH2} \end{align}
Since the elimination of the Goldstino results in a reduction of the rank of $m$, its determinant vanishes and the physical masses of the fermions correspond to the non-zero eigenvalues of $m$.

\subsection{A hidden sector of supersymmetry breaking}

It was already explained in section~\ref{sec:softterms} that the MSSM superfields can not be responsible for supersymmetry breaking
since a VEV of a scalar superpartner of a Standard Model fermion would additionally break a Standard Model gauge symmetry. 
Instead, the origin of supersymmetry breaking should be in a 'hidden sector' of particles. 
Although there are several mechanisms available by which supersymmetry breaking can be communicated to the visible sector (MSSM), we will focus here on gravity mediation \cite{gravitymediation}. 
For other mechanisms, such as gauge mediation~\cite{gaugemediation} of anomaly mediation~\cite{anomalymediation1,anomalymediation2,anomalymediation3,winolike,anomalymediation4}, the reader is referred to the literature.

In a gravity mediated supersymmetry breaking scenario~\cite{gravitymediation}, the chiral multiplets of a model are divided into two sectors: 
An observable sector, which includes the chiral multiplets $\varphi_i$ of the MSSM (as well as its gauge multiplets), 
and a hidden sector where the scalar fields are labeled by $z^\alpha$.
The K\"ahler potential and the superpotential can be decoupled
\begin{align}
 \mathcal K = K(z, \bar z) + \sum \varphi \bar \varphi, \notag \\
 W = W_h(z) + W_{\text{MSSM}}(\varphi),
\end{align}
where the indices labeling chiral multiplets have been omitted. As in the MSSM, the observable sector K\"ahler potential is taken to be canonical and the MSSM superpotential is given in eq.~(\ref{MSSMsuperpot}).
In the limit $\kappa \rightarrow 0$ (or equivalently $M_P \rightarrow \infty$)  one simply has a model with two decoupled sectors.
When gravitational effects are included however, the scalar potential is given by
\begin{align} \mathcal V = e^{\kappa^2 K(z,\bar z) + \sum  \kappa^2 \varphi \bar \varphi} 
\left(-3 \kappa^2 \left|W_h + W_{\text{MSSM}} \right|^2  +g^{z \bar z} |\nabla_z W|^2 + |\nabla_\varphi W|^2 \right), \label{theory:Hiddensector_potential} \end{align}
where
\begin{align}
\nabla_z W &= \partial_z W_h + \kappa^2 \partial_z K(z,\bar z) \left( W_h + W_{\text{MSSM}} \right) , \notag \\
\nabla_\varphi W &= \partial_\varphi W_{\text{MSSM}} + \kappa^2 \bar \varphi \left( W_h + W_{\text{MSSM}} \right) ,
\end{align}
by the definition in eq.~(\ref{nablaW}).

The hidden sector superpotential $W_h$ is constructed such that supersymmetry is broken in this sector and the scalar fields $z^\alpha$ in the hidden sector acquire a VEV (typically of order $\sim m_p = 1/\kappa$).
The couplings in the hidden sector however, carry a high scale (usually assumed to be at the GUT-scale or near the Planck mass), such that its effects 
 are not detectable at low (collider) energies.
We will show below that, when the hidden sector scalars are replaced with their VEVs, one retrieves exactly the soft supersymmetry breaking terms~(\ref{softterms}).
 
The derivative with respect to a visible sector field $\varphi$ of the scalar potential is given by
\begin{align}
 \partial_\varphi \mathcal V =& e^{ \kappa^2 K(z,\bar z) + \sum \kappa^2 \varphi \bar \varphi} 
 \Big[ \kappa^2  \bar \varphi \left(-3 \left|W_h + W_{\text{MSSM}} \right|^2  + g^{z \bar z}  |\nabla_z W|^2 + |\nabla_\varphi W|^2 \right)  \notag \\
 &  - 3 \kappa^2 \bar W W_\varphi + \kappa^2  g^{z \bar z}  K_z W_\varphi (\bar W_{\bar z} + \kappa^2 K_{\bar z} \bar W) \notag \\ 
 &+ (W_{\varphi \varphi}  + \kappa^2 \bar \varphi W_\varphi ) (\bar W_{\bar \varphi} + \kappa^2 \varphi \bar W) + (W_\varphi + \kappa^2 \bar \varphi W) \bar W \Big], 
\end{align}
where $W_\varphi = \partial_\varphi W$, $K_z = \partial_z K$ are shorthand notations.
Under the assumption that the visible sector fields $\varphi$ do not contribute to the supersymmetry breaking, we have $\langle \varphi \rangle = \langle \bar \varphi \rangle = 0$,
and therefore for $W_{\text{MSSM}}$, given by eq.~(\ref{MSSMsuperpot}), we have $\langle W_\varphi \rangle = \langle W_{\bar \varphi} \rangle = 0$. 
The soft supersymmetry breaking masses are then given by
\begin{align}
 \partial_\varphi \partial_{\bar \varphi} \mathcal V\Big|_{\varphi = \bar \varphi = 0} 
 = e^{\kappa^2 K(z,\bar z)} \left( -2 \kappa^4 W_h \bar W_h + \kappa^2 g^{z \bar z}  \left| \nabla_z W_h \right|^2 \right),
\end{align}
where the hidden fields $z$ are replaced with theirs VEVs. 
The $B \mu-$term can be calculated similarly from
\begin{align}
  \partial_\varphi \partial_\varphi \mathcal V\Big|_{\varphi = \bar \varphi = 0} =
  \kappa^2 e^{\kappa^2 K(z,\bar z)} \left[ g^{z \bar z} K_z \left(\bar W_{\bar z} + \kappa^2 K_{\bar z} \bar W \right) - \bar W \right] W_{\varphi \varphi}  , 
\end{align}
as well as the trilinear terms
\begin{align}
 \partial_\varphi \partial_\varphi \partial_\varphi \mathcal V\Big|_{\varphi = \bar \varphi = 0}
 =\kappa^2 e^{\kappa^2 K(z,\bar z)}  g^{z \bar z} K_z \left(  \bar W_{\bar z} + \kappa^2 K_{\bar z} \bar W \right)  W_{\varphi \varphi \varphi}  .
\end{align}

The gaugino mass terms can be calculated from eq.~(\ref{theory:gauginomass}). 
Note however that a constant gauge kinetic function implies a vanishing gaugino mass since eq.~(\ref{theory:gauginomass}) is proportional to the derivative of its respective gauge kinetic function.
We will show in section~\ref{sec:Anomalies-gauginomass} that, even for a constant gauge kinetic function, 
the gaugino mass terms can be generated at one loop.

The MSSM is defined however in rigid supersymmetry, where the scalar potential is given by $\mathcal V = \sum W_{\text{MSSM}, \varphi} \bar W_{\text{MSSM},\bar  \varphi} $.
In the discussion above, this corresponds to the last term in eq.~(\ref{theory:Hiddensector_potential}). 
Note however that, after the hidden sector fields are replaced with their VEVs, this term is proportional to $e^{\kappa^2 K}$. 
Therefore, in order to interpret the last term in eq.~(\ref{theory:Hiddensector_potential}) as the usual MSSM scalar potential at low energies, a rescaling is needed:
\begin{align}  \hat W_{\text{MSSM}} = e^{\kappa^2 K(z,\bar z)/2} W_{\text{MSSM}}, \label{superpotrescaling} \end{align}
such that at low energies the last term in eq.~(\ref{theory:Hiddensector_potential})
reduces to $\sum \hat W_{\text{MSSM}, \varphi}  \hat {\bar  W}_{\text{MSSM},\bar  \varphi}$.

As a result, all trilinear terms $a^{ijk}$ (see eq.~(\ref{softterms})) are the same and are given by $A_0 \hat y_i$, 
where $y_i$ are the Yukawa couplings in the  MSSM superpotential~(\ref{MSSMsuperpot}) (rescaled by eq.~(\ref{superpotrescaling})) and
\begin{align}
 A_0 = \kappa^2 e^{\kappa^2 K(z,\bar z)/2} g^{z \bar z} K_z \left( \bar W_{\bar z} + \kappa^2 K_{\bar z} \bar W \right).
\end{align}
Similarly, the '$B \mu'$-term is given by $B_0 \hat \mu$, where $\hat \mu = e^{\kappa^2 K/2} \mu$ by eq.~(\ref{superpotrescaling}), and
\begin{align}
 B_0 = \kappa^2 e^{\kappa^2 K(z,\bar z)/2} \left[ g^{z \bar z} K_z \left( \bar W_{\bar z} +\kappa^2  K_{\bar z} \bar W \right) - \bar W \right],
\end{align}
Note that one has the relation~\cite{A-B_relation}
\begin{align}
 A_0 = B_0 + m_{3/2}, \label{AB-relation} 
\end{align}
where the gravitino mass term is given by eq.~(\ref{gravitinomass}).

Note however above that we only discussed the F-term contribution to the scalar potential. Indeed, a D-term contribution to the scalar potential can also contribute to the soft supersymmetry breaking terms.
This will be discussed further in section~\ref{sec:other}.

\section{Relation between the minimum of the potential and the cosmological constant}

In this section we show the relation between the minimum of the potential and the cosmological constant $\Lambda$.
In the Einstein field equations, the cosmological constant arises as 
\begin{align}
 R_{\mu \nu} - \frac{1}{2} g_{\mu \nu} R + \Lambda g_{\mu \nu}  =  8 \pi G \ T_{\mu \nu}, \label{Einstein-eq}
\end{align}
where $R_{\mu \nu}$ is the Ricci tensor, $R$ is the curvature scalar $R=g^{\mu \nu} R_{\mu \nu}$, $T_{\mu \nu}$ is the stress-energy tensor, $G$ is the gravitational constant, 
and we put the speed of light in the vacuum $c=1$.

On the other hand, one can write the action of a supergravity theory as the Einstein-Hilbert term plus a term $\mathcal L_M$ involving any matter fields in the theory\footnote{
Any terms involving the gravitino are also assumed to be included in $\mathcal L_M$.}
\begin{align}
 S = \int d^4 x \ \sqrt{- g} \ \left[ \frac{1}{2 \kappa^2} R + \mathcal L_M - V_0 \right] . \label{Einstein-Hilbert}
\end{align}
Here, $g = \det g_{\mu \nu}$ is the determinant of the metric tensor, and $\kappa^2 = 8 \pi G$.
A constant contribution $V_0$ has been added that corresponds to the value at the minimum of the scalar potential. 
We will show below how the constant contribution to the Lagrangian $V_0$ gives rise to a contribution in the Einstein field equations (\ref{Einstein-eq}) 
that can be interpreted as the cosmological constant $\Lambda$.

The variation of the action eq.~(\ref{Einstein-Hilbert}) gives
\begin{align}
 \delta S =  \int d^4 x & \bigg[ 
 \frac{1}{2 \kappa^2} \delta \left( \sqrt{- g} g^{\mu \nu} \right) R_{\mu \nu} 
+  \frac{1}{2 \kappa^2} \sqrt{- g} g^{\mu \nu} \delta R_{\mu \nu}  + \delta \left( \sqrt{- g} \mathcal L_M \right)  \notag \\
  &- \delta \left( V_0 \sqrt{-g} \right) \bigg] . \label{Einstein_Variation}
\end{align}
Since the variation of the Ricci tensor is given by 
\begin{align}
 \delta R_{\mu \nu} = \nabla_\rho \delta \Gamma^\rho_{\mu \nu} - \nabla_\mu \delta \Gamma^\rho_{\nu \rho} ,  \label{Palatini-identity}
\end{align}
the second term in eq.~(\ref{Einstein_Variation}) is a total derivative and vanishes.
Further, we use
\begin{align}
\delta g^{\mu \nu} &= -g^{\mu \rho} \delta g_{\rho \sigma} g^{\sigma \nu} , \notag \\
\delta \sqrt{-g} &= \frac{1}{2} \sqrt{-g} g^{\mu \nu} \delta g_{\mu \nu} = - \frac{1}{2} \sqrt{-g} g_{\mu \nu} \delta g^{\mu \nu},
\end{align}
to reduce eq.~(\ref{Einstein_Variation}) to
\begin{align}
 \delta S =  \frac{1}{2 \kappa^2} \int d^4 x  \ \delta g^{\mu \nu} \sqrt{-g} \ 
\Bigg[ R_{\mu \nu} & - \frac{1}{2} g_{\mu \nu} R 
- \kappa^2 \left( \frac{ - 2 \delta \left( \sqrt{-g} \mathcal L_M \right)}{ \delta g^{\mu \nu} \sqrt{-g} } \right) \notag \\
& + \kappa^2 V_0 g_{\mu \nu}
 \Bigg] .
\end{align}
One now defines the stress-energy tensor
\begin{align}
 T_{\mu \nu} = \frac{ - 2 \delta \left( \sqrt{-g} \mathcal L_M \right)}{ \delta g^{\mu \nu} \sqrt{-g} },
\end{align}
and solve for $\delta S = 0$ to find the Einstein field equations (\ref{Einstein-eq}) with
\begin{align}
 \Lambda = 8 \pi G V_0 .
\end{align}

Alternatively, the constant contribution $V_0$ can be absorbed in the stress-energy tensor by defining
\begin{align}
T_{\mu \nu} = \frac{ - 2 \delta \left( \sqrt{-g} \mathcal L_M \right)}{ \delta g^{\mu \nu} \sqrt{-g} } + T^{\text{vac}}_{\mu \nu},
\end{align}
where\footnote{ 
The stress-energy tensor can be modeled as a perfect fluid,
\begin{align} T_{\mu \nu} = (\rho + p) U_\mu U_\nu + p g_{\mu \nu}, \end{align}
where $\rho$ is the energy density and $p$ is the isotropic pressure in its rest frame, $U^\mu$ is the fluid four-velocity.
One obtains eq.~(\ref{stress-energy_vac}) by putting $\rho_{\text{vac}} = - p_{\text{vac}}$.
}
\begin{align}
 T^{\text{vac}}_{\mu \nu} = - \rho_{\text{vac}} g_{\mu \nu}, \label{stress-energy_vac} \quad \quad \text{with} \quad \quad \rho_{\text{vac}} = V_0 = \kappa^{-2} \Lambda .
\end{align}
The above equivalence shows the relation between the cosmological constant and the energy of the vacuum, and is the reason the terms 'vacuum energy' and 'cosmological constant' are used interchangeably.
Moreover, the latest results from the Planck collaboration~\cite{Planck2015} show
\begin{align}
 \Lambda_{\text{obs}} &= \frac{3 \Omega_\Lambda H_0^2}{c^2} \notag \\
 &\approx 1.1 \times 10^{-52} \text{ m}^{-2} 		,
\end{align}
where the Hubble constant is given by $H_0 = 67.74$~$\frac{\text{km}}{\text{s} \text{ Mpc}}$, the dark energy density $\Omega_\Lambda = 0.6911$, 
and we have explicitly included the speed of light in the vacuum $c$.

\cleardoublepage
  \graphicspath{{\chaptersdir/dS/image/}}%
  \chapter{A model with a tunably small (and positive) cosmological constant}\label{ch:dS}

It was already mentioned in the introduction that experimental evidence suggests  
that the cosmological constant  is very small and positive~\cite{Frieman:2008sn,Ade:2015xua}. 
Therefore, if string theory were to be a realistic theory of nature, its vacuum energy should therefore be positive as well. 
This is called a de Sitter (dS) vacuum.
The hunt for de Sitter vacua in string theory with a tunably small value of the cosmological 
constant is a very interesting and challenging subject (for recent work, see for example~\cite{KKLT1,KKLT2,KKLT3,KKLT4,KKLT5,KKLT6,KKLT7,KKLT8,KKLT9}). 

In this chapter we start the presentation of our own work, where this problem is addressed in the context of supergravity. 
This is interesting since certain supergravity theories appear as the low energy limit of string theory.
The goal of this chapter is to obtain a supergravity theory, based on a chiral multiplet with a gauged shift symmetry, 
which has a tunably small and positive value of the minimum of the scalar potential. 
In the string theory context, this chiral multiplet can be identified with the string dilaton or an appropriate compactification modulus.
We show that for suitable values of the parameters the scalar potential can be identified with the one obtained in~\cite{Antoniadis:2008uk}.

The model~\cite{RK1}, originally proposed in~\cite{Zwirner1,Zwirner2},
consists of (besides the gravity multiplet) a chiral multiplet $S$ and a vector multiplet associated with a gauged $U(1)$ symmetry 
which acts as a shift along the imaginary part of the scalar component of $S$.
By gauge invariance of the theory, the K\"ahler metric can therefore only be a function of the real part of $S$, 
which is taken to be a logarithm $\mathcal K = -\kappa^{-2} p \log( s+ \bar s)$, where the constant $p$ is assumed to be an integer\footnote{
For example, for the total volume-axion one has $p=3$ (which results in no-scale supergravity), or for the dilaton-axion $p=1$.}.
The most general gauge kinetic function, consistent with the shift symmetry, contains a constant and a linear contribution,
while the most general superpotential is either a constant, or an exponential of $S$.
In the case of an exponential superpotential, the gauged shift symmetry becomes a gauged R-symmetry\footnote{
For other studies based on a gauged R-symmetry, see for example~\cite{OtherR1,OtherR2,OtherR3,Dreiner}.}, which is therefore labeled $U(1)_R$ below 
(even in the case where the superpotential is constant).

As a result, the R-symmetry fixes the form of the Fayet-Iliopoulos term, 
leading to a supergravity action with two independent parameters that can be tuned such that the scalar potential possesses a (metastable) de Sitter vacuum 
with a tunably small (and positive) cosmological constant. A third parameter fixes the VEV of the string dilaton, 
while the mass of the gravitino (which is essentially the supersymmetry breaking scale) is separately tunable.

An alternative (dual) description of this model is given in terms of a linear multiplet.
The string dilaton appears naturally in a linear multiplet, 
where it is accompanied by a 2-index antisymmetric tensor potential which is dual to $\im S$ under Poincar\'e duality of its field-strength.
We show that this dual description persists even in the presence of a field dependent (exponential) superpotential.

The model is introduced in section~\ref{sec:model}. 
Our own contributions start in section~\ref{sec:tunability} where we examine and quantify the tunability of the scalar potential in this model.
It turns out that for $p \geq 3$ no (meta)stable de Sitter vacua can be found. 
The cases $p=1$ and $p=2$ are presented in full detail in subsections~\ref{sec:p=1} and \ref{sec:p=2} respectively.
Following, in section~\ref{sec:remarksp2} we make a few remarks on the case $p=2$:
Firstly, since for $p=2$ the vacuum is metastable, quantum tunneling to another vacuum solution can in principle occur. 
It is checked in section~\ref{sec:p2stability} that this tunneling probability is indeed extremely low.
Next, in section~\ref{sec:p2stringtheory} we briefly comment on a possible connection with string theory 
and we close this chapter by rewriting the model in terms of a linear multiplet in section~\ref{sec:linearchiral} 
which is a more natural description from a string theory point of view.

This chapter is based on work together with I.~Antoniadis~\cite{RK1,RK3} and on work with I.~Antoniadis and D.~Ghilencea~\cite{RK2}.
In particular, section~\ref{sec:p=1} is based on work~\cite{RK3}, sections~\ref{sec:p=2}, \ref{sec:p2stringtheory} and \ref{sec:linearchiral} are based on~\cite{RK1},
while section~\ref{sec:p2stability} is based on~\cite{RK2}.

\section{Introduction of a model} \label{sec:model}

In order to obtain a model with a tunably small (and positive) value of the cosmological constant, 
one could try to cancel a (negative) F-term contribution to the scalar potential by a positive D-term contribution.
The simplest model one can look at then has a chiral multiplet $S$ whose superpotential gives rise to an F-term contribution to the scalar potential, 
and a vector multiplet whose corresponding gauge symmetry gives rise to a D-term contribution.
We will assume that the gauge symmetry is abelian, 
and that the scalar component $s$ of the chiral multiplet transforms under this $U(1)_R$ (where the subscript $R$ will be clear below) as a shift\footnote{
In superfields the shift symmetry (\ref{shift}) is given by $\delta S = -ic \Lambda$, where $\Lambda$ is the superfield generalization of the gauge parameter.   
The gauge invariant K\"ahler potential below in eq.~(\ref{model:kahlertransformed}) is then given by 
$\mathcal K(S , \bar S) = - \kappa^{-2} p \log (S+\bar S+cV_R)+\kappa^{-2}b(S+\bar S+c V_R)$, where $V_R$ is the gauge superfield of the shift symmetry.}
\begin{align}s \longrightarrow s - ic\theta, \label{shift} \end{align}
where $\theta$ is the gauge parameter and $c$ is a parameter of the model.
The real scalar component of $s$ can be identified with the string dilaton or a compactification modulus\footnote{
Indeed, we will show below in section~\ref{sec:p2stringtheory} how $s$ corresponds to the string dilaton for a particular value of the parameters in the model.},
while its imaginary part is an axion.
In supersymmetric theories the string dilaton and the axion $a$ can be described as the real and imaginary part of the scalar component $s$ of a chiral multiplet $S = (s,\psi,F)$
\begin{align} s = \frac{1}{g^2} + ia, \label{sgaugecoupling}\end{align}
where $g$ is the four dimensional gauge coupling. In perturbation theory, the axion has an invariance under a Peccei-Quinn symmetry which shifts $s$ by an imaginary constant. 

As outlined in section~\ref{sec:sugra_ingredients}, a supergravity theory is defined by a K\"ahler potential, a superpotential and gauge kinetic functions.
The shift symmetry eq.~(\ref{shift}) forces the K\"ahler potential to be a function of $s + \bar s$. 
Moreover, while keeping the interpretation of the string dilaton (or a compactification modulus) in the back of our mind, the K\"ahler potential is taken to be of the form
\begin{align}
 \mathcal K &= - \kappa^{-2} p \log (s + \bar s), \label{model:kahlerpotential}
\end{align}
where $p$ is a parameter of the model and the factor $\kappa^{-2}$ is needed to restore the appropriate mass dimensions of the K\"ahler potential\footnote{Note that the field $s$ has mass dimension 0.}.
Of course, this K\"ahler potential results in a non-renormalizable theory. It should therefore be viewed either as a classical field theory, or as a quantum field theory with a certain cutoff.

Since the superpotential $W(s)$ is a holomorphic function, it can not be a function of the combination $s + \bar s$ and thus any (non-constant) superpotential will transform under the shift symmetry. 
The only nontrivial possibility is then to allow the superpotential to transform with a phase according to eq.~(\ref{Wtransf}).
The most general superpotential turns out to be an exponential 
\begin{align}  W &= \kappa^{-3}a e^{bs}, \label{model:superpotential} \end{align}
where $a$ (not to be confused with the axion) and $b$ are parameters in the model. Note that the superpotential is not invariant under the shift symmetry. Instead it transforms as
\begin{align}  \delta W = -ibc\theta W, \end{align}
in which case the shift symmetry becomes an R-symmetry and is therefore labeled with a subscript $R$. This is consistent if eq.~(\ref{Wtransf}) is satisfied, which gives
\begin{align} r_A = i \xi \kappa^{-2}, \end{align}
where \begin{align}\xi =  bc      \label{FIconstant}       \end{align} 
and $\kappa^{-2} \xi$ is the Fayet-Iliopoulos constant~\cite{FI} in the theory.
It is interesting that this superpotential appears naturally in the context of gaugino condensation (see for example~\cite{gauginocondensation1,gauginocondensation2,gauginocondensation3,gauginocondensation4}) 
and in recent works on supersymmetric extensions of Starobinsky models of inflation~\cite{Starob1,Starob2,Starob3}. 

Following, the most general gauge kinetic function is either a constant or linear in $s$
\begin{align} f(s) = \gamma + \beta s. \label{model:gauge kinetic function} \end{align}
It is however important to note that if the parameter\footnote{
Below, the parameter $\beta$ will often carry a subscript $R$, 
such that $\beta_R$ can be distinguished from similar parameters ($\beta_A$) corresponding to similar Green-Schwarz contributions for the Standard Model gauge groups $A$.}
$\beta$ is non-zero, the Lagrangian contains a Green-Schwarz \cite{Green-Schwarz_dim10,Green-Schwarz_1,Green-Schwarz_2,Green-Schwarz2_1,Green-Schwarz2_2} contribution
\begin{align} \mathcal L_{GS} &=  \frac{i}{4}\text{Im}\left(f(s) \right) F_{\mu \nu} \tilde{F}^{\mu \nu}  \notag \\
 &= \frac{1}{8}\text{Im}\left(f(s) \right) \ \epsilon^{\mu \nu \rho \sigma} F_{\mu \nu} F_{\rho \sigma},\label{model:Green-Schwarz term} \end{align}
where the dual of an antisymmetric tensor is defined as
\begin{align} \tilde F^{\mu \nu} = - \frac{i}{2} \epsilon^{\mu \nu \rho \sigma} F_{\rho \sigma}. \label{conventions:dual_field_strength} \end{align}
The term (\ref{model:Green-Schwarz term}) however, is not gauge invariant, since its gauge transformation is given by
\begin{align} \delta \mathcal L_{GS} &= -\frac{\theta \beta c}{8} \ \epsilon^{\mu \nu \rho \sigma} F_{\mu \nu} F_{\rho \sigma}. \label{model:Green-Schwarz contribution} \end{align}
To restore gauge invariance, this term should be canceled by another contribution. We postpone this discussion until chapter \ref{ch:anomalies}, where the gauge invariance of this model at the quantum level is studied in full detail.

The scalar potential is given by
\begin{align} \mathcal V =& \frac{-3 |a|^2 }{(s + \bar s)^p} e^{b(s + \bar s)} + \left( b - \frac{p}{s + \bar s} \right)^2 \left( \frac{a^2}{p} \frac{e^{b(s + \bar s)}}{(s + \bar s)^{p-2}} + \frac{c^2}{\beta(s + \bar s) + 2\gamma} \right)\label{scalarpot_generalp} . \end{align}
Below, we will assume $a>0$ and we will drop the absolute value in the scalar potential. This can be done without losing generality with respect to the phenomenological implications in later chapters.

The above model is summarized for future reference as
\begin{align} \mathcal K &=- \kappa^{-2} p \log (s + \bar s) , \notag \\
 W &= \kappa^{-3}ae^{bs}, \notag \\ \label{model:exp_superpot}
 f(s) &= \gamma + \beta s,
 \end{align}
 where the moment map $\mathcal P$, defined in eq.~(\ref{momentmap}), is given by
 \begin{align}
  \kappa^2 \mathcal P = c \left( b - \frac{p}{s + \bar s} \right).
 \end{align}
This can be rewritten by performing a K\"ahler transformation (see eqs.~(\ref{Kahler_tranformation})) with $J=\kappa^{-2} bs$ in the form
\begin{align} \mathcal K &=- \kappa^{-2} p \log (s + \bar s) + \kappa^{-2} b(s+ \bar s), \notag \\
 W &= \kappa^{-3}a, \notag \\ f(s)&=\gamma + \beta s.  \label{model:kahlertransformed}\end{align}
One can therefore absorb the exponential part of the superpotential in the K\"ahler potential. 
In this case the gauge symmetry is not an R-symmetry anymore, which has important consequences that will be discussed in section~\ref{sec:Anomalies-gauginomass}.
Nevertheless, we will continue to refer to the shift symmetry as $U(1)_R$, even in the 'K\"ahler frame' eq.~(\ref{model:kahlertransformed}) where the shift symmetry is technically not an R-symmetry.
It is important to note that indeed both eqs.~(\ref{model:exp_superpot}) and eqs.~(\ref{model:kahlertransformed}) result in the same supergravity Lagrangian, and in particular the same scalar potential.

\section{Tunability of the model} \label{sec:tunability}
In this section we investigate the parameter range of the model defined above by eqs.~(\ref{model:exp_superpot}) 
(or equivalently by eq.~(\ref{model:kahlertransformed})) and (\ref{model:gauge kinetic function})
for which there exists a positive and small minimum of the scalar potential eq.~(\ref{scalarpot_generalp}). 
We then present the relations between the parameters which ensure a (tunably) small value of the cosmological constant.

For $b>0$, there always exists a supersymmetric AdS (anti-de Sitter) vacuum at $\langle s + \bar s \rangle = b/p$, while for $b=0$ supersymmetry is broken in AdS space. Since we are interested in supersymmetry breaking de Sitter vacua, we therefore focus on $b<0$.  
For $p \geq 3$ the scalar potential V is positive and monotonically decreasing \cite{Zwirner1},
while for $p<3$, the F-term contribution $\mathcal V_F$ is unbounded from below when $s + \bar s \rightarrow 0$.
On the other hand, the D-term contribution to the scalar potential $\mathcal V_D$ is positive and diverges when $s + \bar s \rightarrow 0$
 such that for various values of the parameters an (infinitesimally small) positive (local) minimum of the potential can be found.

The parameter $p$ in front of the logarithm in the K\"ahler potential is assumed to be a positive\footnote{
A negative or vanishing $p$ would indeed not lead to consistent kinetic terms in the Lagrangian.} integer.
Below we show that for $p=1$, a tunable vacuum can be found (see subsection \ref{sec:p=1}) when\footnote{
The real part of the gauge kinetic function at the minimum of the potential should be positive, for positivity of the kinetic energy of the gauge field.} 
$\gamma>0$, while for $p=2$ tunable vacua can be found when $\gamma=0$ and $\beta>0$.
We show how the parameters $a$ and $c$ can be tuned such that the scalar potential has a minimum at an infinitesimally small (and positive) value.
Moreover, it will turn out that the parameter $b$ can be tuned independently to determine the VEV of the scalar component $s$ of the chiral multiplet.

\subsection{Tunable dS vacua for $p=1$} \label{sec:p=1}

For $p=1$, a stable de Sitter vacuum can be found on the condition that the constant contribution to the gauge kinetic function is non-zero. 
In this section we present the relations between the parameters $a,b,c$ that are necessary to ensure a positive and tiny cosmological constant for the case $\beta=0$. 
Although a non-zero $\beta$ is in principle allowed, 
it is shown in the Appendix~\ref{Appendix:nonzerobeta} that due to constraints from anomaly cancellations 
its contribution turns out to be very small.
In Appendix~\ref{Appendix:p1_gamma0} we show 
the relations between the parameters to obtain a vanishing cosmological constant for 
a vanishing constant contribution ($\gamma = 0$, $\beta > 0$) to the gauge kinetic function.
However, since this case is inconsistent with anomaly cancellation conditions, similar to the case for $p=2$ in section~\ref{sec:Anomalies:p=2},
it will not be discussed below.

\begin{figure}[!ht]    \centering \medskip     \includegraphics[width=0.9\textwidth]{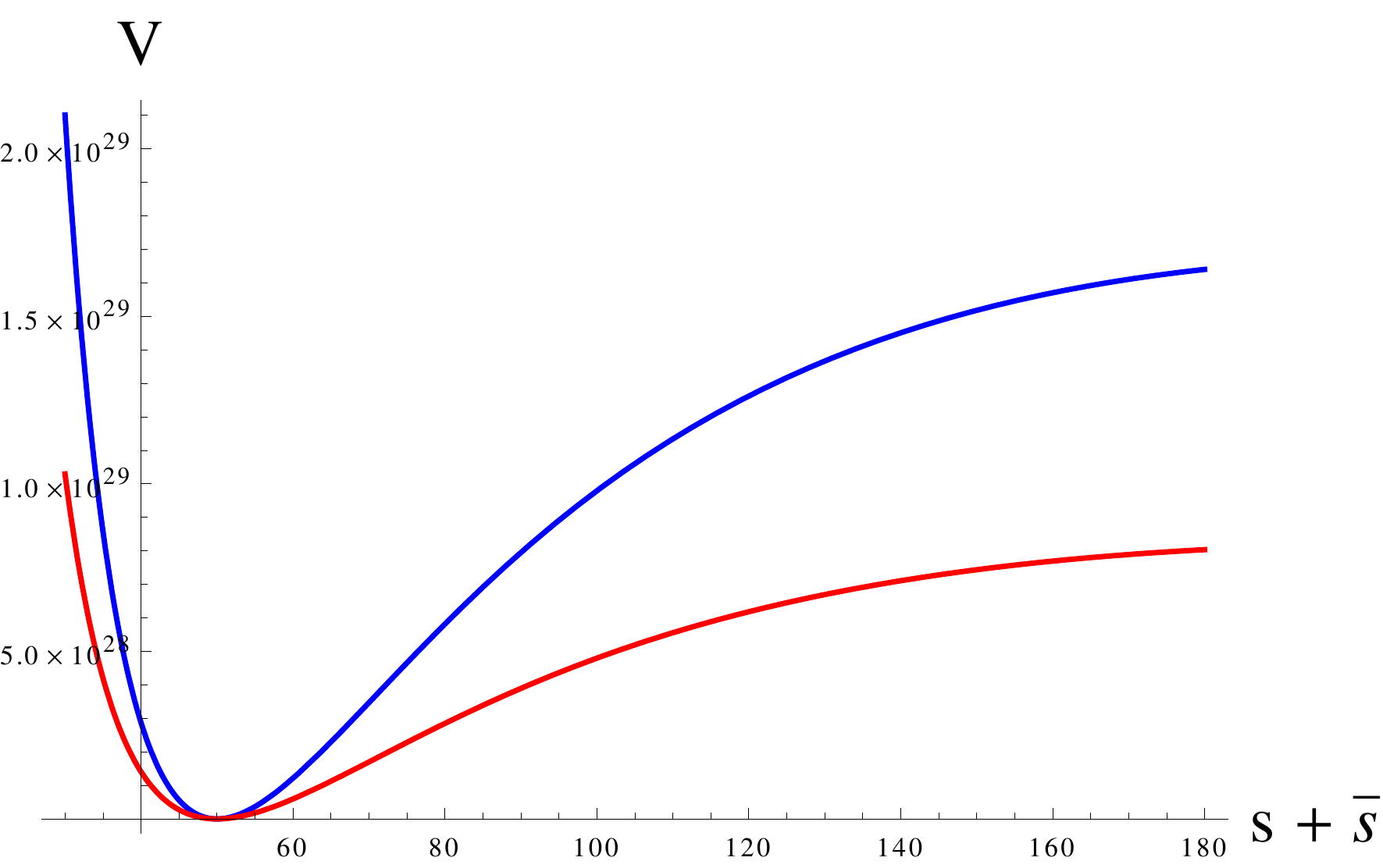}
    \caption{A plot of the scalar potential (\ref{scalarpotp1}) for $p=1$, $b=\alpha/50$, $a = 10^{-15}$ (blue) and $a = 0.7 \times 10^{-15}$  (red),
    and $c$ is determined by eq.~(\ref{model:p1_tunability}) with $\Lambda = 0$.  The VEV of $s + \bar s$ is specified by choosing $b=\alpha/50$, where we have taken $\langle s + \bar s \rangle = 50$ in this plot. The values of $a$ are chosen such that the gravitino mass is of the order of TeV.  
    Note that the scalar potential has a stable dS minimum at $\langle s + \bar s \rangle = \alpha /b$.}      \label{fig:plot_p1} \end{figure}

In the case $\beta=0$, the constant $\gamma$ in eq.~(\ref{model:gauge kinetic function}) can be absorbed in other constants of the theory by an appropriate rescaling of the vector field, 
and we can put $\gamma=1$.
The model is then given by
\begin{align} \mathcal K &= -\kappa^{-2} \log(s + \bar s) + \kappa^{-2} b (s + \bar s), \notag \\ W&= \kappa^{-3} a, \notag \\ f(s) &= 1. \label{model:p1} \end{align}
The scalar potential is
\begin{align} \mathcal V &= \mathcal V_F + \mathcal  V_D, \notag \\ \mathcal V_F &= \kappa^{-4} a^2 \frac{e^{b(s + \bar s)}}{s + \bar s} \sigma_s, && \sigma_s=  -3 + \left( b(s + \bar s) - 1 \right)^2 , \notag \\ \mathcal V_D &= \kappa^{-4}   \frac{c^2}{2} \left( b - \frac{1}{s + \bar s} \right)^2. \label{scalarpotp1}\end{align}
The minimization of the potential $\partial_s \mathcal V = 0$ gives
  \begin{align} \frac{c^2}{a^2} &=  \langle s + \bar s \rangle (2- b^2 \langle s + \bar s\rangle ^2) e^{b \langle s + \bar s \rangle}\, . \end{align}
  By plugging this relation into $ \mathcal V_{min} = \Lambda \approx (10^{-3} \text{eV})^4$, which ensures the minimum of the potential to be at a very small but positive value, one finds
  \begin{align} \kappa^4 e^{-b \langle s + \bar s \rangle} \langle s + \bar s \rangle \frac{\Lambda}{a^2} = -3 + (b \langle s + \bar s  \rangle -1 )^2 \left[ 2 - \frac{b^2 \langle s + \bar s \rangle^2}{2}\right]. \end{align}
  An infinitesimally small cosmological constant $\Lambda$ can then be obtained by tuning the parameters $a,b,c$ such that
  \begin{align} b \langle s + \bar s \rangle &= \alpha \approx -0.233153, \label{model:p1_tunability}  \\ \frac{b c^2}{a^2} &= A(\alpha) + \frac{2 \kappa^4 \Lambda \alpha^2}{a^2 b (\alpha - 1)^2}, & A(\alpha) &=  2 e^\alpha \alpha \frac{ 3 - (\alpha -1)^2}{ (\alpha - 1)^2} \approx -0.359291\, , \notag \end{align}
  where $\alpha$ is the negative root of $-3 + (\alpha -1 )^2(2 - \alpha^2/2) = 0$ close to $-0.23$. The other roots are either imaginary or would not allow for a real solution of the second constraint.
Note that 
\begin{align}  \lim_{s + \bar s \rightarrow \infty} \mathcal V &= \frac{\kappa^{-4} \xi^2}{2} > 0, \label{p1_limit_to_ifty} \end{align}
where $\kappa^{-2} \xi$ is the FI constant defined above in eq.~(\ref{FIconstant}). 
It follows that for $\kappa^{-4} \xi^2/2 > \Lambda$, 
the scalar potential allows for a stable de Sitter (dS) vacuum with an infinitesimally small (and tunable) value for the cosmological constant. 
The gravitino mass (\ref{gravitinomass}) is given by
\begin{align}  m_{3/2} &= \kappa^{-1} \sqrt{|b|}a \frac{e^{\alpha/2}}{\alpha} \notag \\ &\approx 4.42 \sqrt{|b|}a  \times 10^{15} \text{ TeV}. \label{model:p1_gravitino} \end{align}
One can obtain a  TeV gravitino mass by tuning $a\sqrt{|b|} \approx 2.26\times 10^{-16}$. A plot of the scalar potential is given in figure~\ref{fig:plot_p1}.

We conclude that for $p=1$ there can exist indeed a dS vacuum with a tunably small cosmological constant. Moreover, the parameter $b$ can be used to fix the VEV of $\langle s + \bar s \rangle$, while a relation between the parameters $a$ and $c$ ensures a small cosmological constant.
 In chapter \ref{ch:softterms} however, we will show that when this minimal model is coupled to the MSSM, it will lead to tachyonic soft scalar masses for the MSSM fields. 
We therefore postpone a careful treatment of this model, its possible solutions to avoid tachyonic masses and its the resulting energy sparticle spectrum to chapter \ref{ch:softterms}.

\subsection{Tunable dS vacua for $p=2$} \label{sec:p=2}

For $p=2$, metastable de Sitter vacua can be found for various values of the parameters if the gauge kinetic function is linear\footnote{
In fact, a very small $\gamma$ is still consistent with the existence of a dS vacuum, provided $\beta \neq 0$.
The resulting corrections however to the scalar potential and the discussion below are very small. We therefore take $\gamma=0$.
} in $s$ ($\gamma = 0$). 
Moreover, as far as the minimization of the potential is concerned, 
the constant $\beta$ in eq.~(\ref{model:gauge kinetic function}) can be absorbed in other constants of the theory\footnote{
For a non-trivial $\beta$, the D-term contribution to the scalar potential is given by 
\begin{align}  \kappa^4 \mathcal V_D =  \frac{c^2}{\beta} \frac{1}{s + \bar s} \left( \frac{2}{s + \bar s} - b \right)^2. \notag \end{align}
In this case the results of this section hold true with $c$ replaced by $c'$, given by $c'^2 = c^2 / \beta$.
}. 
The model is then given by
\begin{align} \mathcal K &= -2\kappa^{-2}  \log(s + \bar s) + \kappa^{-2} b (s + \bar s), \notag \\  W&= \kappa^{-3} a, \notag \\ f(s) &= s. \label{model:p2} \end{align}
The scalar potential (\ref{scalarpot_generalp}) then reduces to
\begin{equation} \mathcal V = \kappa^{-4} a^2 e^{b(s+\bar s)} \left(  \frac{b^2}{2}  - \frac{2b}{s+\bar s}  - \frac{1}{(s+ \bar s)^2 }    \right) + \kappa^{-4}  \frac{c^2}{s + \bar s} \left( \frac{2}{s+\bar s} - b \right)^2.  \label{scalarpotp2} \end{equation}
As in the previous section, we first look for a Minkowski minimum and solve the equations
\begin{align} \mathcal V(s) &= 0,\notag  \\ \frac{d \mathcal V}{ds}(s) &= 0. \end{align}
This leads to the following relations between the parameters at the minimum of the potential\footnote{
The definition of $A(\alpha)$ is the inverse of the one used in \cite{RK1,RK2,RK4} in order to be consistent with the rest of this work.}:
\begin{align} b \langle s + \bar s \rangle& = \alpha \approx -0.183268, \notag \\ \frac{b c^2}{a^2} &= A(\alpha) \label{model:tunability_p2}, \end{align}
 where $\alpha$ is the root of the polynomial $- x^5 + 7 x^4 - 10 x^3 - 22 x^2 + 40 x + 8$ close to  $-0.18$ and $A(\alpha)$ is given by
\begin{align}  A(\alpha) = e^{\alpha} \alpha \left( \frac{\frac{\alpha^2}{2} - 2 \alpha - 1}{-4 + 4 \alpha - \alpha^2} \right)  \approx -0.01970. \label{model:Aalpha_p2} \end{align}
 Note that the above polynomial has five roots, four of which are unphysical: two roots are imaginary, 
one is positive, which is incompatible with the first line in eqs.~(\ref{model:tunability_p2}), since $ \langle s + \bar s \rangle$ should be positive by eq.~(\ref{sgaugecoupling}), 
and a fourth root gives rise to a positive $A(\alpha)$, which is incompatible with the last line in eqs.~(\ref{model:tunability_p2}) since $b<0$. 

If one were to shift the value of the minimum of the potential to a small positive value $\Lambda$, eq.~(\ref{model:tunability_p2}) gets modified to
\begin{align}  A_\Lambda (\alpha)  = \frac{a^2}{bc^2} - \frac{\kappa^4 \Lambda}{b^3 c^2} \left( \frac{\alpha^2 e^{-\alpha}}{\frac{\alpha^2}{2} - 2\alpha -1}  \right).  \end{align}
It follows that by carefully tuning $a$ and $c$, $\Lambda$ can be made positive and arbitrarily small independently of the supersymmetry breaking scale. 
The gravitino mass term is given by
\begin{align}m_{3/2} = \frac{ \kappa^{-1} a b}{\alpha} e^{\alpha/2} . \label{model:gravitinomassp2} \end{align}

\begin{figure}[!ht]    \centering  \medskip   \includegraphics[width=0.9\textwidth]{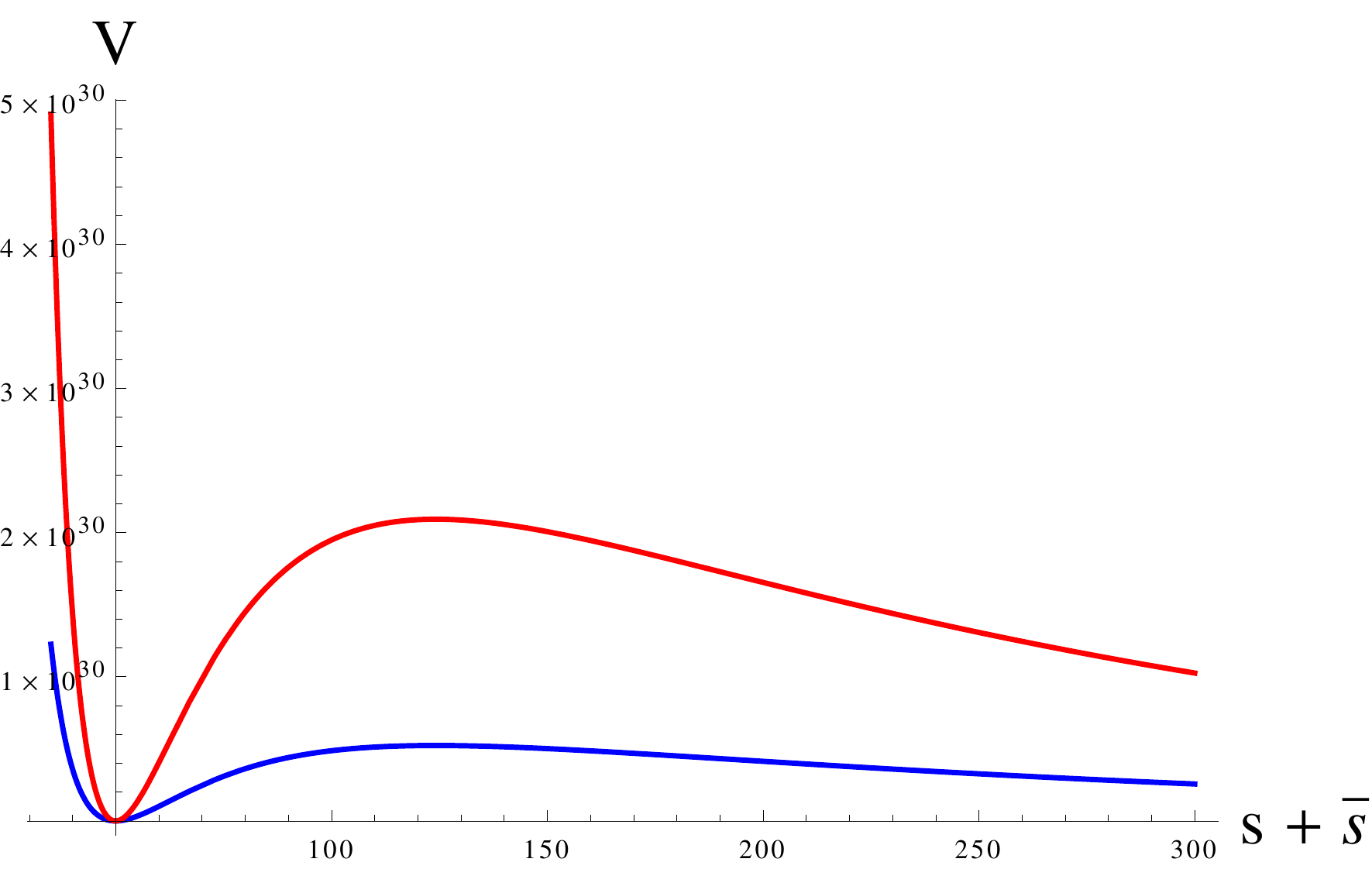}
    \caption{A plot of the scalar potential (\ref{scalarpotp2}) for $p=2$, $b=\alpha/50$, $a = 3\times 10^{-14}$ (blue) and $a = 6 \times 10^{-14}$  (red),
    and $c$ determined by eq.~(\ref{model:tunability_p2}).
 The VEV of $s + \bar s$ is specified by choosing $b=\alpha/50$, where we have taken $\langle s + \bar s \rangle = 50$ in this plot. The values of $a$ are chosen such that the gravitino mass is of the order of TeV. 
Note that the scalar potential has a metastable dS minimum at $\langle s + \bar s \rangle = 50$. The metastability is discussed below.}   \label{fig:plot_p2} \end{figure}

We conclude that one can obtain a tunable de Sitter vacuum with the minimum of the potential at $\langle s + \bar s \rangle = \alpha / b$. As in the previous section, 
the parameter $b$ can be used to determine the value of $\text{Re}(s)$ at the minimum of the potential.
Since the gauge kinetic function is linear in the dilaton $s$, we have $\langle s + \bar s \rangle = 1/g_s^2$, where $g_s$ is the string coupling constant, which can thus be tuned by the parameter $b$, independently of the minimum of the potential and the gravitino mass.
 A TeV gravitino mass can be obtained by tuning $a |b| \sim 8.3 \times 10^{-17}$, and the parameter $c$ should satisfy eq.~(\ref{model:tunability_p2}). 
The scalar potential is plotted in figure~\ref{fig:plot_p2} for $a=3 \times 10^{-14}$ and $a=6\times 10^{-14}$, for $b=\alpha/50$ and $c$ determined by eq.~(\ref{model:tunability_p2}).
 Note that, in contrast with the $p=1$ case in section~\ref{sec:p=1}, the de Sitter vacuum is metastable since $\mathcal V \rightarrow 0$ when $s + \bar s \rightarrow \infty$. 
One should therefore check whether this vacuum is sufficiently long lived. This is confirmed below in section~\ref{sec:p2stability}.

Due to a St\"uckelberg mechanism, the imaginary part of $s$ (the axion) is eaten by the gauge field, which acquires a mass. 
On the other hand, the Goldstino $P_L \nu$, which is a linear combination of the fermion of the chiral multiplet $\chi$ and the gaugino $\lambda$, is eaten by the gravitino (see section~\ref{sec:spontaneous susy breaking}). 
As a result, the physical spectrum of the theory consists (besides the graviton) of a massive scalar, namely the dilaton, a Majorana fermion, 
a massive gauge field and a massive gravitino. A calculation of the masses of these fields is presented in Appendix~\ref{Appendix:masses}. The results are
 \begin{align}
 m_s &=  \kappa^{-1} \frac{b^2 c}{\alpha^2} \sqrt{ \frac{  48+192 \alpha-128 \alpha^2-8 \alpha^3+24 \alpha^4-8 \alpha^5+\alpha^6 } {2+4 \alpha-\alpha^2} }, \notag \\
  m_{A_\mu} &= \kappa^{-1} \frac{2 \sqrt{2}  b^{3/2} c}{\alpha^{3/2}}, \notag \\
 m_f^2 &= \kappa^{-2} a^2 b^2 \frac{e^\alpha \left( 116 + 68 \alpha - 15 \alpha^2  - 4 \alpha^3 + 8 \alpha^4\right)}{144 \alpha^2} .
 \end{align}
  Note that by eqs.~(\ref{model:tunability_p2}) and eq.~(\ref{model:gravitinomassp2}) all masses are proportional to the gravitino mass, 
  which is proportional to the constant $a$ (or $c$ related by eq.~(\ref{model:tunability_p2})). 
Thus, these masses vanish in the supersymmetric limit where the charge of the shift symmetry is taken to zero $c \to 0$, or equivalently, in the limit $m_{3/2} \rightarrow 0$.

\section{A few remarks on $p=2$} \label{sec:remarksp2}

\subsection{Metastability of the de Sitter vacuum} \label{sec:p2stability}

Classically, the above vacua are stable. For the case\footnote{In the case where $p=1$, the ground state is stable when $ \kappa^{-4} \xi^2/2 > \Lambda$ (see eq.~(\ref{p1_limit_to_ifty})) and such analysis is not necessary.}
$p=2$ however, the scalar potential tends to zero for $\re (s) \rightarrow \infty$.
Although the ground state in eq.~(\ref{model:tunability_p2}) is separated by a barrier (see Fig.~\ref{fig:plot_p2}) from this runaway direction at $\re(s) = \infty$, quantum tunneling can in principle occur between these solutions.
One thus needs to estimate the probability ($\Gamma$) for the current vacuum  $ \mathcal V_{\text{min}}= \Lambda = \kappa^{-4} \epsilon_0>0$ 
to decay into the vacuum $\mathcal V_{\text{min}}=0$ along the $\re(s)$ direction, 
through the potential barrier (Fig. \ref{fig:plot_p2}). 
This is to ensure that the ground state is long-lived enough. This probability is (per unit of time and volume) 
\begin{align} {\Gamma}&= \cA \, e^{-\frac{\cB}{\hbar}} \left(1 + O(\hbar) \right), \end{align}
where $\hbar$ is the reduced Planck constant (we set $\hbar = 1$), and $\cA$ and $\cB$ depend on the model. The value of $\cA$ plays a minor role in comparison with the exponential suppression; $\cB$ is fixed  by the Euclidean action  of the instanton (bounce) solution ($S_1$) which in the limit of a very small energy difference  between the two minima is~\cite{Coleman1,Coleman2,Coleman3}
\begin{align} \cB  = \frac{27 \pi^2 S_1^4}{2\,\epsilon^3_0}. \label{B} \end{align}
We redefine the field $s$ into (for standard kinetic terms; $s$ is dimensionless)
\begin{align} \phi_s = \kappa^{-1} \log (s + \bar s), \end{align}
$S_1$ is given by \cite{Coleman1}
\begin{align} S_1 &= \int d \phi_s \sqrt{2 \mathcal V(\phi_s)}=\kappa^{-3} | a| \,\cS(b), \end{align}
where  $\mathcal V$ is the potential of $\phi_s$ and
\begin{align} \cS(b)& =& \sqrt 2 \int_{\ln\left( \frac{\alpha}{b} \right)}^\infty d x \,   
\Big[e^{b\, e^{x}} \left( \frac{b^2}{2} -2\, b e^{-x} - e^{-2x} \right) +   \frac{e^{-x} A(\alpha) }{b } \left( 2\, e^{-x} - b\right)^2  \Big]^{1/2}, \notag \end{align}
which can be computed numerically.  $A(\alpha)$ is given by eq.~(\ref{model:Aalpha_p2}) (where one has set $\Lambda =0$ to a good approximation). 

If we  interpret $\re (s)$ as the (inverse of the) 4D coupling ($1/g_s^2$) to a GUT-like value,  $\re (s)\approx 25$, this  fixes $b$ via $b=\alpha/\langle s+\overline s\rangle\approx -3\times 10^{-3}$.  For this value of $b$ one finds  $\cS \approx 0.0124106$. By demanding that the gravitino mass is of order  TeV (by tuning  $a$, see previous section) one finds  $ \cB \approx 10^{297}$. Therefore $\Gamma$ is extremely  small (largely due to the small difference between the two minima)\footnote{Usually values of $\cB\geq 400$ are regarded as metastable enough \cite{Blum:2009na}.}; the ground state  is long lived enough to use this model as a starting point for building  realistic models, by adding  physical fields, that have such ground state {\it along} this field direction.
However, we will show in section~\ref{sec:Anomalies:p=2} that other constraints coming from quantum anomalies render the model with $p=2$ useless for this purpose.

\subsection{Connection with string theory} \label{sec:p2stringtheory}

As supergravity theories appear as a low energy limit (where the string length goes to zero) of string theory, 
it is an interesting question whether one can find a UV-completion of this model in the context of string theory. 
Indeed, it turns out that for $b=0$, the scalar potential (\ref{scalarpotp2}) reduces to
\begin{equation} \mathcal V =  -  \frac{ \kappa^{-4} a^2}{(s+ \bar s)^2 }   +  \frac{4 \kappa^{-4} c^2}{(s+\bar s)^3}. \label{Vb0} \end{equation}
The potential (\ref{Vb0}) coincides with the one derived in \cite{Antoniadis:2008uk} from D-branes in non-critical strings. Indeed the second term corresponds to a disc contribution proportional to the D-brane tension deficit $\delta \bar T$ induced by the presence of magnetized branes in type I orientifold compactifications, while the first term corresponds to an additional contribution at the sphere-level, induced from going off criticality, proportional to the central charge deficit $\delta c$.  More precisely, in the Einstein frame ($M_p = 1$), the scalar potential in \cite{Antoniadis:2008uk} is given by
\begin{align} \mathcal V_{nc} = e^{2\varphi_4} \delta c + \frac{e^{3 \varphi_4}}{(2 \pi)^3 v_6^{1/2}} \delta \bar T, \label{V_nc} \end{align}
where $\varphi_4$ is the four-dimensional dilaton related to the ten-dimensional dilaton $\varphi$ by $e^{-2 \varphi_4} = e^{-2 \varphi} v_6$, and $v_6$ is the six-dimensional volume given by $v_6 = \frac{V_6}{(4 \pi^2 \alpha ')^3}$.  By identifying $e^{-\varphi_4} = \text{Re}(s)$, one sees that the scalar potential (\ref{V_nc}) is indeed of the form of equation (\ref{Vb0}). One can then identify $\delta c$ and $\delta \bar T$ as 
\begin{align} \delta c &= -\frac{\kappa^{-4} a^2}{4}, \notag \\ \delta \bar T &= \frac{(2 \pi)^3 v_6^{1/2}}{2} \kappa^{-4} c^2 .
\end{align}
Note that $\delta c$ can become infinitesimally small only if it is negative~\cite{Antoniadis:2008uk}, 
as is needed for the existence of an anti-de Sitter vacuum in equation (\ref{V_nc}). 
The potential has a minimum at $ \langle s + \bar s \rangle = \frac{6c^2}{a^2}$, given by $\mathcal V(\langle s\rangle) = -\frac{\kappa^{-4} a^6}{108 c^4} < 0$.

It is interesting to note that a non-zero $b$ implied by the most general superpotential consistent with the shift symmetry allows one to find (metastable) de Sitter vacua for certain values of the parameters, as is shown in section \ref{sec:p=2}.

\subsection{Linear-Chiral duality} \label{sec:linearchiral}

Although the model~(\ref{model:p2}) is formulated in terms of a chiral multiplet and a vector multiplet, 
we show in this section how it can be reformulated in terms of a linear multiplet (see section \ref{sec:ingredients}) and a vector multiplet. 
Because of the shift symmetry~(\ref{shift}) of the axion there exists a dual description where the axion is described by an antisymmetric tensor $b_{\mu \nu}$, 
which has a gauge symmetry given by eq.~(\ref{linear:bgauge}), repeated here for convenience of the reader
\begin{align}
b_{\mu \nu} \longrightarrow b_{\mu \nu}+ \partial_\mu b_\nu - \partial_\nu b_\mu. \notag
\end{align}
 In the context of string theory, it is this antisymmetric tensor that appears in the massless spectrum as the supersymmetric partner of the dilaton.
The antisymmetric tensor is related to a field strength $v_\mu$ by eq. (\ref{linear:vmu}) from which divergenceless of the field strength (which is the Bianchi identity) eq.~(\ref{linear:vmu_constraint}) follows.
Together with a real scalar $l$ and a Majorana fermion $\chi$, this real vector $v_\mu$ belongs to a linear multiplet $L$. 
It should be noted that the shift symmetry (\ref{shift}) of the axionic partner of the dilaton is a crucial ingredient for the duality to work, 
since it is related to the gauge symmetry (\ref{linear:bgauge}) which in turn is responsible for the divergenceless of the field strength eq.~(\ref{linear:vmu_constraint}).

Below, we present the dual version of the model~(\ref{model:exp_superpot}) in the representation where the superpotential is an exponential of $S$ below for $p=2$, 
which is to our knowledge the first time the linear-chiral duality has been performed in the presence of a non-trivial superpotential.
The details of this calculation can be found in Appendix~\ref{ch:appendix_linear_chiral}, 
where a superfield formalism is used in the globally supersymmetric case, and the chiral compensator formalism is used in the locally supersymmetric case. 

The result eq.~(\ref{SUGRAL2}), repeated here for convenience of the reader, is given by
\begin{align} \mathcal L 
&=   \left[ - \frac{1}{2} \left( S_0 e^{-gV_R} \bar S_0 \right)^3 L^{-2} \right]_D - \left[ c L V_R \right]_D \notag  \\ 
& + \frac{1}{b} \left[ \left( T(L) - \alpha \mathcal W^2 \right) - \left( T(L) - \alpha \mathcal W^2 \right) \ln \left( \frac{ T(L) - \alpha \mathcal W^2 }{ab S^3_0} \right) \right]_F + \text{h.c.} \notag \\
& + \left[ \beta \mathcal W^2 \right]_F+ \text{h.c.}, \label{linearchiralresult} \end{align}
where the operations $[ \ \  ]_F$ and $[  \ \ ]_D$ are defined in eqs. (\ref{F-term}) and (\ref{D-term}) respectively.
Although in eq.~(\ref{linearchiralresult}) the multiplet $L$ is an a priori unconstrained multiplet, 
it is shown in the Appendix~\ref{ch:appendix_linear_chiral} that $L$ indeed contains the degrees of freedom of a linear multiplet. 
Moreover, it is shown that the (non-gravitational) spectrum of this theory can be described, except for a gauge multiplet,
in terms of two real scalars $l$ and $w$, where $w$ is defined as $Z = \rho e^{iw}$, with $Z$ the $\theta^2$-component of $L$,
and a Majorana fermion.

In particular, the F-term contribution to the scalar potential is given in terms of $l$ by
\begin{align}
 \mathcal V_F = \kappa^{-4} a^2 e^{b/l} \left( \frac{b^2}{2} - 2bl - l^2 \right), 
\end{align}
which on substitution of $s+ \bar s = 1/l$ (see eq.~(\ref{SUGRAls})) indeed gives the F-term contribution to the scalar potential in eq.~(\ref{scalarpotp2}).

\cleardoublepage
  \graphicspath{{\chaptersdir/anomalies/image/}}%
  \chapter{Anomalies in supergravity theories with FI terms}\label{ch:anomalies}

Symmetries, and in particular gauge symmetries, play an important role in quantum field theories. 
A symmetry of the classical Lagrangian is a field transformation that leaves the Lagrangian, or rather the action,  invariant.
Examples of such symmetries are  the $SU(3)\times SU(2) \times U(1)$ Standard Model gauge group, Poincar\'e transformations or supersymmetry transformations.
It is an important question whether these symmetries still persist at the quantum level. 
If the symmetry is violated in the quantum theory, it is said to be anomalous.
There are several ways to understand how such anomalies can arise.

For example, a violation of the classical action at the quantum level can be seen in the functional integral formulation of quantum field theory. 
Here, the symmetries of the (classical) action can be expressed in the Ward identities for the correlation functions. 
An important assumption, however, is that not only the Lagrangian, but also the integral measure is invariant under the symmetry.
When this is not the case, the symmetry is violated at the quantum level and the symmetry is said to be anomalous and the Ward identities are violated\footnote{
For an introduction to anomalies in quantum field theory, see for example~\cite{Bilal_anomalies}.}. 

Alternatively, one can check whether a classically conserved current remains conserved in the quantum theory. A current that is not conserved at the quantum level indicates the presence of an anomalous contribution to the quantum effective action.
In practice, one can check whether fermion triangle (Feynman) diagrams spoil current conservation at the quantum level \cite{triangle_anomaly}. 

Finally, in computing Feynman diagrams, one is forced to introduce a regularization. Often, it may happen that there exists no regularization that preserves all symmetries. 
In this case, one can not guarantee that the Ward identities (and thus the classical symmetries) are displayed in the renormalized Green's functions. If they do not, the theory is anomalous. 

At this point it is important to distinguish between global and local symmetries:
An anomalous global symmetry of the action does not pose a problem to the consistency of a theory; it merely implies that the classical selection rules are not obeyed in the quantum theory. 
Consequently, certain classically forbidden processes may occur.

The presence of an anomalous local symmetry on the other hand is disastrous for the quantum consistency of any quantum field theory\footnote{
Luckily for the Standard Model, 
where one can a priori have anomalies for each of the triangle diagrams corresponding to $SU(3)^2 \times U(1)$, $SU(2)^2 \times U(1)$, $U(1)^3$ gauge bosons on the external lines, and a mixed gravitational-$U(1)$ anomaly,
some miraculous cancellations occur between the various anomalous contributions of its gauge symmetries within a fermion generation (See for example \cite{anomalies_Standard_Model}, where this was originally shown for a four-quark version of the Standard Model).}. 
We should therefore check, before investigating its low energy phenomenology in chapter \ref{ch:softterms}, 
whether the extra $U(1)_R$ gauge group introduced in the previous chapter in order to obtain a tunable cosmological constant, can be consistently combined with the MSSM in an anomaly-free way.
Moreover, anomaly cancellation conditions can impose strong constraints on a theory.
Luckily, since our model has a scalar field with a shift symmetry, 
we have the terms~(\ref{model:Green-Schwarz term})) whose gauge transformation can cancel certain anomalies, which is the Green-Schwarz mechanism.

The starting point in this chapter will be the work of Elvang, Freedman and K\"ors~\cite{F1,F2} on anomalies in supergravity theories with FI terms by using the Fujikawa method~\cite{Fujikawamethod}.  
We briefly summarize their results in section~\ref{sec:Anomalies-FI}. 
Although these results are correct, they are obscure from a field-theoretic point of view.
We present our work with I.~Antoniadis and D.~Ghilencea~\cite{RK2}, 
where the anomaly cancellation conditions of~\cite{F2} are reformulated in such a way that their interpretation can be interpreted and compared with a naive field-theoretic approach, 
which is presented in Appendix~\ref{ch:Appendix_anomalies}.
These results are then applied  in section~\ref{sec:Anomalies:p=2} to our the model with $p=2$ (see section~\ref{sec:p=2}), and it is shown that the anomaly cancellation condition are inconsistent with a TeV gravitino mass which is necessary to have a viable low energy phenomenology.
We conclude this chapter in section~\ref{sec:Anomalies-gauginomass} where we elaborate on the relation between quantum anomalies and the generation of gaugino mass terms at one loop.

Section~\ref{sec:Anomalies-FI} is based on~\cite{RK2}, while section~\ref{sec:Anomalies:p=2} is based on~\cite{RK2,RK4}, and section~\ref{sec:Anomalies-gauginomass} is a result from~\cite{RK3}.

\section{Anomalies in supergravity theories with FI terms} \label{sec:Anomalies-FI}

In the previous chapter we introduced a model that allows for a tunable cosmological constant, while leaving the gravitino mass separately tunable. 
This discussion however was limited to the classical level. As was elaborated in the introduction, quantum consistency can lead to additional constraints on a theory by imposing anomaly cancellation conditions.
Such conditions can have an important impact on a theory. 

Anomaly cancellation in supergravity with an anomalous $U(1)$ gauge symmetry with Fayet-Iliopoulos terms and a Green-Schwarz mechanism were discussed in the past  in  a general setting in \cite{F2,F1}.  
This subject is certainly not new, 
but it is worth revisiting since supergravity theories with Fayet-Iliopoulos couplings appear 
for example as the effective four-dimensional effective theory in flux compactifications of string theory \cite{gauged_R-symmetry}. 

In supergravity, the metric $g_{\alpha \bar \beta}$ is invariant under K\"ahler transformations~(\ref{Kahler_tranformation}).
In certain theories the K\"ahler potential is not a global scalar. Instead, it can be defined locally in each coordinate chart, 
and the definitions of the K\"ahler potentials in the different charts are related by a K\"ahler transformation.
Such a K\"ahler transformation however is accompanied by an appropriate axial gauge transformation of the fermions in the model, which can be anomalous.
This so-called K\"ahler anomaly does not make the theory inconsistent at the quantum level.
Instead, in this section we focus on the situation where a gauge transformation does not leave the K\"ahler potential invariant,
but leaves a K\"ahler transformation. This typically leads to Fayet-Iliopoulos terms (see section~\ref{sec:sugra_ingredients}).
The accompanying transformations of the fermions can be anomalous and in this section we review their anomaly cancellation conditions.

Here, we review the results of~\cite{F2}. The relation and agreement of such results to the ``naive'' field theory approach however, 
was not examined for the rather special case of  a  gauged $U(1)_R$. 
In fact, in global SUSY $U(1)_R$ can not even be gauged. This can easily be seen from the commutation relations of the supersymmetry algebra eqs.~(\ref{algebra:R-symmetry}).
In this section, we rewrite the results of~\cite{F2} in such a way that they can successfully be interpreted from a field-theoretic point of view 
(which is reviewed in the Appendix~\ref{ch:Appendix_anomalies}).
This section, based on work with D.~Ghilencea and I.~Antoniadis \cite{RK2} is also relevant on its own, independent of the rest of this thesis.

Below, we assume that the field content is that of minimal supergravity, including a gauged $U(1)_R$, and a dilaton multiplet $S$ 
whose shift symmetry gives rise to a Green-Schwarz mechanism (as for example in chapter~\ref{ch:dS}, 
where in the K\"ahler basis with an exponential superpotential eq.~(\ref{model:exp_superpot}) the gauged shift symmetry is indeed a gauged R-symmetry). 
Additionally, we assume the presence of additional MSSM-like superfields, and that the gauge symmetries are those of the Standard Model gauge group times $U(1)_R$. 
The anomaly cancellation conditions  for the cubic anomaly $U(1)_R^3$  ((a) below), and the mixed anomalies of $U(1)_R$  with the K\"ahler connection $K_\mu$ ((b), (c)), 
with the SM gauge group (d) and  gravity (e) are given
\glossary{name=SM,description=Standard Model}
 by in eqs.~(4.4) in Section~4 of \cite{F2}. These are\footnote{
 In~\cite{RK2,F2}, the transformation of a field $z$ with charge $Q$ under an abelian gauge group is defined as $\delta z = \exp(-i \theta Q) z$. 
 In this thesis, we follow the conventions of~\cite{TheBook}, where one uses $\delta z = \exp(i \theta Q) z$. 
 As a result, in eq.~(\ref{an}) the signs of the charges (and FI-term) are flipped compared to~\cite{RK2,F2}.}
\begin{align}  &(a)& C \tilde C  &:  & 0  &= - \Tr\big[(Q+\xi/2)\,Q^2\big] - \xi\,a_{KCC}- c \,b_{C} \notag , \\
&(b)&C\tilde K&:  & 0 &= \Tr\big[(Q+\xi/2) Q\big] + \xi\,a_{CKK}-a_{KCC} - c \,b_{CK} \notag , \\
&(c)&  K\tilde K&:  & 0&= - \Tr\big[Q\big] + \frac{1}{2}\, \xi\,(n_{\lambda}+3-n_\chi) +4\,a_{CKK}-4\,c \,b_{K}, \notag\\
&(d)& (F\tilde F)_A& : & 0&=  \Tr\big[ Q\,(\tau^a \tau^b)_A \big] - \frac{\xi}{2}\,\big[ T_G - T_R \big]_A \delta^{ab} +\frac{1}{3} \,c\,b_{A}\,\delta^{ab} , \notag \\
&(e)& \mathcal{R} \tilde{\mathcal{R}}&: &0 &= -  \Tr\big[Q\big] + \frac{\xi}{2}\,(n_\lambda-21-n_\chi) +8 \, c \, b_{\text{grav}}, \label{an} 
\end{align}
where $\tilde{}\,$  labels the dual field strength defined in eq.~(\ref{conventions:dual_field_strength}). 
Here,  
the gauge field (strength) of $U(1)_R$ is indicated by $C_\mu$ $(C_{\mu\nu})$ to distinguish its role from the Standard Model gauge fields (strengths) $A_\mu$ ($F_{\mu\nu}$)
(with group generators\footnote{
 Refs.~\cite{F1,F2}, as well as~\cite{TheBook} use anti-hermitian  $T^a =-i\tau^a$, so we added a minus  in front of $\Tr$ in line (d) of eqs.~(\ref{an}).} $\tau^a$);  
$A$ is a group index that runs over $U(1)_Y$, $SU(2)_L$, $SU(3)$.  The K\"ahler connection $K_\mu$ essentially fixes the coupling of the gravitino. 
These three fields are involved in the conditions (a), (b), (c), (d). Condition (e) is for the mixed, $U(1)_R$-gravitational anomaly.
There is also the usual condition  $\Tr[\tau^a\{\tau^b,\tau^c\}]=0$ for the Standard Model group generators.

In eqs.~(\ref{an}), $n_\lambda$ is the number of gauginos present (or equivalently, the number of vector multiplets), 
$n_\chi$  is the number of chiral multiplets which include both the matter fermions and the dilatino\footnote{
The dilatino is the fermionic component of the $S$-multiplet.
}.
The $U(1)_R$-charges $Q$ of the (scalar) matter superfields are 
such that the Killing vector corresponding to $U(1)_R$ of a (complex) scalar field $z^\alpha$ in a chiral multiplet is given by
\begin{align}
 k_R^\alpha = i Q z^\alpha \label{Q-killing}.
\end{align}
Moreover in eqs.~(\ref{an}), $\kappa^{-2}\xi$ is the Fayet-Iliopoulos term\footnote{
Note that the contributions to eqs.~(\ref{an}) proportional to $\xi$ are Planck suppressed compared to the FI-term $\kappa^{-2} \xi$.
} corresponding with $U(1)_R$.
Since $U(1)_R$ is an R-symmetry, the Standard Model gauginos,  the $U(1)_R$ gaugino  and the gravitino are
(unlike other types of anomalous $U(1)$'s) all charged under $U(1)_R$ and their contribution to the anomalies should be taken into account.
Next, $\Tr_r(\tau^a \tau^b)=T_R\,\delta^{ab}$ is the trace over the irreducible representations $R$, 
and $f^{acd} f^{bcd} =\delta^{ab} T_G$, with $T_G=N$ for $SU(N)$  and 0 for  $U(1)$.
The terms in eqs.~(\ref{an}) proportional to $\xi$ with coefficients $+3$ and $-21$, in (c) and (e) respectively, 
denote the contributions of the gravitino to those anomalies~\cite{grisaru}. 
Finally, $c$ is the Green-Schwarz coefficient and $a_{KCC}$, $a_{CKK}$  are coefficients of local counterterms that can be present (defined below).

The K\"ahler connection\footnote{The K\"ahler connection differs with a factor $1/3$ from the one in~\cite{TheBook}.
This factor has also been taken into account in the covariant derivatives in eqs.~(\ref{cov_derivatives}).}
\begin{align}
 K_\mu &= \frac{i}{2} \kappa^2 \left(\partial_\mu z^\alpha \partial_\alpha \mathcal K  - \partial_\mu \bar z^{\bar \alpha} \partial_{\bar \alpha} \mathcal K \right) - A_\mu^A \mathcal P_A \notag \\
 &= \frac{i}{2} \kappa^2 \left( \hat \partial_\mu z^\alpha \partial_\alpha \mathcal K - \hat \partial_\mu \bar z^{\bar \alpha} \partial_{\bar \alpha} \mathcal K  \right) + A_\mu^A (r_A - \bar r_A)
\label{Kahler-connection}
 \end{align}
transforms under gauge ($\delta z^\alpha = k^\alpha_A \theta^A$) and K\"ahler transformations (see eqs.~(\ref{Kahler_tranformation})) as
\begin{align} 
\delta K_\mu = -\kappa^2 \partial_\mu \left( \theta^A \text{Im} (r_A) + \text{Im} (J) \right).
\end{align}
In eq.~(\ref{Kahler-connection}), the covariant derivative is defined as
\begin{align}
\hat \partial_\mu z^\alpha = \partial_\mu - A_\mu^A k_A^\alpha. 
\end{align}
To understand the role of the K\"ahler connection it is instructive to write down the covariant derivatives
 of the gravitino $\psi_\mu$, the gauginos $\lambda$ and (MSSM-like) matter fermions $\chi$ 
\begin{align}
 D_\mu \psi_\nu &= \left( \partial_\nu - \frac{i}{2} K_\mu \gamma_* \right) \psi_\nu, \notag \\
 D_\mu \lambda^A &= \left( \partial_\mu - \frac{i}{2} K_\mu \gamma_* \right) \lambda^A - A_\mu^C \lambda^B f_{BC}^A, \notag \\
 D_\mu \chi^\alpha &= \left( \partial_\mu + \frac{1}{2} i K_\mu \right) \chi^\alpha - A_\mu^A \frac{\partial k_A^\alpha}{\partial z^\beta} \chi^\beta, \label{cov_derivatives}
\end{align}
where we have omitted terms involving the $\psi$-torsion, the spin connection and the Christoffel connection.
As was discussed in chapter~\ref{ch:supergravity}, the K\"ahler potential and the superpotential are not invariant under an R-symmetry.
Instead, a $U(1)_R$ gauge transformation leaves a K\"ahler transformation, given by eqs.~(\ref{Ktransf}) and (\ref{Wtransf}).
Indeed, the contributions from the K\"ahler connection $K_\mu$ are taken into account in the anomaly cancellation conditions eqs.~(\ref{an}).
We will however show below that the conditions (a), (b) and (c) can be reduced to the familiar cubic anomaly condition, 
while the remaining two (independent) conditions can be satisfied by a suitable choice of the counterterms $a_{CKK}$ and $a_{KCC}$.

The local counterterms that come with coefficients $a_{KCC}$, $a_{CKK}$ are~\cite{F1,F2}
\begin{align}  \mathcal L_{KC}=\frac{1}{24\pi^2}\,\epsilon^{\mu\nu\rho\sigma}\, 
\left[ a_{CKK}\,C_\mu\,K_\nu\,\partial_\rho K_\sigma + a_{KCC}\,K_\mu\,C_\nu \partial_\rho C_\sigma \right], \end{align}
and $b_C$, $b_{CK}$, $b_K$, etc,  of eqs.~(\ref{an}) are coefficients present in the Chern-Simons terms\cite{Anastasopoulos:2006cz,Anastasopoulos:2007qm,DeRydt:2007vg}
\begin{align}\label{zz} 
 \mathcal L_{CS} &= \frac{1}{48\pi^2}\,\text{Im}(s)\,\epsilon^{\mu\nu\rho\sigma} \partial_\mu\Omega_{\nu\rho\sigma} \\
&= \frac{1}{96 \pi^2} \text{Im}(s) \, \epsilon^{\mu\nu\rho\sigma} \left[ b_C C_{\mu \nu} C_{\rho \sigma} + b_{CK} C_{\mu \nu} K_{\rho \sigma} + b_K K_{\mu \nu} K_{\rho \sigma} \right. \notag \\
& \quad \quad \quad \left. \  + b_A (F_{\mu \nu} F_{\rho \sigma})_A + b_{\text{grav}} \mathcal R_{\mu \nu} \mathcal R_{\rho \sigma} \right]. \notag 
\end{align}
In the context of string theory some of  the coefficients $b_C$, $b_K$, $b_{CK}$, 
$b_{\text{grav}}$ can be related, as is the case for example in heterotic string theory (for $b_C$ and  $b_{\text{grav}}$) and they can therefore not  be adjusted at will. 
However, since it is difficult to derive a $U(1)_R$  from  strings, this applies only to non-R anomalous $U(1)$'s. We therefore relax this constraint and consider them to be independent.
Additionally, we assume that the dilaton ($S$) is the only (super)field that transforms nonlinearly under the gauged $U(1)_R$ and it implements  the Green-Schwarz mechanism. 
The canonical normalization of the gauge kinetic term for the $U(1)_R$ gauge field ($C_\mu$) gives  $b_C=12\pi^2$, while we assume that $b_K$ and $b_{CK}$ can in general be non-zero.

The supergravity anomaly cancellation conditions (\ref{an}) are not transparent from the ``naive'' field theory point of view for the anomalies of $U(1)_R$. 
So let us clarify the link between these conditions and  the  field theory result  in Appendix~\ref{ch:Appendix_anomalies}.
First, the $U(1)_R$-charges of the fields are shown below (see Appendix~\ref{ch:Appendix_anomalies}) and depend on the FI term(s): 
\begin{align} 
R_\chi=Q+\xi/2, \qquad R_\lambda= R_{3/2}= -\xi/2,\qquad R_{\chi_s}=\xi/2, 
\end{align}
where $Q$ are the charges of the scalar components of the matter superfields (see eq.~(\ref{Q-killing}));  
$R_\lambda$, $R_{3/2}$ and $R_{\chi_s}$ are  the charges of the gauginos $\lambda^A$, the gravitino $\psi_\mu$ and the dilatino $\chi_s$, respectively.
Using this information, the first three relations in eqs.~(\ref{an}) can be combined, 
after multiplying them by $4$, $ -4\xi$ and $\xi^2$ respectively, and then adding them by using that $\Tr 1=n_\chi-1$. The result is\footnote{ 
``-1'' in $\Tr 1= n_\chi-1$  isolates the dilatino from matter fermions (the charge $Q$ of the dilaton is 0),
since $n_\chi$ is the total amount of chiral multiplets, which includes the dilaton multiplet $S$.}
\begin{align}\label{fieldtheory} \Tr [R_\chi^3]+ n_\lambda R_\lambda^3 + 3 R_{3/2}^3+R_{\chi_s}^3 = - c \left[ b_{C} - \xi b_{CK} +\xi^2  b_{K}\right]. \end{align}
On the lhs 
\glossary{name=lhs,description=left-hand side}
one recognizes the usual field theory cubic $U(1)_R^3$ anomaly cancellation condition in the presence of FI terms and Green-Schwarz mechanism, \glossary{name=GS,description=Green-Schwarz (mechanism)} 
in which all fermionic contributions are added and compensated by a Green-Schwarz shift on the rhs:
\glossary{name=rhs,description=right-hand side}
The trace includes all contributions from matter fermions.
The number of gauginos is $n_\lambda=1\!+\!(8\!+\!3\!+\!1)\!=\!13$ for the  $U(1)_R\times$ Standard Model gauge group.  Each gaugino has a contribution  $R_\lambda^3$.
The gravitino contribution $(+3) R_{3/2}^3$ is three times larger than that of one gaugino, as was shown in~\cite{grisaru}.
The above result has (with $b_C\!=\!12\pi^2$ and $b_{CK}\!=\!b_{K}\!=\!0$) the same form as  the ``naive'' field theory result,
 eqs.~(\ref{an:cubic}) in Appendix~\ref{ch:Appendix_anomalies}. This is interesting since in global SUSY $U(1)_R$ can not even be gauged.  
Note however the  difference in the rhs  due to   $\xi b_{CK}+\xi^2 b_K$.
The terms in $\mathcal L_{CS}$ of coefficients $b_K$, $b_{CK}$ are not present in the naive field theory case, and give extra freedom in canceling this anomaly.

The two remaining independent conditions of constraints (a), (b),  (c)  in eq.~(\ref{an}), refer to K\"ahler and mixed $U(1)_R$-K\"ahler connection. 
They  can always be respected by a suitable choice of $a_{KCC}$ and $a_{CKK}$ of the local counterterms shown and  are not further discussed.
One finds (by  combining these two remaining constraints with  condition (e) in eq.~(\ref{an})), 
that  $a_{CKK}= - 3\xi  + c (b_K + 2 b_{\text{grav}})$, while $a_{KCC}$ is found from one of equations (a), (b), (c) in eq.~(\ref{an}).

The last two  conditions in eq.~(\ref{an}) can be re-written as 
\begin{align}\label{ee} 
(F\tilde F)_A: & \quad \Tr \left[R_\chi (\tau^a\tau^b)_A \right] + T_G \delta^{ab} R_\lambda = - (1/3)\delta^{ab}  b_{A} c  \notag , \\
\mathcal{R}\tilde{ \mathcal{R}}:&  \quad \Tr \left[R_\chi\right] + n_\lambda R_{\lambda} + (-21) R_{3/2} + R_{\chi_s} =  8  b_{\text{grav}} c.
\end{align}
The lhs of the first equation is exactly the naive field theory contribution from the (MSSM) matter fermions, of $R_\chi=Q+\xi/2$, and the gauginos. 
In the second equation, there are contributions  of: the $n_\lambda$ gauginos of the Standard Model and $U(1)_R$ gauge groups; 
  the gravitino contribution  ($(-21)$ times that of a gaugino \cite{grisaru}), the dilatino,
  and the trace is over all  matter fermions. 
These equations agree with the naive field theory result, eqs.~(\ref{an:mixed}) and (\ref{an:grav}) for the corresponding anomalies. 
The rhs of  the first equation  also contains a contribution $b_{A}$  from the counterterm due to the Green-Schwarz mechanism in eq.~(\ref{zz}). 
By supersymmetry,  the gauge kinetic functions corresponding to the Standard Model gauge groups contain a linear contribution\footnote{
The quantity $b_{A}/12\pi^2=\beta_A$ plays the role of Kac-Moody levels in the heterotic string.} 
$\beta_{A} s$, with  $\beta_A \equiv b_{A}/(12\pi^2)$. 
The rhs of the second condition in eq.~(\ref{ee}) indeed agrees with eqs.~(\ref{an:grav}) for 
$b_{\text{grav}}=12\pi^2 \beta_{\text{grav}}$.
while in supergravity one is free to adjust this coefficient (unlike in heterotic string case).
This ends the relation of the anomaly cancellation conditions to the naive field theoretical results (global SUSY) obtained using the $\Tr$ over the charged states.

\section{Consequences for the case $p=2$}  \label{sec:Anomalies:p=2}

The above results can, in principle, be applied the model introduced in the previous section, 
since in the K\"ahler frame where the superpotential is an exponential of the scalar field $s$ (see eq.~(\ref{model:exp_superpot})), the $U(1)$ gauge symmetry is an R-symmetry and the above results apply.
Moreover, when the gauge kinetic function has a linear contribution in $s$, which transforms under the shift symmetry (\ref{shift}), the constant $c$ can indeed be identified with the constant $c$ in the above section. 

However, as we will show below, the anomaly cancellation conditions are much simpler in the K\"ahler frame with a constant superpotential (see eq.~(\ref{model:kahlertransformed})),
 and we will therefore continue in this frame\footnote{
Note however, that anomalies should cancel in both cases, which in fact has very interesting consequences on the mass terms of the gauginos. This will be discussed in section \ref{sec:Anomalies-gauginomass}.}.
Here, in the case $p=2$ the requirement of the existence of a vanishing (or infinitesimally small and positive) minimum of the potential forces the constant contribution to the gauge kinetic function to vanish, which  leaves only a linear contribution in $s$.
The Lagrangian therefore contains a Green-Schwarz counterterm eq.~(\ref{model:Green-Schwarz term}) and is not gauge invariant: 
The gauge variation of the Lagrangian leaves a non-vanishing contribution given by eq.~(\ref{model:Green-Schwarz contribution}).

One can however introduce another field $z$, which can either be a hidden sector field or a MSSM field.
This extra field $z$ is assumed to have a canonical K\"ahler potential and has a charge $q$ under the $U(1)$ and
its anomalous contribution to the Lagrangian cancels the Green-Schwarz contribution (\ref{model:Green-Schwarz contribution}). The model is given by
\begin{align}  \mathcal K &= - 2 \kappa^{-2} \log(s + \bar s) + \kappa^{-2} b(s + \bar s) + z \bar z, \notag \\  W &= \kappa^  {-3} a , \notag \\ f(s) &= \beta_R s. \end{align}
The scalar potential is given by
\begin{align}  \mathcal V &= \mathcal  V_F + \mathcal  V_D, \notag \\ \mathcal  V_F &= \kappa^{-4} a^2 e^{\mathcal K} \left( \sigma_s + \kappa^2 z \bar z \right), 	\notag \\ \mathcal  V_D &= \frac{1}{2\beta_R (s+ \bar s)}\left( \kappa^{-2}c \left(b - \frac{2}{s + \bar s} \right) - q z \bar z\right)^2. \end{align}
where $\sigma_s$ is defined in eq.~(\ref{scalarpotp2}). The mass of the field $z$ is given by 
\begin{align}  m_z^2 = m_{3/2}^2 (\sigma_s + 1) +  \kappa^{-2} \frac{c q }{\beta (s + \bar s)^2} \left(2-\alpha \right), \label{z-mass} \end{align}
in the notation of section \ref{sec:p=2}, where $\alpha = \langle s + \bar s \rangle b \approx -0.1832$. 
The relation between the parameters eq.~(\ref{model:tunability_p2}), which ensures a vanishing cosmological constant, can be generalized to include a nontrivial $\beta_R$
\begin{align} \frac{b c^2}{\beta_R a^2} = A(\alpha),\end{align}
where $A(\alpha) \approx -0.0197$ is given by eq.~(\ref{model:Aalpha_p2}).

On the other hand, the anomalous contribution to the Lagrangian from cubic $U(1)_R$ anomaly is given by
\begin{align} \delta \mathcal L_{1-loop} &=  - \frac{\theta}{32 \pi^2} \  \frac{ \mathcal C_R}{3} \ \epsilon^{\mu \nu \rho \sigma} F_{\mu \nu} F_{\rho \sigma}, \label{1-loop} \end{align}
where $\mathcal C_R= \Tr[Q^3]$. The total variation of the Lagrangian vanishes, 
i.e. $\delta \mathcal L_{1-loop} + \delta \mathcal L_{\text{GS}} = 0$ (where $\delta \mathcal L_{\text{GS}}$ is given in eq.~(\ref{model:Green-Schwarz contribution})) if\footnote{
Alternatively, this result can be obtained from eq.~(\ref{fieldtheory}) by putting $\xi=0$ (since the symmetry is not an R-symmetry here), 
and by using $\text{Tr}\left[R_\chi^3 \right] = q^3$, $\text{Tr}\left[R_\lambda^3\right] =\text{Tr}\left[R_{3/2}^3\right] =0$, and by identifying $b_C = 12 \pi^2 \beta_R$.
} 
\begin{align} \beta_R = - \frac{q^3}{12 \pi^2 c}. \label{p2:anomalycancellation}\end{align}
This fixes the sign\footnote{
In fact, this fixes the sign of $cq/\beta_R$ in eq.~(\ref{z-mass}), which is indeed the relevant quantity. We however assume below that $c>0$ without loss of generality.}
 of $q$ and results in a negative D-term contribution to the scalar mass squared of $z$, which by using eqs.~(\ref{model:tunability_p2}) and (\ref{p2:anomalycancellation}) is
\begin{align}  m_z^2 &= m_{3/2}^2 (\sigma_s + 1) -  \kappa^{-2/3} m_{3/2}^{\ 4/3} R(\alpha), \notag \\
R(\alpha) &= \left( \frac{12 \pi^2 A(\alpha) (2-\alpha)^3 e^{-2 \alpha} }{ \alpha^2} \right)^{1/3} \approx  2.74. \end{align}
The constraint that the mass of $z$ remains non-tachyonic, 
\begin{align}  m_{3/2}^2 (\sigma_s + 1) > \kappa^{-2/3} m_{3/2}^{\ 4/3} R(\alpha), \end{align}
forces the mass of the gravitino to exceed the Planck scale 
\begin{align} m_{3/2} > \frac{R(\alpha)}{\sigma_s +1} \kappa^{-1} \approx 7.16 \ \kappa^{-1}.\end{align}
One concludes that the model with $p=2$ can not be consistently made gauge invariant by the introduction of an extra field charged under $U(1)_R$.
The above argument can easily be generalized to include several charged fields  (unless their number becomes extremely large).
Since the case $p=1$ is consistent with a constant contribution to the gauge kinetic function, 
this problem does not occur here, and this model can be used to calculate the low energy spectrum of the model in chapter \ref{ch:softterms}.

\section{Anomalies and gaugino masses} \label{sec:Anomalies-gauginomass}

In this section we comment on the relation between the quantum anomalies and the gaugino mass terms that are generated at one loop as a consequence of the necessary counterterms.
Although this work, based on \cite{RK3}, will turn out very useful when calculating the gaugino masses of our model in chapter \ref{ch:softterms},
it is also a very relevant result, independently of the rest of the thesis. 
We explain how at one loop there are two contributions to the gaugino mass parameters. 
On the one hand, one has a contribution necessary to maintain invariance under supersymmetry transformations, accompanying the Green-Schwarz terms that are present to cancel the mixed $U(1)_R$ anomalies with the Standard Model gauge groups.
On the other hand, there are contributions to the mixed anomalies due to a mechanism called anomaly mediation \cite{anomalymediation1,anomalymediation2,anomalymediation3,winolike,anomalymediation4}.
The total one-loop contribution to the gaugino masses is the sum of both contributions.

In a similar way as was done for the cubic $U(1)_R^3$ anomaly and its corresponding Green-Schwarz counterterm in the previous section, 
one also has to take into account the mixed $U(1)_R \times G$ anomalies, where $G$ stands for the Standard Model gauge groups.
The anomalous contributions to the gauge variation of the Lagrangian are proportional to $\mathcal C_A$, 
\begin{align}\label{CA_1}     \mathcal C_A \delta^{ab} &= \Tr \left[ R_\chi (\tau^a \tau^b )_A \right] + T_{G} \delta^{ab} R_\lambda\, ,  \end{align}
  where as above $A=Y,2,3$ labels the Standard Model gauge groups. 
  
Here, we focus on the case where none of the MSSM fields carries a charge under $U(1)_R$, and postpone the discussion on the case where the MSSM superfields carry a non-zero charge to chapter \ref{ch:chargedMSSM}.
In this case, in the K\"ahler frame with a constant superpotential all charges vanish and all $\mathcal C_A=0$. In the K\"ahler frame with an exponential superpotential (\ref{model:exp_superpot}) however,
 the R-charge of the matter fermions is $R_\chi = bc/2$, while the gauginos carry a charge $R_\lambda = -bc/2$, such that eq.~(\ref{CA_1}) can be rewritten as
  \begin{align}   \mathcal C_A = \frac{bc}{2} \left(T_{R} - T_{G} \right).   \end{align}
 As a reminder, $T_G$ is the Dynkin index of the adjoint representation, normalized to $N$ for $SU(N)$ and $0$ for $U(1)$,
 and $T_R$ is the Dynkin index associated with the representation $R$ of dimension $d_R$,   equal to $1/2$ for the $SU(N)$ fundamental.
For $U(1)_Y$ we have $T_G = 0$ and $T_R = 11$, for $SU(2)$ we have $T_G = 2$ and $T_R = 7$, and for $SU(3)$ we have $T_G = 3$ and $T_R = 6$. 

 Anomaly cancellation (see section \ref{sec:Anomalies-FI}) then requires that
  \begin{align}   \beta_A = \frac{  \mathcal C_A} {4 \pi^2 c }, \label{anomalymatching}   \end{align}
where $\beta_A$ appear in the gauge kinetic terms for the Standard Model gauge bosons as 
\begin{align} f_A(s) = \frac{1}{g_A^2} + \beta_A s, \notag \end{align}
and $g_A$ are the gauge couplings.
Since the gaugino masses are proportional to derivatives of their respective gauge kinetic functions (see eq.~(\ref{theory:gauginomass})), the Green-Schwarz counterterms lead to contributions to the gaugino mass terms    
\begin{align}
M_A &= -\frac{g_A^2}{2} e^{\kappa^2 K/2} \partial_\alpha f_{A}(s) g^{\alpha \bar \beta} \bar \nabla_{\bar \beta} \bar W \notag \\
&=-\frac{g_A^2}{2} e^{\kappa^2 K/2} \beta_A g^{s \bar s} \bar \nabla_{\bar s} \bar W . \label{gauginomassGS}
\end{align} 
By using the anomaly cancellation conditions (\ref{anomalymatching}), this can be rewritten as
\begin{align}   M_{A} &= - \frac{g_A^2}{16 \pi^2} b (T_G - T_R)  e^{\kappa^2 \mathcal K/2} g^{s \bar s} \bar \nabla_{\bar s} \bar W\, . \label{gauginoRmass1}   \end{align}
 where it is taken into account that the masses of the MSSM gauginos calculated by (\ref{theory:gauginomass}) need a rescaling proportional to $g_A^2$ 
 due to their non-canonical kinetic terms:
  \begin{align}   \mathcal L/e &= -\frac{1}{2} \text{Re}(f)_A \bar \lambda^A \cancel D \lambda^A \notag \\   
  &= -\frac{1}{2} \left(\frac{1}{g_A^2} + \beta_A \frac{\alpha}{b} \right)  \bar \lambda^A \cancel D \lambda^A, \label{gauginoRmass}   \end{align}
  where $\beta_A \frac{\alpha}{b} < < g_A^{-2}$ if the gauge coupling is in the perturbative region. 
  One concludes that an 'anomalous' $U(1)_R$ can give a contribution (at one loop) to the mass of the gauginos.

It is assumed that the Standard Model superfields are inert under $U(1)_R$.
In this case, their only contribution to the anomaly cancellation conditions comes from the respective R-charges of the fermions in the multiplets.
In the K\"ahler representation with a constant superpotential eqs.~(\ref{model:kahlertransformed})  the shift symmetry is not an R-symmetry.
As a result, $\beta_A = 0$ and the gaugino masses vanish. 
On the other hand, in the representation with an exponential superpotential eqs.~(\ref{model:exp_superpot}) the shift symmetry is an R-symmetry.
In general, the R-charges of the Standard Model fermions contribute to the anomaly cancellation conditions which results in a non-zero $\beta_A$, and consequently non-vanishing gaugino masses.
  
It is curious that the gaugino masses vanish for the model (\ref{model:kahlertransformed}), while the classically equivalent model (\ref{model:exp_superpot}) which is related by a K\"ahler transformation has non-zero gaugino masses. This creates a puzzle on the quantum equivalence of these models. 
The answer to this puzzle is based on the fact that gaugino masses are present in both representations and are generated at one loop level by an effect 
called anomaly mediation \cite{anomalymediation1,anomalymediation2,anomalymediation3,winolike,anomalymediation4}.  
Indeed, it has been argued that gaugino masses receive a one-loop contribution due to the super-Weyl-K\"ahler and sigma-model anomalies\footnote{This can also be seen in~\cite{KL}: 
since a K\"ahler transformation is anomalous, there are in general additional contributions to the effective action at the quantum level. }. 
These contributions are different in both representations, and we will show below that the difference accounts exactly for the contributions~(\ref{gauginoRmass}).  

The contribution from anomaly mediation to the gaugino masses is given by~\cite{anomalymediation4}
\begin{align}   \frac{g^2}{16 \pi^2} \left[ (3 T_G - T_R) m_{3/2} + (T_G - T_R) \mathcal K_\alpha F^\alpha  
\vphantom{+ 2 \frac{T_R}{d_R} (\log \det \mathcal K|_R \ '')_{,\alpha} F^\alpha}  
  + 2 \frac{T_R}{d_R} (\log \det \mathcal K|_R \ '')_{,\alpha} F^\alpha \right] , \label{gauginomass:anomalymediated}  \end{align}
where an implicit sum over all matter representations $R$ is understood.  The expectation value of the auxiliary field $F^\alpha$, evaluated in the Einstein frame is given by   
  \begin{align}   F^\alpha = - e^{ \kappa^2 \mathcal K/2} g^{\alpha \bar \beta} \bar \nabla_{\bar \beta} \bar W.   \end{align}
In eq.~(\ref{gauginomass:anomalymediated}), 
the first term proportional to the one-loop beta function coefficient $3T_G - T_R$ is always present in a gravity mediated supersymmetry breaking scenario, 
and is in fact the main ingredient of anomaly mediated supersymmetry breaking scenarios~\cite{anomalymediation1,anomalymediation2,anomalymediation3,winolike,anomalymediation4}.
\glossary{name=mAMSB,description=minimal Anomaly Mediated Supersymmetry Breaking}
\glossary{name=mSUGRA,description=minimal (super)gravity mediated (supersymmetry breaking)}
The second term is only present when a hidden sector field acquires a Planck scale VEV.
In the third term, $\mathcal K_R ''$ is the K\"ahler metric restricted to the representation $R$, 
and $(\log \det \mathcal K|_R \ '')_{,\alpha}$ therefore vanishes for any MSSM-like fields with a canonical K\"ahler potential.

Since the K\"ahler potential in eqs.~(\ref{model:exp_superpot}) and (\ref{model:kahlertransformed}) differ by a linear term $b(s + \bar s)$, the contribution of the second term in  eq.~(\ref{gauginomass:anomalymediated}) differs by a factor
  \begin{align}   \delta M_{A} = - \frac{g_A^2}{16 \pi^2} (T_G - T_R) b e^{\kappa^2 \mathcal K/2} g^{\alpha \bar \beta} \bar \nabla_{\bar \beta} \bar W,   \end{align}
  which exactly coincides with eq.~(\ref{gauginoRmass1}). 

  We conclude that even though the models (\ref{model:exp_superpot}) and (\ref{model:kahlertransformed}) differ by a (classical) K\"ahler transformation, 
they generate the same gaugino masses at one loop  given by the sum of the Green-Schwarz contribution eq.~(\ref{gauginoRmass1}) and the anomaly mediated contribution eq.~(\ref{gauginomass:anomalymediated}).
   While the one-loop gaugino masses for the model (\ref{model:kahlertransformed}) are generated entirely by eq.~(\ref{gauginoRmass1}), 
the gaugino masses for the model (\ref{model:exp_superpot}) which is related by a K\"ahler transformation  have a contribution from eq.~(\ref{gauginomass:anomalymediated})
 as well as from a field dependent gauge kinetic term whose presence is necessary to cancel the mixed $U(1)_R \times G$ anomalies.
 This is due to the fact that the extra $U(1)$ has become an R-symmetry giving an R-charge to all fermions in the theory.

\cleardoublepage

  \graphicspath{{\chaptersdir/softterms/image/}}%
  \chapter{Soft terms and phenomenology}\label{ch:softterms}

In chapter~\ref{ch:dS}, a model was introduced based on a gauged shift symmetry of the dilaton that allows for a tunably small and positive value for the cosmological constant.
Moreover, in section~\ref{sec:remarksp2}, we made contact between this model and a possible UV-completion in the context of string theory for a particular value of the parameters.
By studying the quantum consistency of such models in chapter~\ref{ch:anomalies}, we showed that for $p=2$ the gravitino mass becomes unacceptably large (since it exceeds the Planck scale) due to the anomaly cancellation conditions.
For $p=1$ however, a value of the gravitino mass of the TeV scale is allowed, and we therefore continue with $p=1$.

In this chapter, we calculate the low energy phenomenology and look for possible experimental signatures of this model.
We assume that the MSSM fields are inert under the shift symmetry, and postpone a discussion on a charged MSSM to chapter~\ref{ch:chargedMSSM}.
The model which was introduced in chapter~\ref{ch:dS} will be used as a hidden sector for supersymmetry breaking, 
where the breaking of supersymmetry is communicated to the visible sector (MSSM) via gravity mediation.
In the simplest case however the resulting scalar masses are tachyonic.
We show that this can be avoided,  without modifying the main properties of the model, 
by introducing either a new `hidden sector' field participating in  the  supersymmetry  breaking,  similar  to  the  Pol\'onyi  field~\cite{polonyi},    
or by introducing   dilaton  dependent matter kinetic terms for the MSSM fields. 
In both cases an extra parameter is introduced with a narrow range  of  values  in  order  to  satisfy  all  required  constraints.   
All  scalar  soft  masses  and trilinear A-terms are generated at the tree-level and are universal under the assumption
that matter kinetic terms are independent of the `Pol\'onyi' field, while gaugino masses are generated  at  the  quantum  level,  via  the  so-called  anomaly  mediation  contributions, and are naturally suppressed compared to the scalar masses.

It follows that the low energy spectrum is very particular and can be distinguished from  other  models  of  supersymmetry  breaking  and  mediation,  such  as  mSUGRA  and
mAMSB. It consists of light neutralinos, charginos and gluinos, where the experimental bounds on the (mostly bino-like) LSP and the gluino force the gravitino mass to be above
15 TeV and the squarks to be very heavy (above 13 TeV), with the exception of the stop squark which can be as light as 2 TeV.

This chapter is organized as follows: In section~\ref{sec:tachyonicmasses} we show that the simplest model with $p=1$ results in negative scalar soft masses squared.
A first solution to this problem, based on an extra Pol\'onyi-like field is presented in section~\ref{sec:polonyi}.
The resulting low energy spectrum is then calculated in section~\ref{sec:low_energy_spectrum} for various values of the extra parameter $\gamma$ 
and it is shown that the spectrum can be distinguished from other minimal models of supersymmetry breaking.
Finally, in section~\ref{sec:non-can_Kahler}  another possible solution to the tachyonic masses is presented. 
However, since the low energy spectrum is expected to be very similar, we do not elaborate on its low energy phenomenology.
This chapter is based on work together with I.~Antoniadis~\cite{RK3}.

\section{Problem of tachyonic masses} \label{sec:tachyonicmasses}

The model with $p=1$ was introduced in section \ref{sec:p=1}.
There it was shown that the relations between the parameters $a,b,c$ in eqs.~(\ref{model:p1_tunability}) ensure an infinitesimally small and positive cosmological constant, 
while the gravitino mass, given by eq.~(\ref{model:p1_gravitino}), is separately tunable.

If one now adds an MSSM-like field $\varphi$ with a canonical K\"ahler potential,  and invariant under $U(1)_R$,
 \begin{align} \mathcal K &= - \kappa^{-2} \log (s + \bar s) + \kappa^{-2} b (s + \bar s) + \sum \varphi \bar \varphi, \notag \\ W &= \kappa^{-3} a + W_{MSSM} ,  \label{model0MSSM} \end{align}  
where $W_{MSSM}$ is the MSSM superpotential defined in eq.~(\ref{MSSMsuperpot}). The soft scalar mass squared at $\langle \varphi \rangle = \langle \bar \varphi \rangle = 0$ is negative, given by
  \begin{align}  \left. \partial_\varphi \partial_{\bar \varphi} \mathcal V \right|_{\langle \varphi \rangle = 0}=   |a|^2 b \frac{e^{\alpha}}{\alpha} \left(\langle \sigma_s \rangle +1 \right) < 0\ , \label{tach} \end{align}
  where $\sigma_s$ is defined in eq.~(\ref{scalarpotp1}), $\alpha\approx -0.233$ and $b$ is assumed to be negative which is necessary to allow for a de Sitter minimum. 
Since $\langle \sigma_s \rangle \approx -1.48$, any non-zero solutions $\langle \varphi \rangle \neq 0$ of $\partial_\varphi V = 0$ would mean that the field $\varphi$ contributes in general to the supersymmetry breaking. 
We conclude that the model on its own can not be consistently extended to include the MSSM with canonical kinetic terms.  To circumvent this problem, one can add an extra hidden sector field which contributes to (F-term) supersymmetry breaking. 

This will be worked out in full detail in the following sections.  However, we will show in section \ref{sec:non-can_Kahler} that the above problem of tachyonic soft masses can also be solved if one allows for a non-canonical K\"ahler potential in the visible sector, which gives an additional contribution to the masses through the D-term.    

\section{An extended model with extra Pol\'onyi-like field} \label{sec:polonyi}

 \subsection{Tuning of the parameters}
  As described above, the model (with $p=1$ and a field independent gauge kinetic function) presented there would give a tachyonic mass to any MSSM-like fields (that are invariant under the shift symmetry and have a canonical K\"ahler potential).  
  In this section we add an extra hidden sector field $z$ (similar to the so-called Pol\'onyi field~\cite{polonyi}) to circumvent this problem. Note that this choice is not unique and that the problem can also be circumvented by allowing a non-canonical K\"ahler potential for the MSSM fields (see section \ref{sec:non-can_Kahler}).

 The K\"ahler potential, superpotential and gauge kinetic function are given by
  \begin{align} \mathcal K &= -\kappa^{-2} \log (s + \bar s) + \kappa^{-2} b (s + \bar s) + z \bar z, \notag \\ W &= \kappa^{-3} a (1+ \gamma \kappa z) , \notag \\ f(s) &= 1\, , \label{model:polonyi} \end{align}
  with $\gamma$ an additional constant parameter.  The scalar potential is
  \begin{align} \mathcal V &= \mathcal V_F + \mathcal V_D, \notag \\ \mathcal V_F &= \kappa^{-4} |a|^2 \frac{e^{b(s + \bar s) + \kappa^2 z \bar z}}{s + \bar s} \left( \sigma_s A(z,\bar z) + B(z, \bar z) \right),  \notag \\ \mathcal V_D &= \kappa^{-4}   \frac{c^2}{2} \left( b - \frac{1}{s + \bar s} \right)^2, \end{align}
  where
  \begin{align}  A(z, \bar z) &= \left| 1 + \gamma \kappa z \right|^2, \notag \\  B(z, \bar z) &= \left| \gamma + \kappa \bar z + \gamma \kappa^2 z \bar z \right|^2.  \end{align}
  We focus on real $z=\bar z = \kappa^{-1} t$ (we will confirm below that the imaginary part of the VEV of $z$ indeed vanishes): 
  \begin{align} A(t) &= (1+\gamma t)^2, \notag \\ B(t) &= (\gamma + t + \gamma t^2)^2\, ; \label{AB} \end{align} 
    $\partial_t \mathcal V = 0$ then gives
  \begin{align} 0 =& \gamma( \sigma_s  + 1) + (\sigma_s + 1 + \gamma^2 (\sigma_s + 2) ) t + \gamma (2 \sigma_s +5) t^2  \notag \\ &+ ( 1 + \gamma^2 (\sigma_s + 4) ) t^3 + 2 \gamma t^4 + \gamma^2, \notag \\ \sigma_s =& -3 + (\alpha - 1)^2,  \ \ \ \ \ \ \ \ \  \alpha = b \langle s + \bar s \rangle. \label{Vt} \end{align} 
  As before, $\partial_s \mathcal V = 0$ implies
  \begin{align}  \frac{c^2}{a^2 } =\frac{\alpha}{b} e^{\alpha + t^2} \left[ A( t ) ( 2 - \alpha^2 ) - B(t) \right]. \label{ca} \end{align}
  This can be combined with $\mathcal V = 0$
  \begin{align}  \frac{c^2}{a^2} = -2 \frac{\alpha}{b} e^{\alpha + t^2} \left[\frac{\sigma_s A(t) + B(t)}{(\alpha-1)^2} \right], \label{model:polonyi_tunability} \end{align}
  to give
  \begin{align} 0 &= A(t) \left( \sigma_s - \frac{1}{2} (\alpha - 1)^2 (\alpha-2) \right) + B(t) \left(1 - \frac{1}{2} (\alpha - 1)^2 \right). \label{Vs} \end{align}
  In principle for any value of $\gamma$, a Minkowski minimum can be found by solving eqs.~(\ref{Vt}) and (\ref{Vs}) for $\alpha$ and $t$, and then tuning the parameters $a$,$b$ and $c$ by using the relation (\ref{model:polonyi_tunability}). 

  The role of the extra hidden sector field $z$ is to give a (positive) F-term contribution to the  scalar potential, which in turn gives a positive contribution (proportional to $\left| \nabla_z W\right|^2$) to the soft mass squared of any MSSM-like field in eq.~(\ref{tach}). It turns out that the addition of the extra hidden sector field $z$ indeed results in positive soft masses squared. 

  It is however necessary that $z$ contributes to the supersymmetry breaking. 
  The existence of any minimum of the potential with $\left| \nabla_z W \right|^2 =0$ can be troublesome\footnote{
  The Pol\'onyi-like field $z$ was originally introduced to circumvent the problem of the tachyonic soft scalar masses outlined in section~\ref{sec:tachyonicmasses}.
  Its role is to give an extra contribution to the scalar potential, proportional to $\left| \nabla_z W \right|^2$, 
  such that the scalar soft masses squared (see eq.~(\ref{tach})) become positive. 
  The existence of a minimum of the potential with $\left| \nabla_z W \right|^2 =0$ can therefore be troublesome since in this vacuum one has tachyonic soft scalar masses for any MSSM-like fields.
  } and we therefore require
  \begin{align} \nabla_z W = \partial_z W + \kappa^2 \mathcal K_z W = a\left(\gamma + \bar z (1+ \gamma z) \right) \neq 0 . \label{noAdS} \end{align}
  Since $\gamma$ is real, any root of $\nabla_z W =0$ is also real. To ensure the condition (\ref{noAdS}) we must ensure that the roots $ \text{Re}(z) =( -1 \pm \sqrt{1-4 \gamma})/4\gamma$ are complex. This requires $|\gamma| > 1/2$.

  Also, for any $\gamma$ the solution $(\alpha, t)$ of the set of equations  (\ref{Vt}) and (\ref{Vs}) should give a positive right-hand side of eq.~(\ref{ca}) (or equivalently, eq.~(\ref{model:polonyi_tunability})). This constraint leads to $\gamma < 1.707$. We conclude that
  \begin{align} \gamma \in \left[  0.5 ,1.707 \right]. \label{gammarange} \end{align} 

  For example, for $\gamma = 1$, we have $b \langle s + \bar s \rangle = \alpha \approx -0.134014$, $\langle t \rangle = 0.39041$. The (negative) constant $b$ can be chosen freely to fix the VEV of Re$(s)$. The parameters $a$ and $c$ should be tuned carefully according to
  \begin{align} \frac{b c^2}{a^2} = -2 \alpha e^{\alpha + t^2} \left[\frac{\sigma_s A(t) + B(t)}{(\alpha-1)^2} \right]	\approx -0.1981	. \label{ca1} \end{align}
  Note that the number on the right-hand side changes when $\gamma$ is varied. The remaining free parameter $a$ can be used to tune the supersymmetry breaking scale and (as shown below) the soft masses for the MSSM-like fields compared to the gravitino mass depend slightly on $\gamma$  (provided $c$ and $a$ are also tuned according to eq.~(\ref{ca})). We summarize the VEVs of $\alpha$ and $t$, together with the above constraint on the parameters for the particular  choice $\gamma=1$ below for future reference
  \begin{align}   \gamma = 1, \ \ \alpha \approx -0.134014, \ \ \langle t \rangle \approx 0.39041, \ \ \frac{bc^2}{a^2} \approx -0.1981 \ . \label{parameterchoice}  \end{align} 

  For $\gamma$ in the allowed parameter range (\ref{gammarange}), the scalar potential is positive definite for all Re$(s)>0, z, \bar z$, including the imaginary part of $z$, which justifies our assumption to look for a Minkowski minimum with Im$(z) = 0$. 
  In fact, for the allowed values of $\gamma$, the solution of the set of equations (\ref{Vt}) and (\ref{Vs}) together with $\partial_{\text{Im}(z)}V=0$ gives $\text{Im}(z)=0$ as a solution. 

  Finally, note that this Minkowski minimum can be lifted to a dS vacuum with an infinitesimally small cosmological constant by a small increase in $c$. A cosmological constant $\Lambda$ can be obtained by replacing the condition (\ref{ca1}) with
  \begin{align} \frac{c^2}{a^2 } =-2 \frac{\alpha}{b} e^{\alpha + t^2} \left[\frac{\sigma_s A(t) + B(t)}{(\alpha-1)^2} \right] + \frac{2 \alpha^2}{(\alpha-1)^2} \frac{\kappa^4 \Lambda}{a^2 b^2} . \end{align}

  \subsection{Masses of the hidden sector fields}

  The gravitino mass is given by
  \begin{align} m_{3/2} =  \kappa^{-1} a \sqrt { \frac{b}{ \alpha} } e^{\alpha/2 + t^2/2} \left( 1 + \gamma t \right).  \label{model:polonyi_gravitino} \end{align}
  Note that this can be arranged to be at the TeV scale (to obtain a viable low energy sparticle spectrum) by suitably tuning $a$. 
  For example, for $\gamma=1$, such that $\alpha$ and $t$ are given by eq.~(\ref{parameterchoice}) and $m_{3/2} =1 \text{ TeV}$, we have
  \begin{align} a \sqrt b \approx 3.53 \times 10^{-17}. \label{a_numerics}     \end{align}
Since the VEV of Im$(z)$ vanishes, it does not mix with the other hidden sector scalars and its mass is given by
\begin{align}  m^2_{\text{Im}(z)} &= m_{3/2}^2 \  f_{\text{Im}(z)}, \\
   f_{\text{Im}(z)} &=\frac{2  \left(1+2 t^3 \gamma +t^4 \gamma ^2+\sigma_s +2 t \gamma  (2+\sigma_s )+\gamma ^2 (3+\sigma_s )+t^2 \left(1+\gamma ^2 (4+\sigma_s )\right)\right)}{(1 + \gamma t)^2 }.\notag \end{align}
However, the masses of the scalars Re$(s)$ and Re$(z)$ mix, so one should diagonalize their mass matrix (with eigenvalues $m_{ts1}$ and $m_{ts2}$) while taking in account the non-canonical kinetic term for $s$.  We omit the details and merely state the result for the particular choice of parameters $\gamma = 1$ in eq.~(\ref{parameterchoice}):
  \begin{align}   m_{\text{Im}(z)} \approx 1.21\ m_{3/2}, \notag \\   m_{ts1} \approx 4.34\ m_{3/2}, \notag \\  m_{ts2} \approx 1.08\ m_{3/2} . \label{scalarmasses}   \end{align}

  The imaginary part of $s$ is eaten by the $U(1)$ gauge boson, which becomes massive. Its mass is given by\footnote{
  The result (\ref{polonyi_gaugebosonmass}) is obtained by substituting $\langle s + \bar s \rangle = \alpha/b$ in eq.~(\ref{appendix:gaugebosonmass}) and by putting $p=1$. For more details, see the Appendix~\ref{sec:appendix_gauge}.}:
 \begin{align} m_{A_\mu} &=  \kappa^{-1} \frac{ \sqrt{2} b c }{\alpha} \notag \\ 
 &\approx 1.23 \ m_{3/2},\label{polonyi_gaugebosonmass} \end{align} 
where the last line was obtained by the relation between the parameters eq.~(\ref{ca1}) and by substituting the numerical values for $\gamma=1$ eq.~(\ref{parameterchoice}).

  The Goldstino, which is a linear combination of the gaugino, the z-fermion and the s-fermion, is eaten by the gravitino, which in turn becomes massive. 
  The masses of the remaining two hidden sector fermions are calculated in Appendix \ref{Appendix:Fermions_polonyi} and their values for $\gamma=1$ are given by
\begin{align}  m_{\chi_1} & \approx   2.57 \ m_{3/2}, \notag \\  m_{\chi_2} & \approx  0.12 \ m_{3/2}. \end{align}
  
  \subsection{Soft supersymmetry breaking terms}

  We are now ready to couple the model above, which allows for a TeV gravitino and an infinitesimally small cosmological constant, to the MSSM and to calculate its soft breaking terms.

As was already mentioned, for simplicity, we take the MSSM-like fields $\varphi_\alpha$ to be chargeless\footnote{For a charged MSSM, see section~\ref{sec:B-L}} under $U(1)_R$. 
These can then be coupled to the above model in the following way:
  \begin{align} \mathcal K &= - \kappa^{-2} \log (s + \bar s) + \kappa^{-2} b (s + \bar s) + z \bar z + \sum_\alpha \varphi \bar \varphi , \notag \\ W &= \kappa^{-3} a (1+ \gamma \kappa z) + W_{\text{MSSM}}(\varphi) , \notag \\ f_R(s) &= 1, \ \  \ \ \ \ \ \ \ \ \ \  \ f_A(s) = 1/g_A^2. \label{modelMSSM} \end{align}
  The various chiral multiplets in the MSSM are labeled by an index $\alpha$, which is omitted for simplicity. 
  As before, the Standard Model gauge groups are labeled by an index A, while the extra $U(1)$ is referred to with an index $R$. All gauge kinetic functions are taken to be constant. 

  The scalar potential is now given by
      \begin{align} \mathcal  V &= \mathcal V_F + \mathcal  V_D, \notag \\ \mathcal  V_F &= \kappa^{-4} \frac{e^{b(s + \bar s) + z\bar z + \varphi \bar \varphi} }{s + \bar s} \left( \sigma_s A(z,\bar z, \varphi, \bar \varphi) + B(z, \bar z, \varphi, \bar \varphi) + \kappa^4 \sum_\alpha \left| \nabla_\alpha W \right|^2 \right),  \notag \\ \mathcal V_D &=  \kappa^{-4} \frac{c^2}{2} \left( b - \frac{1}{s + \bar s} \right)^2, \label{scalarpotMSSM} \end{align}
      where
      \begin{align}   A(z,\bar z, \varphi, \bar \varphi) &= \left| a + a \gamma \kappa z + \kappa^3 W_{\text{MSSM} } \right|^2 \notag \\  B(z, \bar z, \varphi, \bar \varphi) &= \left| a \gamma + \kappa \bar z (a + a \gamma z + \kappa^3 W_{\text{MSSM} } ) \right| ^2 \notag \\   \left| \nabla_\alpha W \right|^2 &= \left| \partial_\alpha W_{\text{MSSM}} + \kappa^2 \bar \varphi W \right|^2\, .  \end{align}
  It can be easily seen that the resulting scalar potential has a minimum at $\langle \varphi \rangle = \langle W_{\text{MSSM}} \rangle =  0$, in which case the potential of last section is reproduced and its conclusions are still valid. 
  For example, $A(z,\bar z, \varphi,\bar \varphi)|_{\langle z \rangle = \kappa t, \langle \varphi \rangle = 0} = a^2 A(t)$
  and $B(z,\bar z, \varphi,\bar \varphi)|_{\langle z \rangle = \kappa t, \langle \varphi \rangle = 0} = a^2 B(t)$, where $A(t)$ and  $B(t)$ are defined in eqs.~(\ref{AB}).   
  The second derivatives of the potential, evaluated on the ground state are given by 
  \begin{align}   \partial_\varphi \partial_{\bar \varphi} \mathcal V &= \frac{\kappa^{-2} a^2 b e^{\alpha + t^2}}{\alpha} \left[ (\sigma_s + 1 ) A(t) + B(t)  + \kappa^2 W_{\varphi \varphi} \bar W_{\bar \varphi \bar \varphi} \right], \notag \\  \partial_\varphi \partial_\varphi \mathcal V &= \frac{ \kappa^{-1} ab W_{\varphi \varphi} e^{\alpha + t^2}}{\alpha}   \left[ (\sigma_s +2) (1+ \gamma t) + t (\gamma +t + \gamma t^2)  \right].  \end{align}
  There is no mass mixing between the different $\varphi^\alpha$ (except of course for the $B_0$ term defined below) and between the MSSM fields with $z$ and $s$. The MSSM superpotential is defined in eq.~(\ref{MSSMsuperpot}).
  Note that in the scalar potential eq.~(\ref{scalarpotMSSM}) the MSSM F-terms $\sum_\alpha \left| \nabla_\alpha W \right|^2 $ come with a prefactor $\exp (\alpha  + t^2) b/\alpha$ (where the fields have been replaced by their VEVs). To bring this into a conventional form, one should rescale the MSSM superpotential
  (see also eq.~(\ref{superpotrescaling}))
  \begin{align}  \hat W_{MSSM} = \sqrt{\frac{b} {\alpha}}  e^{\alpha/2 + t^2/2 } \ W_{MSSM}.  \label{MSSMhat}  \end{align}
  Then the squark and slepton soft masses are given by
  \begin{align}  m^2_{\tilde Q} &= m^2_{\tilde {\bar u}}= m^2_{\tilde {\bar d}}= m^2_{\tilde L} = m^2_{\tilde {\bar e}} = m_0^2 \ \mathbb{I}, \notag \\  m_0^2 &= \kappa^{-2} b a^2 \frac{e^{\alpha + t^2 } }{\alpha} \left[ A(t) \left( \sigma_s + 1 \right) + B(t)\right].  \end{align}
  Here, $\mathbb{I}$ is the unit matrix in family space.  The trilinear couplings are given by 
  \begin{align}  &a_u = A_0 \hat y_u, \ \ \ \ \ \ \  a_d = A_0  \hat y_d, \ \ \ \ \ \ \ a_e = A_0 \hat y_e, \notag \\  &A_0= \kappa^{-1}a \sqrt{\frac{b}{\alpha}} e^{(\alpha + t^2)/2 }  \left[  (\sigma_s +3) (1+\gamma t) + t(\gamma + t +\gamma t^2) \right],  \end{align}
  where $\hat y_u, \hat y_d $ and $ \hat y_e$ are the Yukawa couplings of the MSSM superpotential after the rescaling of eq.~(\ref{MSSMhat}).  Also,
  \begin{align} &m^2_{H_u } = m^2_{H_d} = m_0^2, \end{align}
  and 
  \begin{align}  &B_0 = \kappa^{-1}a \sqrt{\frac{b}{\alpha}} e^{(\alpha  + t^2)/2}  \left[  (\sigma_s + 2)(1+\gamma t) + t (\gamma + t + \gamma t^2)  \right],  \end{align}
  where $B_0$ generates a term proportional to $- \hat \mu B_0  H_u \cdot H_d + \text{h.c.}$, where $\hat \mu$ is the rescaled $\mu$-parameter (in the sense of eq.~(\ref{MSSMhat})).  
Summarized, in terms of the gravitino mass (eq.~(\ref{model:polonyi_gravitino})), the MSSM soft terms are given by 
  \begin{align}
  m_0^2 &= m_{3/2}^2  \left[ \left( \sigma_s + 1 \right) + \frac{(\gamma + t + \gamma t^t)^2}{(1 + \gamma t)^2}\right], \notag \\   A_0&= m_{3/2} \left[  (\sigma_s +3)  + t \frac{ (\gamma + t +\gamma t^2) } {1+\gamma t} \right], \notag \\   B_0 &= m_{3/2}  \left[  (\sigma_s + 2) + t \frac{ (\gamma + t + \gamma t^2) }{(1+\gamma t)}  \right]. \label{softterms_polonyi}   \end{align}
  Note the relation~\cite{A-B_relation}
  \begin{align}   A_0 = B_0 + m_{3/2}, \notag  \end{align}
 as in eq.~(\ref{AB-relation}).
 
  At tree level, the gaugino mass terms are given by eq.~(\ref{theory:gauginomass}), which is proportional to the derivative of the gauge kinetic function.
  Since the gauge kinetic functions are constant, they vanish at tree-level
  \begin{align} \left. m_{AB} \right|_{\text{tree}} = 0. \end{align}
At one loop however, the 'anomaly mediated' gaugino mass  contribution, given by eq.~(\ref{gauginomass:anomalymediated}), becomes
   \begin{align}   M_A = - \frac{g^2}{16 \pi^2} m_{3/2} \left[(3 T_G - T_R) - (T_G - T_R) \left( (\alpha-1)^2 + t \frac{\gamma + t + \gamma t^2}{1 + \gamma t} \right) \right] , \notag \end{align}
where the relation for the gravitino mass eq.~(\ref{model:polonyi_gravitino}) was used.
The different gaugino mass parameters this gives (in a self-explanatory notation) are\footnote{
The gaugino masses eqs.~(\ref{model:polonyi_m1m2m3}) can equivalently be calculated in the K\"ahler representation of the model where the superpotential is an exponential, as in eqs.~(\ref{model:exp_superpot}).
Since in this representation the shift symmetry becomes an R-symmetry, all fermions in the model (including Standard Model fermions) carry an R-charge and one has to take into account the anomaly cancellation conditions of chapter~\ref{ch:anomalies}.
A Green-Schwarz mechanism is necessary to cancel anomalies, which requires a linear contribution to the gauge kinetic functions.
This field-dependent gauge kinetic function results in a contribution to the gaugino mass parameters, given by~\cite{RK3}
   \begin{align}   \hat M_1 = \frac{11}{16\pi^2} b g_Y^2 e^{\alpha/2} (\alpha-1),  \quad  \hat M_2 = \frac{5}{16\pi^2} b g_2^2 e^{\alpha/2} (\alpha-1),    \quad
  \hat M_3 = \frac{3}{16\pi^2} b g_3^2 e^{\alpha/2} (\alpha-1). \notag \end{align}
The total gaugino mass parameters are then given by the sum of the above contribution and eq.~(\ref{gauginomass:anomalymediated}),
which give the same result given in eq.~(\ref{model:polonyi_m1m2m3}). It was shown in section~\ref{sec:Anomalies-gauginomass} that this equivalence is true in a more general case.
  } 
 \begin{align}
  M_1 &= 11 \frac{g_Y^2}{16 \pi^2} m_{3/2} \left[ 1 - (\alpha -1)^2 -  \frac{ t(\gamma + t + \gamma t)}{1 + \gamma t^2}  \right], \notag \\
  M_2 &=  \frac{g_2^2}{16 \pi^2} m_{3/2} \left[1 - 5 (\alpha-1)^2 -5 \frac{ t (\gamma + t + \gamma t^2)}{1 + \gamma t} \right], \notag \\
  M_3 &= - 3 \frac{g_3^2}{16 \pi^2} m_{3/2} \left[ 1 + (\alpha - 1)^2 + \frac{ t (\gamma + t + \gamma t^2) }{1 + \gamma t} \right]. \label{model:polonyi_m1m2m3}
  \end{align}
  For example, if we choose $\gamma=1$ (as in eq.~(\ref{parameterchoice})) the above equations give
  \begin{align}   M_1 & \approx 0.05 \  g_Y^2 \  m_{3/2}, \notag \\   M_2 & \approx 0.048 \ g_2^2 \ m_{3/2}, \notag \\   M_3 & \approx 0.052 \ g_3^3 \  m_{3/2}.   \end{align}
If we now assume that the gauge couplings unify at some unification scale $\frac{5}{3} g_Y^2\equiv g_1^2 = g_2^2 = g_3^2 = 0.51$, we get the gaugino masses at this scale 
\begin{align} M_1 & \approx 0.015\ m_{3/2}, \notag \\  M_2 & \approx 0.025\ m_{3/2},\notag \\ M_3 & \approx 0.026\ m_{3/2}. \end{align}
The gaugino masses for other values of $\gamma$ are listed in table \ref{tanbeta} below.

Note that in a similar way, the trilinear terms $A_0$ also receive corrections proportional to   
\begin{align}   \delta A_{ijk} = - \frac{1}{2} \left( \gamma_i + \gamma_j + \gamma_k \right) m_{3/2},   \end{align}  where the $\gamma$'s are the anomalous dimensions of the corresponding cubic term in the superpotential. These contributions however are small compared to the tree-level value in eq.~(\ref{softterms_polonyi}).
  
Although the gaugino masses are generated at one-loop, our model is very different from an mAMSB~\cite{anomalymediation1,anomalymediation2,anomalymediation3,winolike,anomalymediation4} scenario: In mAMSB, the second and third term in eq.~(\ref{gauginomass:anomalymediated}) are missing due to the absence of hidden sector fields with a Planck scale VEV.  In our model however, the second term in eq.~(\ref{gauginomass:anomalymediated}) is present because of the non-vanishing F-terms of the $s$ and $z$ fields,  and has the effect that it raises the gaugino masses slightly to the order $M_{1/2} \approx 2 \times 10^{-2}  \ m_{3/2}$ compared to $M_{1/2} \approx 10^{-2} - 10^{-3} \ \ m_{3/2}$ for a mAMSB where only the first term in eq.~(\ref{gauginomass:anomalymediated}) is non-vanishing. Another important difference is that we have $M_{1} < M_{2}$ which results in a mostly bino-like LSP, compared with a mostly wino-like LSP in mAMSB. Note also that we do not have any danger of tachyonic scalar soft masses because of the presence of a tree-level soft mass $m_0$ in eqs.~(\ref{softterms_polonyi}).  We also have tree-level trilinear couplings $A_0$, which are not present in the mAMSB.

Our model is also different from the minimal supergravity mediated scenario (mSUGRA)~\cite{gravitymediation}. Indeed, in mSUGRA gaugino masses are imposed to be equal at tree-level at the GUT unification scale $M_3 \ : \ M_2 \ : \ M_1 = g_3^2 \ : \ g_2^2 \ : \ g_1^2$ of the order $m_0$ (plus or minus an order of magnitude), while our model has vanishing tree-level gaugino masses. They are generated at one-loop and do not satisfy the above relation. Since the gaugino masses are generated at one-loop they are much smaller than the other soft terms.

We conclude that although the soft terms $m_0, A_0$ and $B_0 = A_0 - m_{3/2}$ are similar to an mSUGRA scenario, the anomaly mediated gaugino masses  (which have on top of the usual AMSB contribution proportional to the beta function another contribution from the Planck scale VEVs of $s$ and $z$) are not universal and are much smaller.  Therefore, the particle spectrum will resemble much more the spectrum of a mAMSB scenario, with the important difference that the lightest neutralino is bino-like instead of wino-like (See section \ref{sec:low_energy_spectrum}). 

\section{Low energy spectrum} \label{sec:low_energy_spectrum}

 The results for the soft terms calculated in the previous section, evaluated for different values of the parameter $\gamma$, are summarized in table~\ref{tanbeta}.   For every $\gamma$, the corresponding $t$ and $\alpha$ are calculated by imposing a vanishing cosmological constant by eqs.~(\ref{ca}) and (\ref{model:polonyi_tunability}).   The scalar soft masses and trilinear terms are then evaluated by eqs.~(\ref{softterms_polonyi}) and the gaugino masses by eqs.~(\ref{model:polonyi_m1m2m3}).   Note that the relation~(\ref{AB-relation}), namely $ A_0 = B_0 - m_{3/2} $, is valid for all $\gamma$. We therefore do not list the parameter $B_0$.

\begin{table}[!ht] \centering \begin{tabular}{| l | ll | lllll | rr |} \hline
$\gamma$ & t     & $\alpha$ & $m_0$ & $A_0$ & $M_1$ & $M_2$ & $M_3$ & $\tan \beta$ & $\tan \beta $ \\ 
	& 	&	&	&	&	&	&	&$\mu> 0$	& $\mu < 0$ \\
\hline 
0.6      & 0.446 & -0.175   & 0.475         & 1.791         & 0.017         & 0.026         & 0.027         &                                   &                               \\
1        & 0.409 & -0.134   & 0.719         & 1.719         & 0.015         & 0.025         & 0.026         &                                   &                                \\
1.1      & 0.386 & -0.120   & 0.772         & 1.701         & 0.015         & 0.024         & 0.026         & 46                                & 29                             \\
1.4      & 0.390 & -0.068   & 0.905         & 1.646         & 0.014         & 0.023         & 0.026         & 40                                & 23                             \\
1.7      & 0.414 & -0.002   & 0.998         & 1.588         & 0.013         & 0.022         & 0.025         & 36                                & 19                              \\ \hline 
\end{tabular} \caption{The soft terms (in terms of $m_{3/2}$) for various values of $\gamma$. If a solution to the RGE exists, the value of $\tan \beta$ is shown in the last columns for $\mu >0$ and $\mu <0$ respectively.} \label{tanbeta} \end{table} 

In most phenomenological studies, $B_0$ is substituted for $\tan \beta$, the ratio between the two Higgs VEVs, as an input parameter for the RGEs that determine the low energy spectrum of the theory. Since $B_0$ is not a free parameter in our theory, but is fixed by eq.~(\ref{AB-relation}), this corresponds to a definite value of $\tan \beta$. 
For more details see~\cite{Ellis:2003pz,Ellis:2004qe} (and references therein). The corresponding $\tan \beta$ for a few particular choices for $\gamma$ are listed in the last two columns of table \ref{tanbeta} for $\mu>0$ and $\mu<0$ respectively. No solutions were found for $\gamma \lesssim 1.1$, for both signs of $\mu$. 

\begin{figure}[!ht]  \centering \medskip \begin{minipage}[b]{0.45\textwidth}  \includegraphics[width=\textwidth]{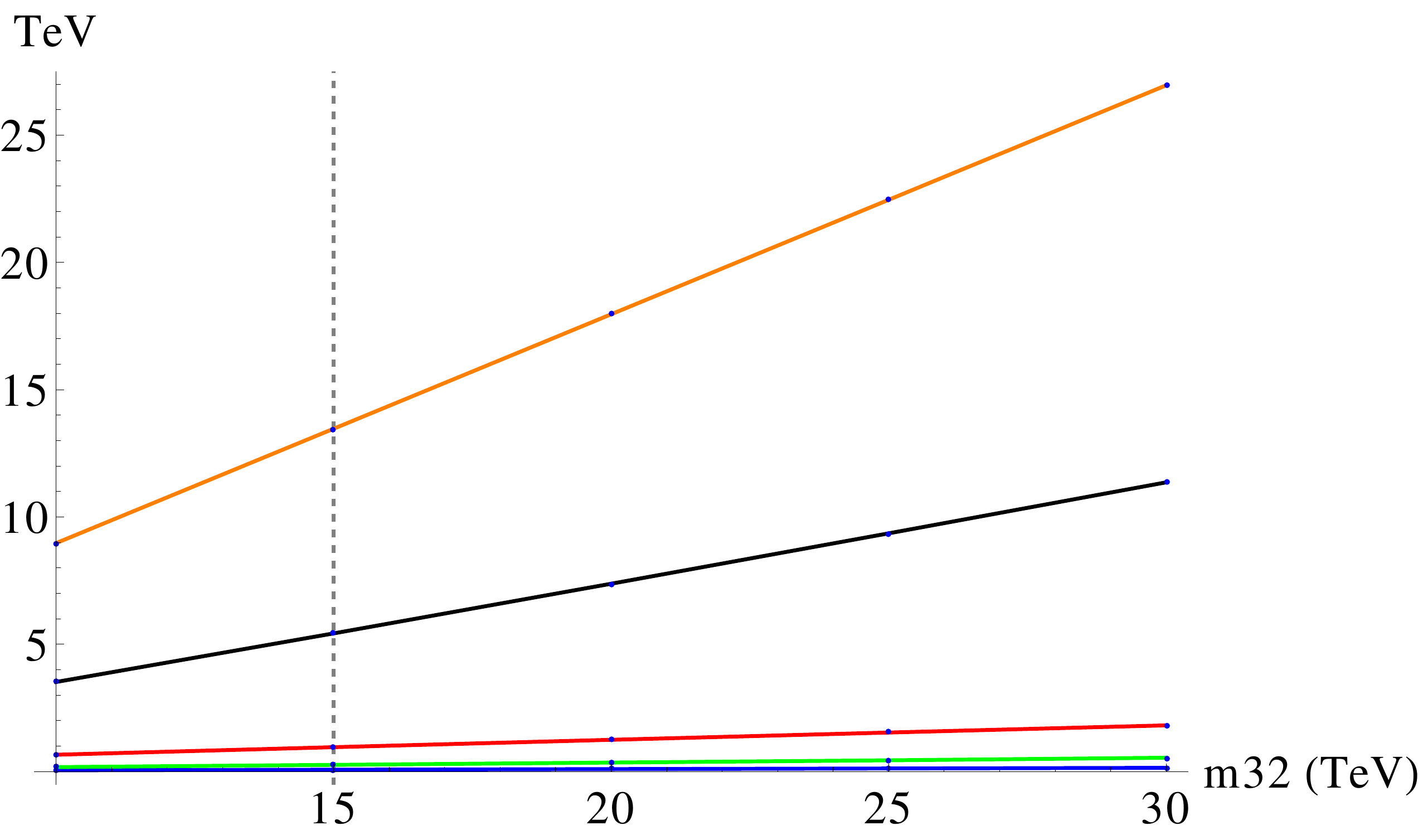}  \caption*{$\mu>0$} \end{minipage} \hfill
\begin{minipage}[b]{0.45\textwidth}  \includegraphics[width=\textwidth]{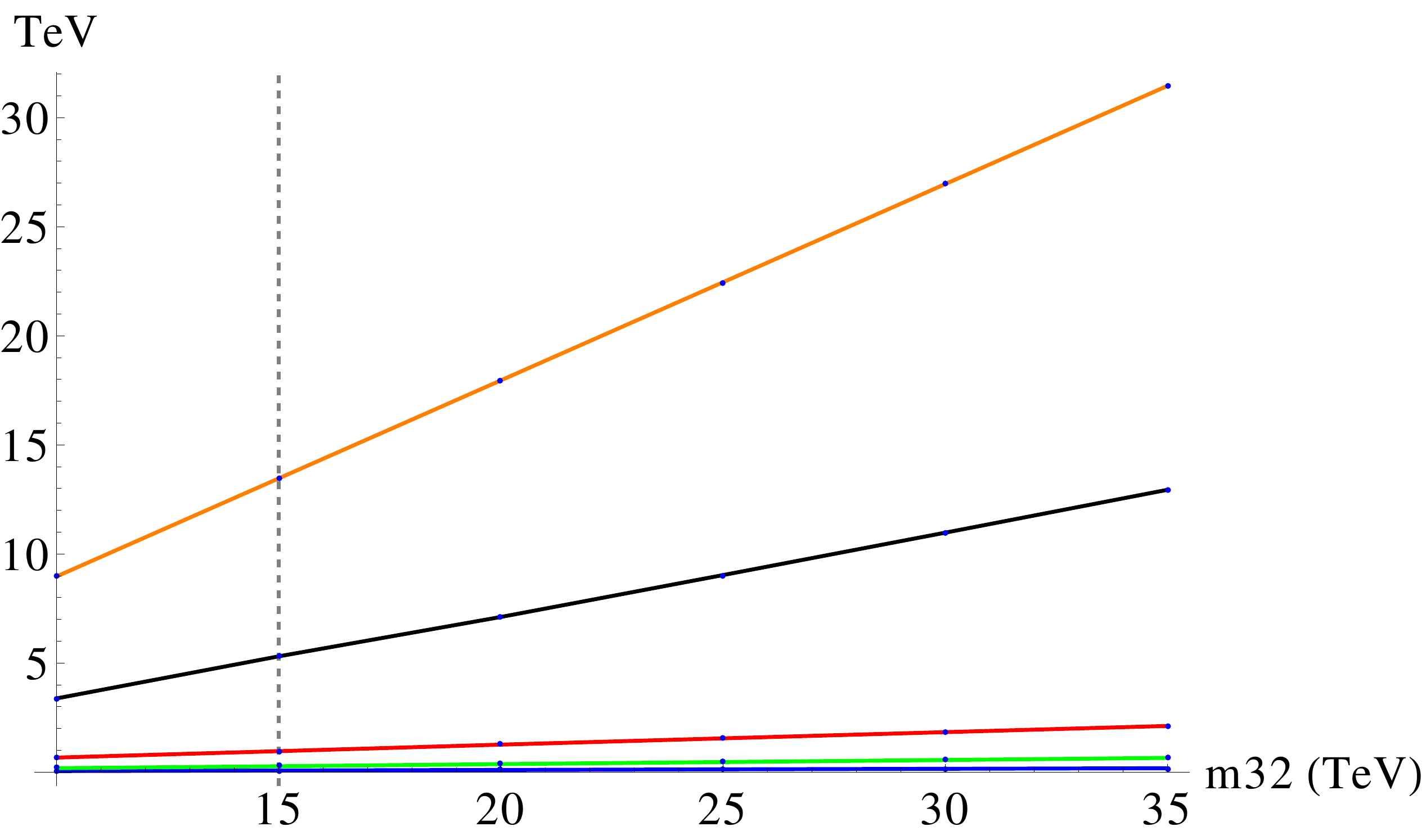} \caption*{$\mu<0$} \end{minipage}
\caption{The masses (in TeV) of the sbottom squark (orange), the stop squark (black), the gluino (red), the lightest chargino (green) and the lightest neutralino (blue) as a function of the gravitino mass for $\gamma=1.4$ and for $\mu >0$ (left) and $\mu<0$ (right).
The mass of the lightest neutralino varies slightly between 42 GeV (46 GeV) for $m_{3/2}=10$ TeV and 138 GeV (149 GeV) for  $m_{3/2}=30$ TeV for $\mu>0$ ($\mu<0$). 
The vertical dashed line at $m_{3/2} \approx 15$ TeV indicates the exclusion limit (lower bound) on the gluino mass. }\label{plotm32} \end{figure}

Some characteristic masses~\cite{softsusy} for $\gamma=1.4$ as a function of the gravitino mass are shown in figure~\ref{plotm32}. A lower experimental bound of 1 TeV for the gluino mass (vertical dashed line) forces $m_{3/2} \gtrsim 15$ TeV. On the other hand, for $\mu > 0$ ($\mu <0$) no viable solution for the RGE was found when $m_{3/2}\gtrsim 30$ TeV ($m_{3/2}\gtrsim 35$ TeV). We conclude that (for $\gamma = 1.4$) 
\begin{align} & 15 \text{ TeV} \lesssim m_{3/2} \lesssim 30 \text{ TeV} && \text{for }\mu > 0, \notag \\ & 15 \text{ TeV} \lesssim m_{3/2} \lesssim 35 \text{ TeV} && \text{for }\mu < 0. \end{align}
As we will see below, these upper bounds can differ for different choices of $\gamma$. 

\begin{figure}[!ht]  \centering \medskip \begin{minipage}[b]{0.45\textwidth}  \includegraphics[width=\textwidth]{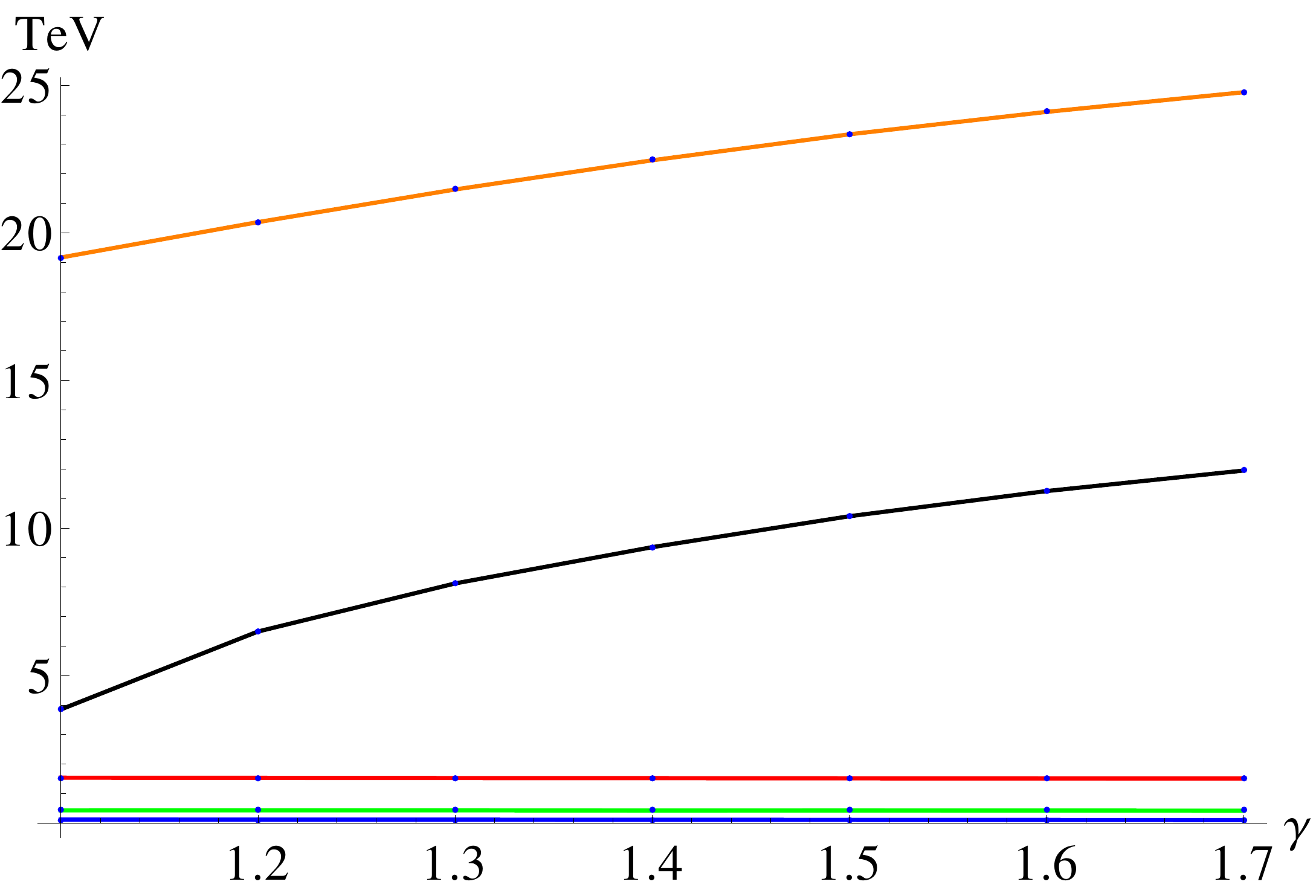} \caption*{$\mu>0$} \end{minipage} \hfill
\begin{minipage}[b]{0.45\textwidth}  \includegraphics[width=\textwidth]{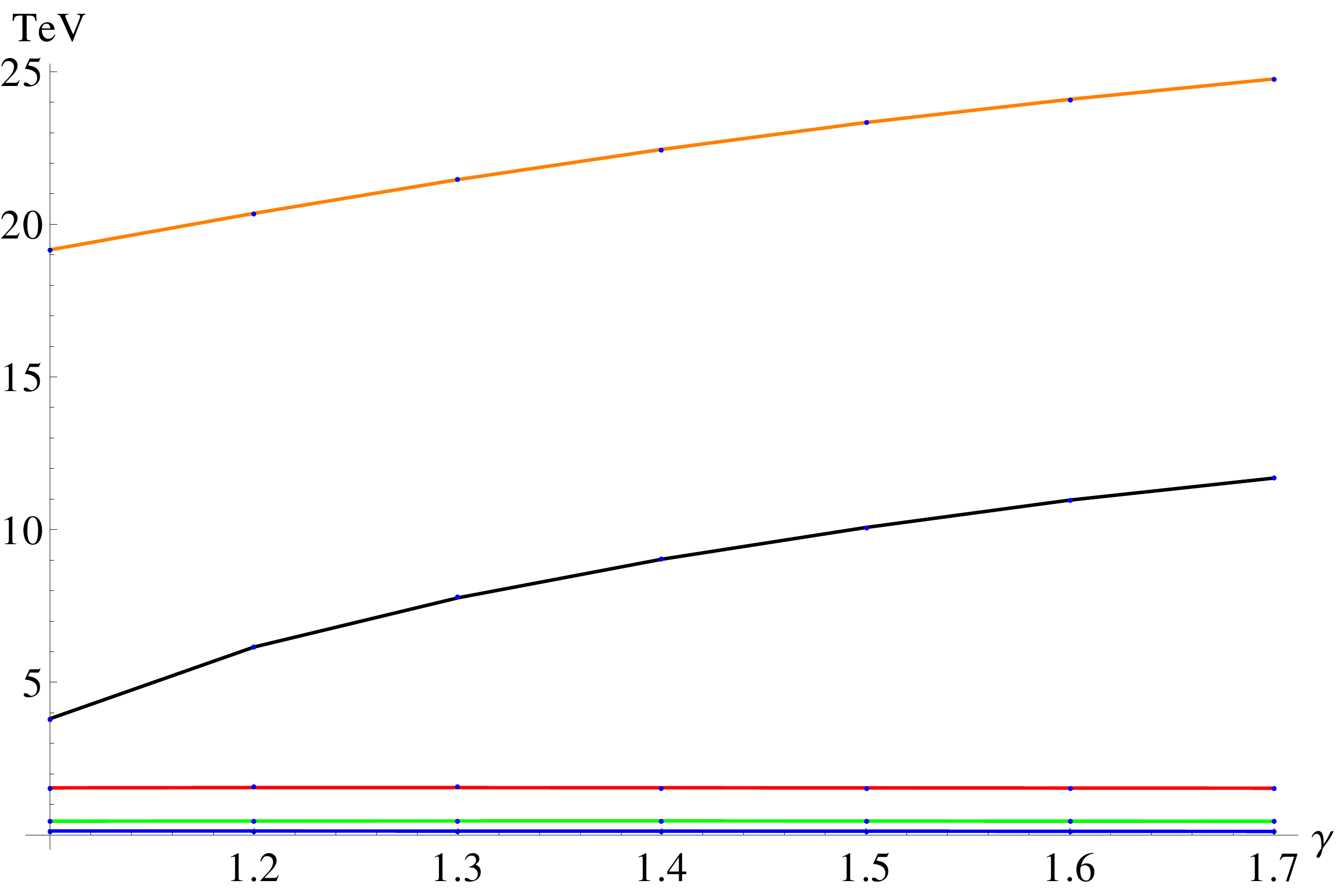} \caption*{$\mu<0$} \end{minipage}
\caption{The masses (in TeV) of the sbottom squark (orange), the stop squark (black), the gluino (red), the lightest chargino (green) and the lightest neutralino (blue) as a function of $\gamma$ for $m_{3/2}=25$ TeV and for $\mu >0$ (left) and $\mu<0$ (right). No solutions for the RGE were found for $\gamma <1.1$. Notice that for $\gamma \rightarrow 1.1$ the stop mass becomes relatively light.} \label{plotgamma} \end{figure}

In figure~\ref{plotgamma}, the same spectrum is plotted  as a function of $\gamma$ for $m_{3/2}=25$ TeV. As one can see, the stop mass varies heavily with $\gamma$, and can become relatively light when $\gamma \approx 1.1$.
For all values of $\gamma$ the LSP is given by the lightest neutralino and since $M_1 < M_2$ (see table \ref{tanbeta}) the lightest neutralino is mostly bino-like, in contrast with a typical mAMSB scenario, where the lightest neutralino is mostly wino-like~\cite{winolike}.

\begin{figure}[!ht]  \centering \medskip \begin{minipage}[b]{0.45\textwidth}  \includegraphics[width=\textwidth]{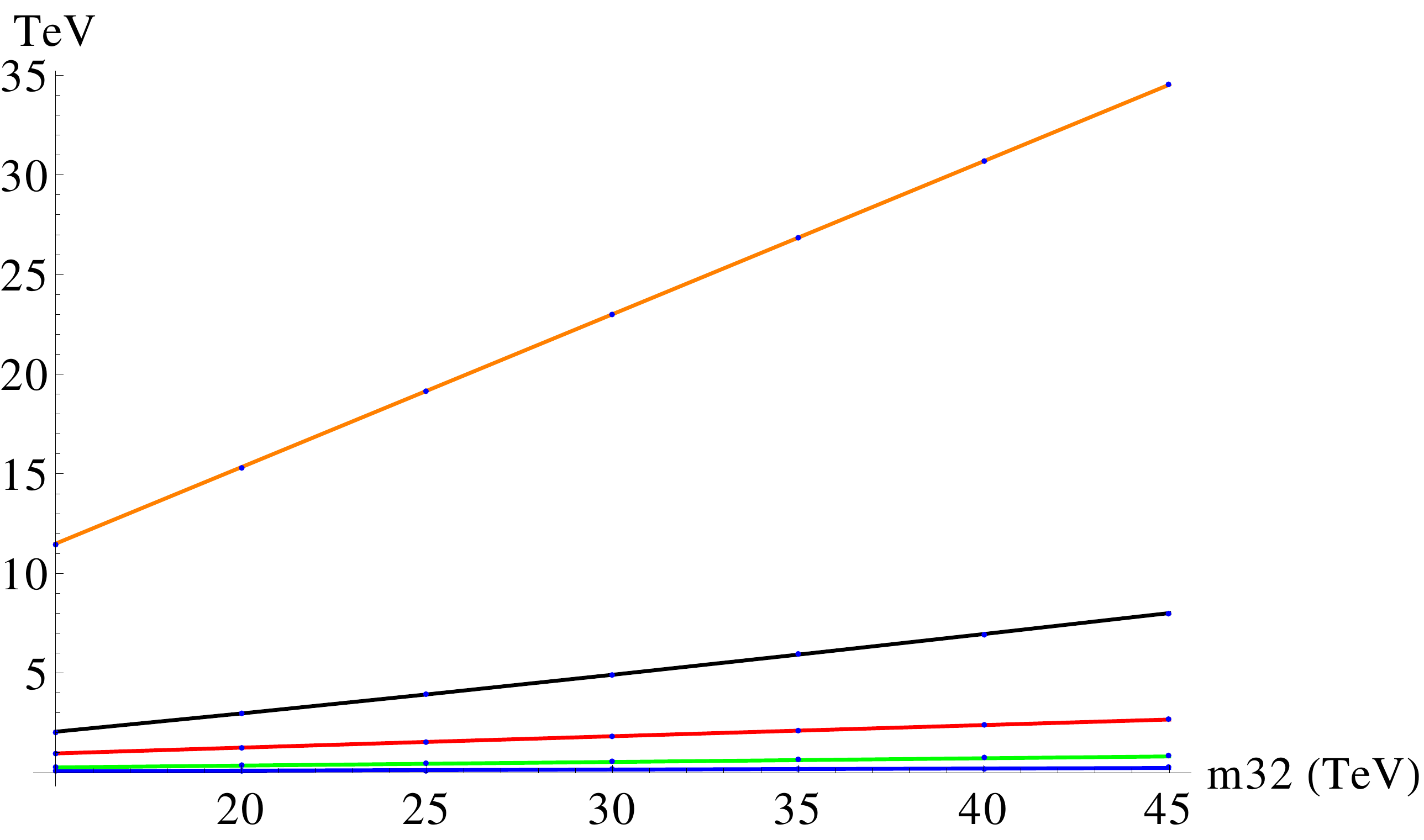} \caption*{$\gamma=1.1$ and $\mu >0$} \end{minipage} \hfill
\begin{minipage}[b]{0.45\textwidth}  \includegraphics[width=\textwidth]{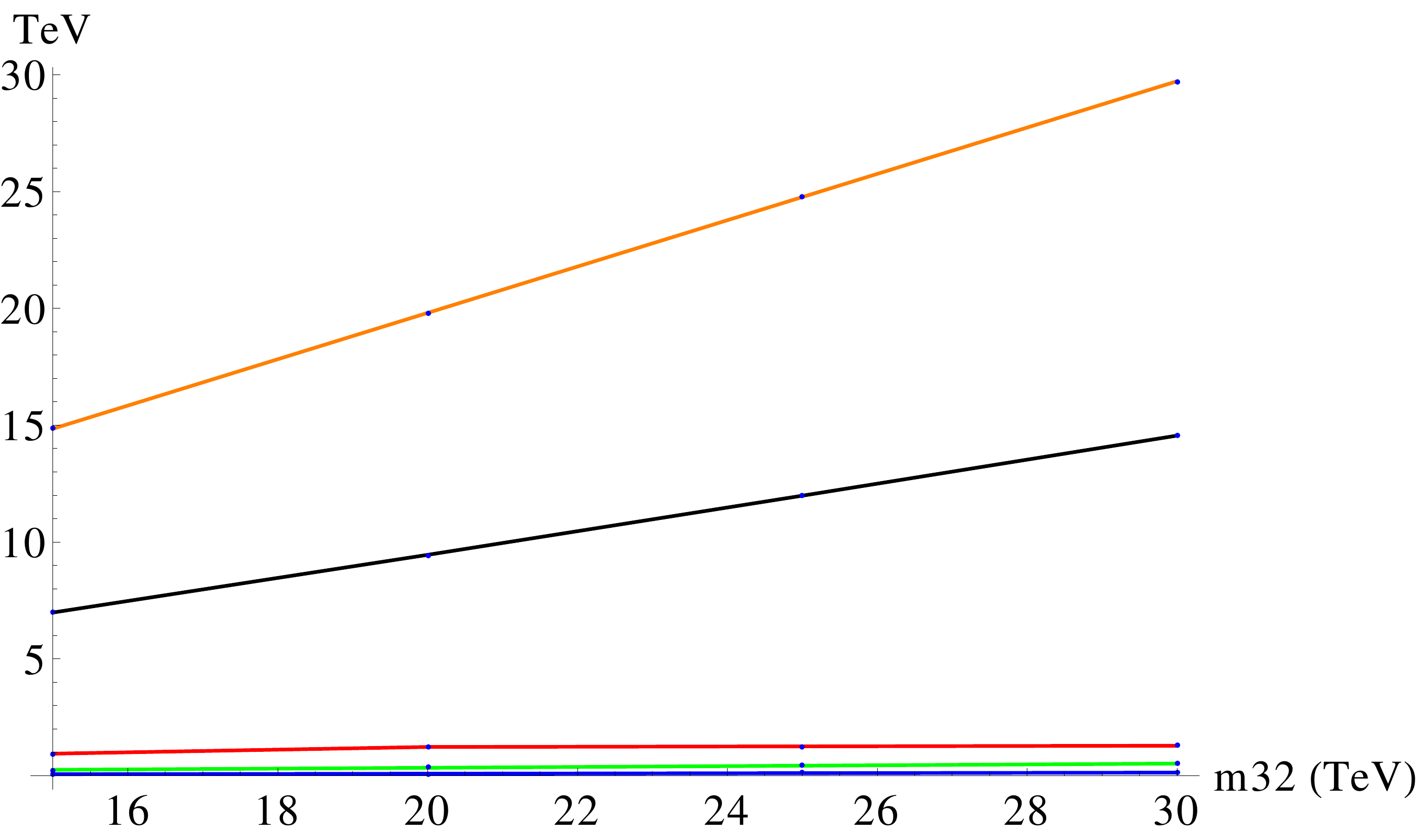} \caption*{$\gamma=1.7$ and $\mu>0$} \end{minipage}
\caption{The masses (in TeV) of the sbottom squark (yellow), the stop squark (black), the gluino (red), the lightest chargino (green) and the lightest neutralino (blue) as a function of $m_{3/2}$ for $\gamma=1.1$ (left) and for $\gamma=1.7$ (right), for $\mu>0$. 
For $\gamma=1.1$ (left) no solutions to the RGE were found when $m_{3/2} \gtrsim 45$ TeV, while for $\gamma=1.7$ (right) no solutions were found when $m_{3/2} \gtrsim 30$ TeV. The lower bound corresponds in both cases to a gluino mass of 1 TeV. } \label{fig_11} \end{figure}

To get a lower bound on the stop mass, the sparticle spectrum is plotted in figure~\ref{fig_11} (left) as a function of the gravitino mass for $\gamma=1.1$ and $\mu >0$ (for $\mu <0$ the bound is higher). As above, the experimental limit on the gluino mass forces $m_{3/2} \gtrsim 15$ TeV. In this limit the stop mass can be as low as 2 TeV. To obtain an upper bound on the stop mass on the other hand, the sparticle spectrum is plotted in figure~\ref{fig_11} (right) for $\gamma =1.7$ and $\mu>0$. Above a gravitino mass of (approximately) 30 TeV, no solutions to the RGE were found. In this limit the stop mass is about 15 TeV. 

To conclude, the lower end mass spectrum consists of (very) light charginos (with a lightest chargino between 250 and 800 GeV) and neutralinos, with a mostly bino-like neutralino as LSP ($80-230$ GeV), which would distinguish this model from the mAMSB where the LSP is mostly wino-like. These upper limits on the LSP and the lightest chargino imply that this model could in principle be excluded in the next LHC run. In order for the gluino to escape experimental bounds, the lower limit on the gravitino mass is about 15 TeV. The gluino mass is then between 1-3 TeV. This however forces the squark masses to be very high ($10-35$ TeV), with the exception of the stop mass which can be relatively light ($2-15$ TeV).

\section{Non-canonical kinetic terms} \label{sec:non-can_Kahler}

As was shown in section~\ref{sec:tachyonicmasses}, the model (\ref{model:p1}) has tachyonic soft scalar masses for the MSSM fields. In section~\ref{sec:polonyi} we proposed a solution by adding an extra field to the hidden sector. However, we will show in this section that the problem can also be circumvented by allowing non-canonical kinetic terms for the MSSM fields.

 We consider the following model
 \begin{align} \mathcal K &= - \kappa^{-2} \log (s + \bar s) + \kappa^{-2} b (s + \bar s) + (s + \bar s)^{-\nu} \sum \varphi \bar \varphi , \notag \\ W &= \kappa^{-3} a + W_{MSSM} , \notag \\  f(s) &= 1, \ \ \ \ \ \ \ f_A(s)=1/g_A^2, \label{model_noncan} \end{align}
 where a sum over all visible sector fields $\varphi$ is understood in the K\"ahler potential. 
 Here, $\nu$ is considered to be an additional parameter in the theory, where $\nu=1$ corresponds with the leading term in the Taylor expansion of $-\log(s + \bar s - \varphi \bar \varphi)$ and the model resembles 
 a no-scale model~\cite{Cremmer:1983bf,Ellis:1984bm,Lahanas:1986uc}. 
 The gauge kinetic functions for the Standard Model gauge groups $f_A(s)$ are taken to be constant.

The scalar potential is given by
  \begin{align} \mathcal  V &= \mathcal  V_F + \mathcal V_D, \notag \\ 
  \mathcal V_F &= 
  \kappa^{-4}  \frac{e^{b(s + \bar s) + \sum \kappa^2 \left( s + \bar s\right)^{-\nu} \varphi \bar \varphi}}{s + \bar s}  
  \left(-3W\bar W + g^{s \bar s} \left|\nabla_s W \right|^2+ \sum_\varphi (s + \bar s)^{\nu} \left| \nabla_\varphi W \right|^2 \right), \notag \\  
  \mathcal V_D &= \frac{c^2}{2} \left(\kappa^{-2} b - \frac{\kappa^{-2}}{s + \bar s} - \nu (s  + \bar s)^{-\nu -1} \sum \varphi \bar \varphi \right)^2 . \label{scalarpotnoncan}\end{align}
Since the visible sector fields appear only in the combination $\varphi \bar \varphi$, their VEVs vanish provided that the scalar soft masses squared are positive. Moreover, for vanishing visible sector VEVs, the scalar potential reduces to eq.~(\ref{scalarpotp1}) and the non-canonical K\"ahler potential for the visible sector fields does not change the discussion on the minimization of the potential in section \ref{sec:p=1}.  Therefore, the non-canonical K\"ahler potential does not change the fact that the F-term contribution to the soft scalar masses squared is negative. One has as in eq.~(\ref{tach})
\begin{align}  \left. \frac{\partial^2 \mathcal V_F}{\partial \varphi \partial \bar \varphi} \right|_{\langle \varphi \rangle = 0} = \kappa^{-2}a^2 e^{\alpha} \left(\frac{b}{\alpha} \right)^{\nu + 1} (\langle \sigma_s \rangle  +1) <0. \end{align}
However, the visible fields will enter in the D-term scalar potential through the derivative of the K\"ahler potential with respect to $s$. Even though this has no effect on the ground state of the potential, the $\varphi$-dependence of the D-term scalar potential does result in an extra contribution to the scalar masses squared
\begin{align}  \left. \frac{\partial^2 \mathcal  V_D}{\partial \varphi \partial \bar \varphi} \right|_{\langle \varphi \rangle = 0} = \nu \kappa^{-2} c^2 \left( \frac{b}{\alpha} \right)^{\nu + 2} \left(1-\alpha \right). \end{align}
The total soft mass squared is then the sum of these two contributions
\begin{align} m_0^2 &=  \kappa^{-2} a^2  \left(\frac{b}{\alpha} \right) \left( e^\alpha (\sigma_s +1) + \nu \frac{A(\alpha)}{\alpha} (1-\alpha)  \right), \end{align}
where eq.~(\ref{model:p1_tunability}) has been used to relate the constants $a$ and $c$, and corrections due to a small cosmological constant have been neglected. A field redefinition due to a non-canonical kinetic term $g_{\varphi \bar \varphi} = (s + \bar s)^{-\nu}$ is taken into account. The soft mass squared is now positive if
\begin{align}  \nu > -\frac{e^\alpha (\sigma_s +1) \alpha }{ A(\alpha) (1-\alpha)} \approx 2.6. \end{align}
The gravitino mass is given by eq.~(\ref{model:p1_gravitino}). In the hidden sector, the imaginary part of $s$ is eaten by the gauge boson corresponding to the shift symmetry, which becomes massive (similar to eq.~(\ref{appendix:gaugebosonmass}))
 \begin{align} m_{A_\mu} = \frac{\kappa^{-1} b c }{\alpha} \approx 1.39 \ m_{3/2}.  \end{align}
The mass of the real part of $s$ squared is given by $2 ( \alpha/b)^2 \partial_s \partial_s V$ evaluated at the ground state, where the factor $2(\alpha/b)^2$ comes from the non-canonical kinetic term,
\begin{align}  m_s^2 &=2 \left(\alpha ^4-2 \alpha ^2+4 \alpha +\frac{e^{-\alpha } (3-2 \alpha ) A}{\alpha }-4\right)  m_{3/2}^2 \notag \\ &\approx 3.48 \ m_{3/2}^2. \end{align}
Finally, the Goldstino is given by a linear combination of the fermionic superpartner of $s$ and the gaugino, which is eaten by the gravitino by the BEH mechanism. The mass of the remaining fermion is given by (see Appendix \ref{Appendix:Fermionsp1})
\begin{align}  m_f^2 &\approx 3.81 \  m_{3/2}^2. \end{align}

 Note that in the scalar potential eq.~(\ref{scalarpotnoncan}) the MSSM F-terms $\sum_\varphi \left| \nabla_\varphi W \right|^2 $ come with a prefactor $e^{\kappa^2 \mathcal K} g^{\varphi \bar \varphi} $  (where the hidden fields are replaced by their VEVs). To bring this into a more recognizable (globally supersymmetric) form where $\mathcal L \sim - g_{\varphi \bar \varphi} \partial_\mu \varphi \partial^\mu \bar \varphi - g^{\varphi \bar \varphi} W_\varphi \bar W_{\bar \varphi}$, one should rescale the MSSM superpotential (defined in eq.~(\ref{MSSMsuperpot}))
  \begin{align}   \hat W_{MSSM} = \exp (\alpha) \left(b/\alpha \right) \ W_{MSSM}.  \label{MSSMhatnoncan}   \end{align}
However, another rescaling is needed to take into account the non-canonical K\"ahler potential for the visible sector\footnote{ After the rescaling (\ref{MSSMhatnoncan}), the Lagrangian contains (very schematically) the following terms \begin{align}  \mathcal L = & -(s + \bar s)^{-\nu} \partial_\mu \bar \varphi \partial^\mu \varphi  -(s + \bar s)^{-\nu} \partial_\mu \bar h \partial^\mu  h + \hat \mu^2 \bar h h + \hat y \hat \mu \bar h \varphi \varphi + \hat y^2 \bar \varphi \bar \varphi \varphi \varphi + \dots \notag \\  & + \frac{1}{6} A_0 \hat y \varphi \varphi \varphi + \frac{1}{2} B_0 \hat \mu h h, \end{align} where $h$ stands for the Higgsinos and $\varphi$ labels the other scalar superpartners and all indices are suppressed for clarity. $y$ stands for the Yukawa couplings and $\mu$ is the usual $\mu$-parameter. The first line contains the kinetic terms and the F-terms coming from $\hat W_{MSSM}$. The last line contains the trilinear supersymmetry breaking terms (A-terms) and the B-term. In order to obtain canonical kinetic terms, one needs a rescaling $\varphi \rightarrow \varphi'=(s+ \bar s)^{-\nu/2} \varphi$ (and similarly for $h$). However, to bring the MSSM superpotential back into its usual form one also needs to redefine $\hat \mu \rightarrow \hat \mu'=(s+\bar s)^{\nu/2} \hat \mu$ and $\hat y \rightarrow \hat y'=(s+\bar s)^{\nu} \hat y $. 
One then obtains \begin{align}  \mathcal L = & - \partial_\mu \bar \varphi ' \partial^\mu \varphi '  - \partial_\mu \bar h' \partial^\mu h ' + \hat \mu '^2 \bar h' h' + \hat y' \hat \mu ' \bar h' \varphi ' \varphi ' + \hat y'^2 \bar \varphi ' \bar \varphi ' \varphi ' \varphi ' + \dots \notag \\  & + \frac{1}{6} (s + \bar s)^{\nu/2} A_0 \hat y ' \varphi ' \varphi ' \varphi ' + \frac{1}{2} (s + \bar s)^{\nu/2} B_0 \hat \mu ' h' h'. \end{align} }.  The trilinear couplings are given by 
  \begin{align}     &A_0= m_{3/2}(s + \bar s)^{\nu/2} \left(\sigma_s + 3 \right),   \end{align}
 and
  \begin{align}  &B_0 = m_{3/2} (s + \bar s)^{\nu/2} \left(\sigma_s + 2 \right).  \end{align}
  
The main phenomenological properties of this model are not expected to be different from the one we analyzed in section \ref{sec:low_energy_spectrum} with the parameter $\nu$ replacing $\gamma$. Gaugino masses are again generated at one-loop level while mSUGRA applies to the soft scalar sector. We therefore do not repeat the phenomenological analysis for this model.

\cleardoublepage

  \graphicspath{{\chaptersdir/chargedMSSM/image/}}%
  \chapter{Gauging global symmetries of the MSSM}\label{ch:chargedMSSM}

In the previous chapter, the soft supersymmetry breaking terms were calculated for the model introduced in chapter~\ref{ch:dS}. 
It was shown that the minimal model suffers from tachyonic scalar soft masses, and two possible solutions to this problem were presented:
One is based on the introduction of an extra hidden sector (Pol\'onyi-like) field $z$, which introduces an extra parameter $\gamma$ in the model.
A second solution involves a $s$-dependent K\"ahler potential for the MSSM fields and an extra parameter $\nu$.
Since the low energy spectra of both cases are very similar, only the spectrum of the former case was calculated. 
While the gaugino masses vanish at tree level, they are generated at one loop due to anomaly mediation, and are therefore suppressed compared to the scalar soft masses.
It was shown that the low energy spectrum is very particular, and how it can be distinguished from other minimal supersymmetry breaking scenarios.

In the discussion in chapter~\ref{ch:softterms} however, the Standard Model fields were assumed to be inert under the shift symmetry.
In this chapter we study the possibility that the gauged shift symmetry is identified with a known global symmetry of the Standard Model, or more generally its supersymmetric
extension.
This is done while keeping the desirable properties of the model, namely the existence of a metastable dS vacuum 
with a tunable cosmological constant and a viable low energy spectrum of the superpartners.
A particular attractive possibility is to use a symmetry that contains the usual R-parity, or matter-parity (depending on the K\"ahler basis) of the MSSM (see section~\ref{sec:R-symmetry}).
We find that this is indeed possible and analyze explicitly the anomaly free symmetry $B-L$. 
A charge $q$ proportional to $B-L$ is introduced for the Standard Model fields, 
that allows to extrapolate between the current analysis and the one of chapter~\ref{ch:softterms}.  
It turns out though that $q$ is bounded from the requirement of existence of the electroweak vacuum.
Moreover, the spectrum is very similar to the one where the MSSM fields are inert (see section~\ref{sec:low_energy_spectrum}), 
with the exception that the stop squark mass can now be as low as $1.5$ TeV.

We also address the question if the problem of tachyonic scalar masses can be solved without adding extra field or modifying the matter kinetic terms, 
by appropriately choosing the transformations of the MSSM fields, due to the extra D-term contribution in the scalar potential since SM fields are now charged under the $U(1)$.
However,  the answer turns out to be negative due to constraints arising from the existence of the usual electroweak vacuum.  
Finally, we analyze the phenomenological implications of the extra $U(1)$ and  we  find  that  its  coupling  is  too  small  to  have  possible  experimental  signatures  in
colliders at present energies.

This chapter is organized as follows. In section~\ref{sec:B-L}  we analyze the possibility of identifying the  gauged  shift  symmetry  with  the $B-L$.
We work out the model and its phenomenology; we also comment on the case of $3B-L$, but we do not repeat the analysis since the results are very similar.
In Section~\ref{sec:other}, we consider the most general global symmetry and address the question of tachyonic scalar masses without extra field or modification of the matter kinetic terms.  

This chapter is entirely based on work with I.~Antoniadis~\cite{RK4}.

\section{The model extended with B-L charges} \label{sec:B-L}

As discussed above, we now explore the possibility to give the MSSM superfields (denoted by $\varphi_\alpha$) a charge $q_\alpha$ under the extra $U(1)_R$, 
proportional to $B-L$, extending the model~(\ref{model:polonyi}).  This means that $Q, \bar u$ and $\bar d$ have charges $q/3, -q/3$ and $-q/3$, respectively. 
The Higgs superfields do not carry a charge and the leptons $L$ and $\bar e$ carry a charge $-q$ and $+q$ respectively.

First, this gives contributions to the D-term part of the scalar potential, and one should check that this does not ruin its stability. The scalar potential is now given by
\begin{align} \mathcal V &= \mathcal  V_F + \mathcal V_D, \notag \\ \mathcal  V_D &= \frac{1}{2} \left( -\frac{\kappa^{-2} c}{s + \bar s}  
+ \kappa^{-2} bc - \sum q_\alpha \varphi_\alpha \bar \varphi_\alpha \right)^2, \label{polonyi:scalarpot2} \end{align}
where $ \mathcal V_F$ is the same as in eq.~(\ref{scalarpotMSSM}). The D-term part will give an extra contribution to the soft scalar masses of the matter fields $\varphi_\alpha$. The restriction that these remain non-tachyonic gives
\begin{align}  0&< \left. \partial_{\varphi_\alpha} \partial_{\bar \varphi_\alpha} \mathcal V \right|_{\varphi = 0}  \notag \\&= \kappa^{-2} a^2 \frac{e^{b (s + \bar s) + t^2} }{s + \bar s} \left(A(t) (\sigma_s + 1) + B(t) \right) + \kappa^{-2} q_\alpha c \left(\frac{1}{s + \bar s} - b \right),\notag \end{align}
where $A(t)$ and $B(t)$ are given in eqs.~(\ref{AB}). Since $q_\alpha$ can be either positive or negative, and the first term on the r.h.s. is positive for the VEVs of $t$ and $s$ (see section~\ref{sec:polonyi}), it follows that
\begin{align}  |q| < \frac{a^2}{c} e^{\alpha + t^2} \frac{A(t) (\sigma_s+1) + B(t) }{1-\alpha},  \end{align}
which can be rewritten as (by use of eqs.~(\ref{model:polonyi_tunability}) and (\ref{model:polonyi_gravitino}))
\begin{align}  |q| < q_{\text{max}} = \kappa m_{3/2} \frac{A(t) (\sigma_s+1) + B(t)}{\sqrt{|A(t) \sigma_s + B(t)|}} \frac{1}{\sqrt{2} (1 + \gamma t) }.  \label{qconstraint} \end{align}
However, we will show below that one actually needs $|q|/q_{\text{max}}< 0.013$ in order to find a viable solution to the RGE.
Note that the constraint~(\ref{qconstraint}) can be rewritten  (by using the relation~(\ref{model:polonyi_tunability})) as
\begin{align}   |q| < q_{\text{max}} = bc \ \frac{A(t)(\sigma_s +1) + B(t)}{A(t) \sigma_s + B(t)} \frac{1-\alpha}{\alpha}, \end{align} 
where $\kappa^{-2} bc$ is the Fayet-Iliopoulos constant in the scalar potential~(\ref{polonyi:scalarpot2}).

While the mixed $U(1)_R \times U(1)_Y^2$,  $U(1)_R \times SU(2)$ and  $U(1)_R \times SU(3)$ anomalies vanish, the cubic anomaly vanishes only upon the inclusion of three right-handed neutrinos which are singlets under the Standard Model gauge groups. Otherwise, the cubic anomaly is proportional to
\begin{align}  \text{Tr} Q^3 = -3q^3, \end{align}
but the mixed anomalies still vanish. In this case, the cubic anomaly should be canceled by a Green-Schwarz counterterm (\ref{model:Green-Schwarz term}), provided
\begin{align}   f(s) &= 1 + \beta_R s, \notag \\  \beta_R &= - \frac{q^3}{4\pi^2 c}. \end{align} 

The gaugino masses are generated at one loop from anomaly mediaton, given by eqs.~(\ref{model:polonyi_m1m2m3}), while the other soft supersymmetry breaking terms are given by
\begin{align}   m_{0,i}^2 &= m_{3/2}^2  \left[ \left( \sigma_s + 1 \right) + \frac{(\gamma + t + \gamma t^2)^2}{(1 + \gamma t)^2}\right] + \kappa^{-2} q_\alpha bc \left(\frac{1}{\alpha} - 1 \right) , \notag \\   A_0&= m_{3/2} \left[  (\sigma_s +3)  + t \frac{ (\gamma + t +\gamma t^2) } {1+\gamma t} \right], \notag \\   B_0 &= m_{3/2}  \left[  (\sigma_s + 2) + t \frac{ (\gamma + t + \gamma t^2) }{(1+\gamma t)}  \right].    \end{align}
Or, by using
\begin{align} bc = m_{3/2} \kappa \frac{ \sqrt{-2\left( A(t) \sigma_s + B(t) \right)} }{1+ \gamma t} \frac{\alpha}{1-\alpha}, \label{bc}\end{align}
the soft terms can be written as
\begin{align}   m_{0,i}^2 &= m_{3/2}^2  \left[ \left( \sigma_s + 1 \right) + \frac{(\gamma + t + \gamma t^2)^2}{(1 + \gamma t)^2}\right] + \kappa^{-1} m_{3/2} q_\alpha \frac{\sqrt{-2 \left(A(t) \sigma_s + B(t) \right)}}{1+\gamma t} , \notag \\   A_0&= m_{3/2} \left[  (\sigma_s +3)  + t \frac{ (\gamma + t +\gamma t^2) } {1+\gamma t} \right], \notag \\   B_0 &= m_{3/2}  \left[  (\sigma_s + 2) + t \frac{ (\gamma + t + \gamma t^2) }{(1+\gamma t)}  \right]. \label{softtermsB-L}   \end{align}
Note that the relation (\ref{AB-relation}) still holds.

In chapter~\ref{ch:softterms} the special case $q=0$ was analyzed in full detail; it was shown that for $\gamma < 1.1$ no solutions to the RGEs exist that satisfy eq.~(\ref{AB-relation}). 
Moreover, it was shown that for $\gamma \rightarrow 1.1$ the mass of the lightest stop $m_{\tilde t}$ can become very small. By imposing a lower bound of about $m_{3/2}\geq 15$ TeV on the gravitino mass, which originates from a lower bound of about 1 TeV on the gluino mass, it was shown that the mass of the lightest stop can be as low as about 2 TeV, while the masses of the other squarks remain high ($>10$ TeV).

\begin{figure} \centering \medskip \includegraphics[width=0.5\textwidth]{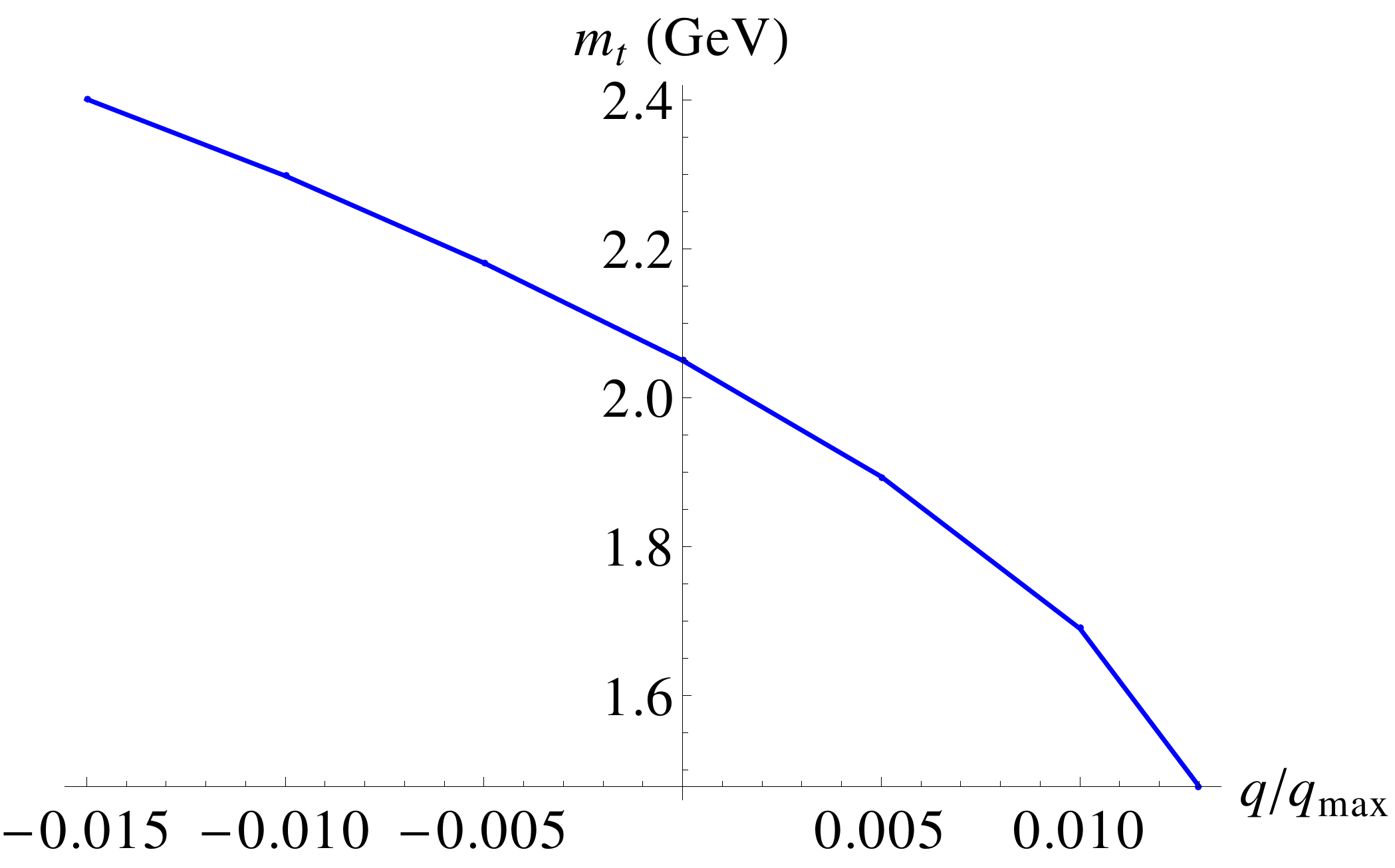} \label{B-L_mt_qqmax}  \caption{The mass of the lightest stop squark as a function of the charge $q/q_{\text{max}}$ for $\gamma=1.1$ and $m_{3/2}=15 \text{ TeV}$. The gravitino mass is chosen such that the gluino mass is right above the experimental bound of 1 TeV (while other experimental bounds such as the neutralino and charginos are also satisfied). A positive $q$ corresponds to the scalar soft masses $m_Q$ and $m_e$ being heavier than $m_L$ and $m_d = m_u$ (see eq.~(\ref{softtermsB-L})). For $q/q_{\text{max}} > 0.013$ no solutions to the RGE were found.} \label{fig:stop} \end{figure}

As it turns out, the only considerable effect of a non-zero charge $q$ to the sparticle spectrum is for the lightest stop, whose dependence on the input parameter $q/q_{\text{max}}$ for $m_{3/2}=15 \text{ TeV}$ and $\gamma=1.1$ is plotted~\cite{softsusy} in figure~\ref{fig:stop}.
For $q/q_{\text{max}} > 0.013$, no solutions to the RGE were found. A lower limit for the mass of the lightest stop of about $1.5$ TeV is found when $q/q_{\text{max}} \rightarrow 0.013$. 

\begin{figure} \centering \medskip \includegraphics[width=0.5\textwidth]{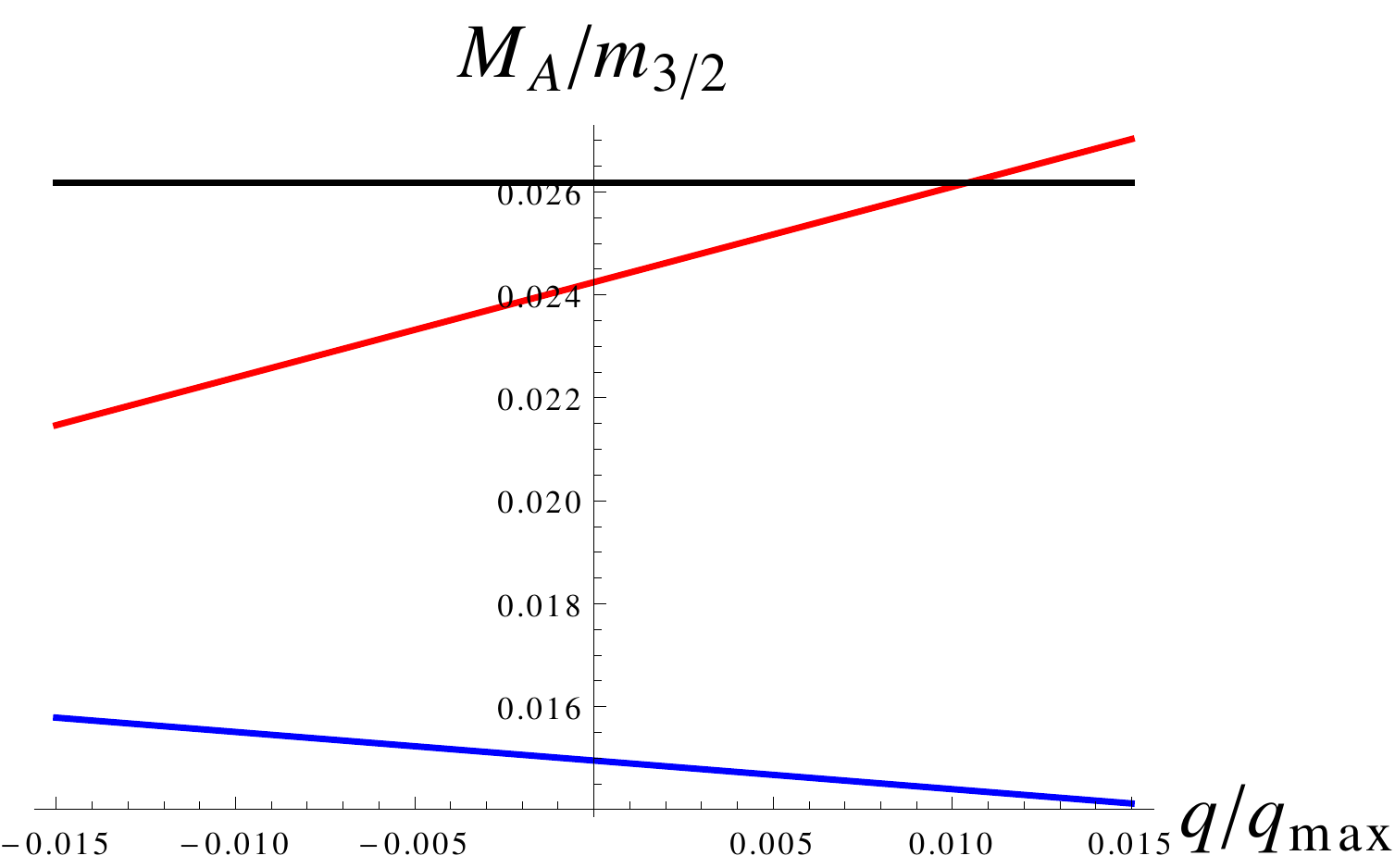} 
\caption{The gaugino masses $M_1/m_{3/2}$ (blue), $M_2/m_{3/2}$ (red), $M_3/m_{3/2}$ (black) in the $3B-L$ model, as a function of $q/q_{\text{max}}$, where $\frac{5}{3} g_Y^2 = g_2^2 = g_3^2=0.51$ is assumed at the GUT scale. Note that the gaugino masses vary only very little for $\left| q/q_{\text{max}} \right| < 0.013$} 
\label{fig:3B-L}  \end{figure}

It should however be noted that, since anomalies are canceled by a Green-Schwarz mechanism, one can in principle choose different charge allocations for the MSSM fields which allow the terms in the superpotential (\ref{MSSMsuperpot}), while forbidding the Baryon and Lepton violating terms (\ref{deltabl}) and the dimension five operators (\ref{dim 5}).
As was mentioned in section~\ref{sec:R-symmetry}, a gauged $B-L$ forbids the terms in eq.~(\ref{deltabl}), but it still allows certain dimension five operators. 
This can be solved by gauging $3B-L$. A gauged $3B-L$ is anomalous and its $U(1)_{3B-L} \times U(1)_Y^2$ and $U(1)_{3B-L} \times SU(2)$ anomalies are proportional to
\begin{align} \mathcal C_1 &= -3q, \notag \\ \mathcal C_2 &= 6q,\end{align}
while the $U(1)_{3B-L}^2 \times U(1)_Y$ and $U(1)_{3B-L} \times SU(3)$ anomalies vanish.
As was explained in section~\ref{sec:Anomalies-gauginomass}, this results in a contribution to the gaugino masses eq.~(\ref{gauginomassGS}) given by
\begin{align}
 M_1 &= - g_Y^2 \frac{3 \alpha (\alpha-1)}{8 \pi^2} \frac{q}{bc}{m_{3/2}}, \notag \\
 M_2 &=  g_2^2 \frac{6 \alpha (\alpha-1)}{8 \pi^2} \frac{q}{bc}{m_{3/2}}.
\end{align}
The total gaugino masses are then given by
\begin{align}   M_1 &=  \frac{g_Y^2}{16 \pi^2} m_{3/2} \left( 11 \left[ 1 - (\alpha -1)^2 -  \frac{ t(\gamma + t + \gamma t^2)}{1 + \gamma t}  \right] -\alpha (\alpha-1) \frac{6q}{bc}  \right), \notag \\   
M_2 &=  \frac{g_2^2}{16 \pi^2} m_{3/2} \left( \left[1 - 5 (\alpha-1)^2 -5 \frac{ t (\gamma + t + \gamma t^2)}{1 + \gamma t} \right]  +\alpha (\alpha-1) \frac{12q}{bc} \right), \notag \\   
M_3 &= - 3 \frac{g_3^2}{16 \pi^2} m_{3/2} \left[ 1 + (\alpha - 1)^2 + \frac{ t (\gamma + t + \gamma t^2) }{1 + \gamma t} \right]. \label{3B-L:Gauginomasses}   \end{align}  
By using (from eqs.~(\ref{qconstraint}) and (\ref{bc}))
\begin{align}
 \frac{q}{bc} = \frac{q}{q_{\text{max}}} \frac{A(t)(\sigma_s +1) + B(t)}{A(t) \sigma_s + B(t)} \frac{\alpha -1}{2 \alpha},
\end{align}
the corrections to the gaugino masses proportional to $q/q_{\text{max}}$ can be calculated for every $\gamma$. It turns out that these corrections are very small, as can be seen in figure~\ref{fig:3B-L}, where the gaugino masses are plotted as a function of $q/q_{\text{max}}$ for $\gamma =1.1$.
The low energy spectrum is then expected to be similar to that of the $B-L$ case described above and we therefore do not perform a separate analysis for the $3B-L$ case.

The kinetic terms of the $U(1)_R$ gauge boson are given by\footnote{Alternatively, one can define the gauge kinetic function as $f(s)= 1/q^2$, such that the charge of the fermions is given by (instead of being proportional to) $B-L$.}
\begin{align} \mathcal L_{\text{kin}}/e = -\frac{1}{4} F_{\mu \nu} F^{\mu \nu}.\end{align}
Its mass is given by eq.~(\ref{appendix:gaugebosonmass}),
\begin{align} M_R &= \kappa^{-1} \frac{bc}{\alpha} \notag \\ &= m_{3/2} \left[ (1+\gamma t) e^{\alpha^2 t^2} \sqrt{\frac{\sigma_s A(t) + B(t)}{(\alpha-1)^2}} \right]. \end{align}
In the allowed parameter range this corresponds to $M_R \in \left]25.4 , 99.4 \right[$ TeV. The covariant derivative of the Standard Model fermions $\chi^\alpha$ (with charge $q$) is
\begin{align}  D_\mu \chi^\alpha = \left( \partial_\mu -  iq A_\mu \right) \chi^\alpha, \end{align}
where we have omitted the spin connection and the K\"ahler connection. The charge $q$ of the MSSM fermions satisfies $|q|<0.013  q_{\text{max}} \approx \mathcal O(10^{-17})$.  
We conclude that the $U(1)_R$ gauge boson is (unfortunately) well beyond the current experimental $Z'$ bounds\footnote{
For a review on $Z'$s, see for example~\cite{Langacker:2008yv,Rizzo:2006nw}.} or the corresponding compositeness limits~\cite{Agashe:2014kda}.

\section{D-term contributions to the scalar soft masses} \label{sec:other}

In this section we show that another possible solution to the problem of negative soft scalar masses squared in section~\ref{sec:tachyonicmasses}, 
based on a D-term contribution to the scalar soft masses, does not lead to consistent electroweak vacua. 
As in the other solutions proposed in chapter~\ref{ch:softterms}, solving the problem of tachyonic masses comes at the cost of introducing an extra parameter, $b_1$. 
In this case the model is given by\footnote{
Note that the coefficient $\beta$ in the gauge kinetic function has a slightly different definition than in the rest of this thesis. 
The reason lies in the fact that this definition allows one to eliminate the parameter $b$ from the vacuum conditions (\ref{betaprime}),
similar to the case with a constant gauge kinetic function in section~\ref{sec:p=1}.}
\begin{align} \mathcal K &= -\kappa^{-2}\log(s+\bar s) +\kappa^{-2} b(s+ \bar s) + K_{MSSM}, \notag \\  W &= \kappa^{-3} a +  e^{b_1 s} W_{MSSM}, \notag \\  f(s) &= 1 + \beta b s. \label{model:other} \end{align}
Because of the shift symmetry (\ref{shift}), the MSSM superpotential needs to transform (with gauge parameter $\theta$) as
\begin{align}  W_{MSSM} \longrightarrow W_{MSSM} e^{ib_1 c \theta}. \label{WMSSMtransform} \end{align}
The scalar potential is given by
\begin{align}\mathcal V &= \mathcal V_F + \mathcal V_D, \notag \\ \mathcal  V_F &= e^{\mathcal K} \left[ -3 W \bar W + g^{s \bar s} |\nabla_s W |^2 
+ |\nabla_\varphi W|^2 \right] \notag \\ \mathcal V_D &= \frac{1}{2} \left(\kappa^{-2}bc - \frac{\kappa^{-2} c}{s+ \bar s} - q_\alpha \varphi_\alpha \bar \varphi_\alpha \right)^2, \end{align}
where $\varphi_\alpha$ stands for the various MSSM fields and the linear part in the gauge kinetic function has been neglected. 
Indeed, it is shown in Appendix~\ref{Appendix:nonzerobeta} that $\beta  \ll 1$.

The F-term contribution to the scalar soft masses squared is negative. 
This can however be compensated by a (positive) contribution proportional to the charge $q_\alpha$ from the D-term scalar potential. 
This implies that all MSSM fields must have a positive charge under this extra $U(1)$, which is the motivation behind the transformation (\ref{WMSSMtransform}) 
and the factor $e^{b_1 s}$ in eq.~(\ref{model:other}). 
The soft supersymmetry breaking terms can be calculated (with respect to a rescaled superpotential $\hat W_{\text{MSSM}} = e^{\mathcal K/2} e^{b_1 s} W_{\text{MSSM}}$) to be
\begin{align}  m_0^2 &= m_{3/2}^2 (\sigma_s + 1) +\kappa^{-2} q_\alpha bc \frac{ 1 - \alpha}{\alpha}, \notag \\  
A_0 &= m_{3/2} \ \rho_s,   \\  
B_0 &= m_{3/2} \ (\rho_s +1), && \rho_s &= -1 + (\alpha-1)(\alpha - 1 + b_1 (s + \bar s) ) \notag , \end{align}
where $\alpha$ and $\sigma_s$ are defined in eqs.~(\ref{Vt}). The gravitino mass is given by 
\begin{align} m_{3/2} &=  \kappa^{-1} a e^{\alpha/2} \sqrt{\frac{b}{\alpha}}. \label{p1:gravitino_mass} \end{align}
By using the relation for the gravitino mass eq.~(\ref{model:polonyi_gravitino}), the relations (\ref{model:polonyi_tunability}) can be rewritten as
 \begin{align} bc = - \kappa m_{3/2} e^{-\alpha/2} \sqrt{\alpha A(\alpha)}, \label{bc=alphaA} \end{align}
 which can be used to rewrite the D-term contribution to the mass in the form 
\begin{align}  m_0^2 &= m_{3/2}^2 (\sigma_s + 1) + m_{3/2} \kappa^{-1} q_\alpha \ Q(\alpha), \notag \\  
Q(\alpha) &= \frac{\alpha-1}{ \alpha} e^{-\alpha/2} \sqrt{\alpha A(\alpha)} \approx 0.8598. \label{Qalpha} \end{align} 
To avoid tachyonic masses, the charges of the MSSM fields should satisfy
\begin{align} q_\alpha > q_0 = - \kappa  m_{3/2} \frac{\sigma_s +1}{Q(\alpha)} \approx \  0.558 \ \kappa  m_{3/2},   \label{qlimit} \end{align}
which corresponds to $q \gtrsim 0.5 \times 10^{-15}$ for $m_{3/2} \approx 20$ TeV. 
The (non-universal) scalar soft masses and trilinear terms can be summarized as
\begin{align}
 m_{0,i}^2 &= m_{3/2}^2 \left[ (\sigma_s +1) \left( 1 - \frac{q_\alpha}{q_0} \right) \right], \notag \\
 A_0 &= m_{3/2} \left[ -1 + (\alpha -1)\left(\alpha -1 + \frac{q}{q_0} P(\alpha)\right) \right], \notag \\
 P(\alpha) &=   \frac{2 \alpha e^{\alpha/2} (\sigma_s +1)}{\sqrt{\alpha A(\alpha)} Q(\alpha)} \approx 0.799,
\end{align}
where eq.~(\ref{model:polonyi_tunability}) was used and $q = b_1 c /2$. The charges $q_\alpha$ are given in terms of three independent parameters $\theta$, $\theta_Q$ and $\theta_L$ by
\begin{align} q_{H_u} &= \theta q, \notag \\ q_{H_d} &= (2-\theta)q, \notag \\ q_L &= \theta_L q, \notag \\ q_{\bar e} &= (\theta - \theta_L) q, \notag \\
q_{Q} &= \theta_Q q, \notag \\q_{\bar u} &= (2-\theta - \theta_Q)q, \notag \\ q_{\bar d} &= (\theta - \theta_Q)q, \label{charges}\end{align}
such that 
\begin{align}  q_{H_u} + q_{H_d} &= 2q, \notag \\  q_{\bar e} + q_{L}+ q_{H_d} &= 2q, \notag \\  q_{\bar d} + q_Q + q_{H_d} &= 2q, \notag \\  q_{\bar u} + q_Q + q_{H_u} &= 2q,\end{align}
and eq.~(\ref{WMSSMtransform}) is satisfied for the MSSM superpotential (\ref{MSSMsuperpot}).

Next, the cubic and mixed anomalies are proportional to
\begin{align} \mathcal C_R &= q^3 f(\theta,\theta_L,\theta_Q), \notag \\  \mathcal C_1 &=  \frac{3q}{2} \left(6- 3 \theta_Q - \theta_L  \right), \notag \\
 \mathcal C_2 &= q \left(2+ 3 \theta_L + 9 \theta_Q \right), \notag \\ \mathcal C_3 &= 6q,\end{align}
where
\begin{align} f(\theta,\theta_L,\theta_Q) &=  3 \left( 6 \theta_Q^3 + 3(2-\theta_Q - \theta_L)^3+ 3(\theta-\theta_Q)^3 + 2 \theta_L^3 + (\theta- \theta_L)^3 \right) \notag \\
& \qquad + 2 \left( (2-\theta)^3 + \theta^3 \right). \label{ftheta}\end{align}
The Green-Schwarz counterterms are then proportional to 
\begin{align}
 \beta &= - q^3 \frac{f(\theta,\theta_L,\theta_Q) }{12 \pi^2 b c}, \notag \\  \beta_1 &= - q\frac{  3 \left(6- 3 \theta_Q - \theta_L  \right)}{8 \pi^2 c}, \notag \\
 \beta_2 &= - q\frac{ \left(2+ 3 \theta_L + 9 \theta_Q \right)}{4 \pi^2 c}, \notag \\ \beta_3 &= - q\frac{3}{2 \pi^2 c}. \label{other:anomalycancellation}
\end{align}
This results in contributions to the gaugino masses
\begin{align} M_A &= -\frac{g_A^2}{2} \frac{\alpha (\alpha - 1)}{b} m_{3/2} \beta_A. \end{align}
The Green-Schwarz contributions to the gaugino masses are given by
\begin{align}
 M_1 &= - \frac{g_Y^2}{16\pi^2} 3(6-3\theta_Q-\theta_L) \frac{ \alpha (\alpha - 1) e^\alpha}{\sqrt{\alpha A(\alpha)} } \kappa^{-1}q, \notag \\
 M_2 &= - \frac{g_2^2}{16\pi^2}2 (3 + \theta_L + 9 \theta_Q) \frac{2 \alpha (\alpha - 1) e^\alpha}{\sqrt{\alpha A(\alpha)}} \kappa^{-1}q, \notag \\
 M_3 &= - \frac{g_3^2}{16\pi^2} \frac{ 12 \alpha (\alpha - 1) e^\alpha}{\sqrt{\alpha A(\alpha)}} \kappa^{-1}q.
\end{align}
The anomaly mediated contribution to the gaugino masses are given by eq.~(\ref{gauginomass:anomalymediated}). As was explained in section~\ref{sec:Anomalies-gauginomass}, the total one-loop gaugino mass is the sum of these contributions 
\begin{align}
 M_1 &= - \frac{g_Y^2}{16\pi^2}  \left(  11 m_{3/2} \left[-1 + (\alpha -1)^2 \right] -  \frac{3}{2} (6-3\theta_Q-\theta_L) \frac{ (\alpha - 1)^2}{Q(\alpha) } \kappa^{-1}q \right), \notag \\
 M_2 &= - \frac{g_2^2}{16\pi^2} \left( m_{3/2} \left[ -1 + 4 (\alpha - 1)^2 \right] - (3 + \theta_L + 9 \theta_Q)    \frac{ (\alpha - 1)^2}{Q(\alpha) } \kappa^{-1}q \right), \notag \\
 M_3 &= - \frac{g_3^2}{16\pi^2} \left( 3 m_{3/2} \left[ 1+ (\alpha-1)^2\right] -6  \frac{ (\alpha - 1)^2}{Q(\alpha) } \kappa^{-1}q \right).
\end{align}
The soft terms can be summarized as (where $\xi= q/q_0 > 2$, and eqs.~(\ref{Qalpha}) and (\ref{bc=alphaA}) were used to rewrite the gaugino masses)
\begin{align}
 m_{0,i}^2 &= m_{3/2} \left[ (\sigma_s +1) \left( 1 - \theta_\alpha \xi \right) \right], \notag \\
 A_0 &= m_{3/2} \left[ -1 + (\alpha -1)\left(\alpha -1 + \xi P(\alpha)\right) \right], \notag \\
 B_0 &= A_0 + m_{3/2}, \notag \\
M_1 &= - \frac{g_Y^2}{16\pi^2} m_{3/2}  \left(  11 \left[-1 + (\alpha -1)^2 \right] +   \frac{3}{2} (6-3\theta_Q-\theta_L)  \frac{ (\alpha - 1)^2}{Q(\alpha)^2 }(\sigma_s +1)  \right), \notag \\
 M_2 &= - \frac{g_2^2}{16\pi^2}  m_{3/2} \left(  \left[ -1 + 4 (\alpha - 1)^2 \right] +  (3 + \theta_L + 9 \theta_Q)  \frac{ (\alpha - 1)^2}{Q(\alpha)^2 } (\sigma_s +1) \right), \notag \\
 M_3 &= - \frac{g_3^2}{16\pi^2}  m_{3/2} \left( 3  \left[ 1+ (\alpha-1)^2\right] +6  \frac{ (\alpha - 1)^2}{Q(\alpha)^2 } (\sigma_s +1) \right).
\end{align}
The above soft terms depend on five parameters, namely $m_{3/2}$, $\xi$, $\theta$, $\theta_L$ and $\theta_Q$. 
Following, a parameter scan has been performed~\cite{softsusy} for $m_{3/2} \in [15 \text{ TeV},40 \text{ TeV}]$, $\xi \in [2,10]$, $\theta$, $\theta_L$, $\theta_Q$ $\in ]0,2[$, tan$\beta \in [1,60]$. For $m_{3/2} < 15$ TeV, the gaugino masses are expected to be experimentally excluded, $\xi > 2$ in order to satisfy the constraint (\ref{qlimit}). $A_0$ is negative and monotonically decreasing with $\xi$, such that for $\xi >10$ the trilinear term becomes $A_0 < - 9 m_{3/2}$. 
In principle, the value of tan$\beta$ (which is the ratio between the two Higgs VEVs) is fixed by $B_0$ \cite{Ellis:2003pz,Ellis:2004qe} (as in chapter~\ref{ch:softterms}), however we performed a scan over all possible values of tan$\beta$ instead. Such a high value for $|A_0|$ would contribute to the RGE for the stop mass parameter, so that it runs to a negative value before the electroweak symmetry breaking scale is reached\footnote{
Or the stau becomes tachyonic for a very small region of the parameter space.}.

In this parameter range, no viable electroweak symmetry breaking conditions were found. We conclude that, even though the above idea is very appealing from a theoretical point of view, one can not (at least in this model) use a D-term contribution to the scalar soft masses proportional to the charge of a MSSM field under an extra $U(1)$ factor to solve the problem with tachyonic masses.

\cleardoublepage

  \graphicspath{{\chaptersdir/conclusion/image/}}%
  \chapter{Conclusions}\label{ch:conclusions}

Supersymmetry is one of the most promising theories that extends the Standard Model. 
Not only does it provide a possible solution to the hierarchy problem, 
it naturally contains a dark matter candidate in certain supersymmetric theories like the MSSM, 
and the gauge couplings seem to unify at a certain energy when supersymmetric corrections are taken into account.

If a theory is invariant under local supersymmetry variations, it contains gravity and is called a supergravity theory.
Moreover, since certain supergravity theories appear in the low energy limit of string theory, 
one can address certain open questions in string theory in supergravity.
One such active topic of research in string theory is the endeavor to find de Sitter vacua. 
In terms of a supergravity theory one can address this problem by searching for theories where the minimum of the scalar potential has a positive but small value.

In this work we introduced a model based on a single chiral multiplet $S$, which can be identified with the string dilaton or a compactification modulus.
The scalar component $s$ of this chiral multiplet is invariant under a gauged shift symmetry, with an associated vector multiplet. 
From gauge invariance, the K\"ahler potential should be a function of $s+\bar s$, which is chosen to be $\mathcal K = -\kappa^{-2} p \log(s + \bar s)$.
The most general superpotential is an exponential of the chiral multiplet $W = a \exp(bs)$, while the most general gauge kinetic function has a linear as well as a constant contribution.
As a result, the scalar potential contains a positive D-term contribution including a Fayet-Iliopoulos term, and an F-term contribution which can be negative. 
For positive $b$ there always exists a supersymmetric AdS vacuum, while for negative b
an infinitesimally small and positive value for the minimum of the potential can be found for $p<3$. 
In chapter~\ref{ch:dS}, we quantified the relations between the parameters in the model in order to obtain a tunably small value for the cosmological constant for the two cases $p=1$ and $p=2$,
and showed that the gravitino mass term (which is essentially the supersymmetry breaking scale) is separately tunable.

Further in chapter~\ref{ch:dS}, the case $p=2$ is studied in great detail. 
While for $p=1$ the dS vacuum is stable, for $p=2$ the scalar potential has another minimum at the runaway direction $\text{Re}s\rightarrow \infty$, 
in principle allowing for tunneling to this other minimum and rendering our desired vacuum metastable.
It was shown that the tunneling rate is very low and exceeds the lifetime of the universe.
Following, it was shown that in the limit where $b \rightarrow 0$, the scalar potential corresponds with the one derived in~\cite{Antoniadis:2008uk} from D-branes in non-critical strings.
While $b=0$ results in a supersymmetry breaking vacuum in AdS space, a non-zero $b$ allows for a positive cosmological constant. 
It is therefore a very interesting (and open) question how a non-zero $b$ can be realized in string theory.
Finally, the presence of a shift symmetry indicates that the theory can in principle be reformulated in terms of a linear multiplet. 
In the literature however, this duality was never performed in the presence of a field dependent superpotential. 
We showed that the above model can be reformulated in terms of a (unconstrained) vector multiplet, which indeed has the degrees of freedom corresponding to a linear multiplet.

Quantum consistency can impose stringent constraints on quantum field theories in terms of anomaly cancellation conditions.
In chapter~\ref{ch:anomalies} we extend the work by Elvang, Freedman and K\"ors~\cite{F2}
on quantum anomalies in supergravity theories with Fayet-Iliopoulos terms and we show that their results can be interpreted from a field-theoretic point of view.
Moreover, we show that in the case $p=2$ the anomaly cancellation conditions are inconsistent with a TeV gravitino mass term, which excludes this case since it does not lead to a viable low energy phenomenology.
Finally, we demonstrated the relation between quantum anomalies and the gaugino mass terms that are generated at one loop.
We showed that two contributions should be included in the calculation of the gaugino masses. Firstly, there is a contribution due to anomaly mediation.
Secondly, there is a contribution due to the anomaly cancellation conditions, since anomaly cancellation can require the presence of a Green-Schwarz counter term which originates from a linear contribution to the gauge kinetic function.
This contribution however, also results in a contribution to the gaugino mass terms.
Finally, we showed that by an appropriate K\"ahler transformation, the latter contribution to the gaugino mass terms can be reformulated in terms of an anomaly mediated contribution.

An interesting question arises whether this model leads to a distinguishable low energy spectrum for the sparticles in the theory.
In chapter~\ref{ch:softterms} the case with $p=1$ is 
used as a hidden sector where supersymmetry breaking occurs. 
This supersymmetry breaking is communicated to the visible sector (MSSM) via gravity mediation. 
However, it is shown that the resulting scalar soft supersymmetry breaking terms are tachyonic.
This can be cured by the introduction of an extra Pol\'onyi-like hidden sector field,
or by allowing for non-canonical kinetic terms of the Standard Model fields, while maintaining the desirable features of the model.
This however introduces an extra parameter $\gamma$ (or $\nu$ in the second case), which turns out to be heavily constrained:
$\gamma$ should be in the range $[1.1;1.707]$, where  the  lower
bound is to prevent a tachyonic stop squark mass, and the upper bound follows from the tunability of the scalar potential.
Since  the  soft  SUSY-breaking  parameter $B_0$ is  related  to  the  trilinear  coupling  by $B_0 = A_0 - m_{3/2}$,
the ratio between the two Higgs VEVs $\beta$ is not an independent parameter and the model turns out to be very predictive.
The low energy spectrum of the theory consists of (very) light neutralinos, charginos and gluinos, where the experimental bounds on the
(mostly  bino-like)  LSP,  the  lightest  chargino  and  the  gluino  mass  force  the  gravitino mass  to  be  above  15  TeV.  This  in  turn  implies  that  the  squarks  are  very  heavy,  with
the exception of the stop squark which can be as light as 2 TeV when the parameter approaches its lowest limit  $\gamma \rightarrow 1.1$.
It follows that the resulting spectrum can be distinguished from other models of supersymmetry breaking and mediation such as mSUGRA and mAMSB.

In chapter~\ref{ch:chargedMSSM} it is shown that the shift symmetry can be identified with known global symmetries of the MSSM.
We analyzed the phenomenological implications in great detail for the particular case where the global symmetry is $B-L$, or $3B-L$
which contains the known matter parity of the MSSM as a subgroup.  
The latter combination has also the advantage of forbidding all dimension-four and dimension-five operators violating
baryon or lepton number in the MSSM. 
We showed that the phenomenology is similar to the one obtained in chapter~\ref{ch:softterms}, where the MSSM fields are inert under the shift symmetry, 
with the exception of the stop mass which can become lighter to about $1.5$ TeV (compared to 2 TeV).
Finally, we  showed  that  the  problem  of  tachyonic scalar masses can not be solved from the presence of the (positive) D-term contributions
to the scalar potential, proportional to the charge of the MSSM superfields under the shift symmetry, because of incompatibility with the existence of a viable electroweak vacuum.

\cleardoublepage

\appendix

 \chapter{Masses of the scalar, the fermion and the gauge field} \label{Appendix:masses}

In this appendix we calculate the masses of the fermions (section~\ref{sec:appendix_fermions}) for the cases $p=1$, $p=2$, and for the model involving an extra Pol\'onyi-like field in section~\ref{sec:polonyi}.
We demonstrate how the $U(1)_R$ gauge boson acquires a mass by the Stueckelberg mechanism \cite{Stueckelberg1,Stueckelberg2,Stueckelberg3,Stueckelberg4,Stueckelberg5}  for general $p$ in section~\ref{sec:appendix_gauge}.
Finally, we show how the scalar mass is calculated in the case $p=2$ in section~\ref{sec:appendix_scalars}.

\section{Calculation of the fermion mass matrices in the various models} \label{sec:appendix_fermions}

For convenience of the reader, we here repeat the fermion mass matrices and its corrections due to the super-BEH mechanism, which were discussed in sections~\ref{sec:sugra_ingredients} and~\ref{sec:spontaneous susy breaking}. 
The fermion mass matrix is given by
\begin{align}  m =   \begin{pmatrix}  m_{\alpha \beta} + m_{\alpha \beta}^{(\nu)}  & m_{\alpha B} + m_{\alpha B}^{(\nu)}    \\  m_{A \beta} + m_{A \beta}^{(\nu)}      &     m_{AB} + m_{AB}^{(\nu)}       \end{pmatrix}, \tag{\ref{theory:fermion_matrices_BEH}} \end{align}
where $m_{\alpha \beta}$, $m_{\alpha B}$ and $m_{AB}$ are given by eqs.~(\ref{theory:chiralmass}-\ref{theory:gauginomass}).
The corrections to the fermion mass terms due to the elimination of the Goldstino $P_L \nu$ are given by
\begin{align} m_{\alpha \beta}^{(\nu)} &= - \frac{4 \kappa^2}{3 m_{3/2}} (\delta_s \chi_\alpha)(\delta_s \chi_\beta), \notag \\  m_{\alpha A}^{(\nu)} &=- \frac{4 \kappa^2}{3 m_{3/2}} (\delta_s \chi_\alpha)(\delta_s P_R \lambda_A ),\notag \\ m_{AB}^{(\nu)} &= - \frac{4 \kappa^2}{3 m_{3/2}} (\delta_s P_R \lambda_A)(\delta_s P_R \lambda_B ),
 \tag{\ref{theory:fermion_matrices_BEH2}} 
\end{align}
and the fermion shifts are given by
\begin{align} \delta_s \chi_\alpha &= - \frac{1}{\sqrt 2} e^{\kappa^2 \mathcal K /2} \nabla_\alpha W, \notag \\ \delta_s P_R \lambda_A &= -\frac{i}{2} \mathcal P_A. 
\tag{\ref{theory:fermion_shifts}}
\end{align}

The kinetic involving the fermions in the model however, are (in general) not canonically normalized. Instead, they are given by
\begin{align}
 \mathcal L_{\text{kin}} / e = -\frac{ g_{\alpha \bar \beta}}{2}  \chi^{\alpha} \cancel D^{(0)} \bar \chi^{\bar \beta} - \frac{\text{Re}(f_{AB})}{2} \bar \lambda^A  \cancel D^{(0)} \lambda^B, \label{theory:fermion_kin}
\end{align}
where $D^{(0)}$ is the fermion covariant derivative.
For diagonal $g_{\alpha \bar \alpha}$ and diagonal gauge kinetic functions $f_A$,
canonically normalized kinetic terms for the fermions can be obtained by making a field redefinition $\chi'^\alpha = \sqrt{g_{\alpha \bar \alpha}} \chi^\alpha$ 
and $\lambda'^A =  \sqrt{\text{Re}(f_A)} \lambda^A$ (note that there is no summation over indices here).
As a result, the components in the fermion mass matrix above should be rescaled accordingly. 
The resulting mass matrix then has one zero eigenvalue corresponding to the Goldstino, while the other eigenvalues of $\text{Tr}(m^\dagger m)$ correspond to the masses of the physical fermions.

\subsection{Fermion masses for $p=1$} \label{Appendix:Fermionsp1}

The fermion mass matrix for case $p=1$ in section~\ref{sec:p=1} (as well as the model in the model in section~\ref{sec:non-can_Kahler})
for the fermionic superpartner of $s$ and the gaugino corresponding to the shift symmetry (\ref{shift}) is given by
\begin{align}  m = \kappa^{-1} \left( \begin{array}{cc}   \left( \frac{\alpha}{b}\right)^2 \frac{ a e^{\alpha /2} \left(\alpha ^2+4 \alpha -2\right)}{3 \left(\alpha/b \right)^{5/2}} & - \left( \frac{\alpha}{b}\right) \frac{i \sqrt{2} b^2 c \left(\alpha ^2-2 \alpha -2\right)}{3 \alpha ^2} \\  - \left( \frac{\alpha}{b}\right) \frac{i \sqrt{2} b^2 c \left(\alpha ^2-2 \alpha -2\right)}{3 \alpha ^2} & \frac{c^2 e^{-\frac{\alpha }{2}} (\alpha -1)^2}{3 a \left(\alpha/b \right)^{3/2}} \\ \end{array} \right), \end{align}
where the factors $\left( \frac{\alpha}{b}\right)$ have been taken into account due to non-canonical kinetic terms for the chiral fermions. The gaugino already has canonical kinetic terms since $f(s)=1$. The hidden sector fermions do not mix with the fermions of the MSSM. Also, the determinant of $m$ is proportional to $(2 + 8 \alpha - 3 \alpha^2 - 2 \alpha^3 + \alpha^4)$, which indeed has a root at $\alpha \approx -0.23315$. 
The mass squared of the physical fermion is then given by the non-zero eigenvalue of $m^\dagger m$. 
Since one of the eigenvalues (the Goldstino)of the $2 \times 2$ matrix vanishes, the mass (squared) of the physical fermion is given by its trace
\begin{align}  m_f^2 &=  \Tr \left[ m^\dagger m \right] = m_{3/2}^2 f_\chi, \end{align}
where 
\begin{align}f_\chi&=  \frac{\left(e^{2 \alpha } \alpha ^2 \left(\alpha ^2+4 \alpha -2\right)^2+(\alpha -1)^4 A(\alpha)^2+4 e^{\alpha } \alpha  \left(\alpha ^2-2 \alpha -2\right)^2 A(\alpha)\right)}{9 \alpha ^2 e^{2 \alpha } } \notag \\ &\approx 3.807, \end{align}
and we have used the relations between the parameters and the numerical values for $\alpha$ and $A(\alpha)$ in eqs.~(\ref{model:p1_tunability}).

\subsection{Fermion masses for $p=2$} \label{Appendix:Fermionsp2}

For $p=2$, the fermion mass terms are given by
\begin{align}
m_{\alpha \beta} &=  \kappa^{-1} \frac{a e^{bs}}{(s + \bar s)^3} \left( b^2 (s + \bar s)^2 - 2 b (s + \bar s) + 2 \right), \notag  \\
 m_{\alpha A} &=  \kappa^{-1} i \sqrt{2} c \left( \frac{3}{(s + \bar s)^2} - \frac{b}{2} \frac{1}{(s + \bar s)} \right) = m_{A \alpha}, \notag \\
m_{AB} &= \kappa^{-1} \frac{-a}{4} e^{b \bar s} (s + \bar s) \left( b - \frac{2}{s + \bar s} \right) ,\end{align}
 where $(s + \bar s)$ should be evaluated in the minimum of the potential, i.e. eqs.~(\ref{model:tunability_p2}) hold.  
The corrections due to the BEH-effect are given by
\begin{align}
m_{\alpha \beta}^{(\nu)} &= \kappa^{-1} \left( \frac{2}{s + \bar s} - b \right)^2 \frac{-2}{3} \frac{a}{s + \bar s} e^{\frac{b}{2} (s + \bar s)}, \notag \\
m_{\alpha A}^{(\nu)} &= \kappa^{-1} \left( \frac{2}{s + \bar s} - b \right)^2  \frac{-\sqrt 2}{3} ic, \notag \\
m_{AB}^{(\nu)} &= \kappa^{-1} \left( \frac{2}{s + \bar s} - b \right)^2 \frac{1}{3} \frac{c^2}{a} (s + \bar s) e^{-\frac{b}{2} (s + \bar s)},
\end{align}
Next, due to a field redefinition (see the discussion below eq.~(\ref{theory:fermion_kin})) in order to obtain canonically normalized kinetic terms, the 
entries in the fermion mass matrix corresponding to $m_{\alpha \bar \beta}$, $m_{\alpha A}$ and $m_{AB}$ should be multiplied 
by a factor $\alpha^2/2b^2$, $\sqrt{\alpha/2b}$ and $b/\alpha$ respectively.
As a result, the full ($2 \times 2$) mass matrix is $\kappa^{-1}$ times
\begin{align}
\begin{pmatrix} \frac{\alpha^2}{2b^2} \frac{ab^3}{3 \alpha^3} e^{\alpha/2} \left( - 2 + 2 \alpha + \alpha^2 \right) & 
- \sqrt{\frac{\alpha}{2b}}  \frac{ic b^2}{3 \sqrt 2 \alpha^2} \left( -10 - 5 \alpha + 2 \alpha^2  \right)   \\ 
- \sqrt{\frac{\alpha}{2b}} \frac{ic b^2}{3 \sqrt 2 \alpha^2} \left( -10 - 5 \alpha + 2 \alpha^2  \right)     &
 \frac{b}{\alpha}  \frac{e^{-\alpha/2}\left( \alpha - 2  \right)}{12 a \alpha}  \left( -3 a^2 \alpha e^{\alpha} + 4 b c^2 \left( \alpha - 2\right) \right)        \end{pmatrix} 
 \notag 
 \end{align}
By inserting equation (\ref{model:tunability_p2}), the determinant of $m$ is 
\begin{align}
\text{Det}(m) &= -\kappa^{-2} \frac{b^4 c^2 \left( -40 - 240 \alpha - 102 \alpha^2 + 94 \alpha^3 + 15 \alpha^4 - 15\alpha^5 + 2 \alpha^6 \right)  }{6 \alpha^4 \left( -2 -4 \alpha + \alpha^2 \right)} \notag \\  &= 0
\end{align}
The determinant has a root at the earlier defined value of $\alpha = -0.183268... $, so that the mass matrix has indeed a zero mode, corresponding to the reduction of the rank of the full mass matrix $m$, because the Goldstino disappears.
The (squared) mass of the fermion is thus given by the trace of the mass matrix squared 
\begin{align}
 m_f^2 &= \text{Tr}(m^\dagger m ) \notag \\ &= \kappa^{-2} a^2 b^2 \frac{e^\alpha \left( 116 + 68 \alpha - 15 \alpha^2  - 4 \alpha^3 + 8 \alpha^4\right)}{144 \alpha^2}.
\end{align}
By using the numerical value of $\alpha$ and eq.~(\ref{model:gravitinomassp2}), the fermion mass can be written in terms of $m_{3/2}$
\begin{align}
m_f  &\approx 0.846 \ m_{3/2} .
\end{align}

\subsection{Fermion masses for the model including an extra Pol\'onyi-like field} \label{Appendix:Fermions_polonyi}
We now calculate the fermion masses for the model with the extra hidden sector field $z$ in section \ref{sec:polonyi}. 
This model contains one extra hidden sector fermion. Its mass matrix is given by 
\begin{align} \notag \hspace{-1.5cm} 
m =
\left( \begin{array}{ccc}  
m_{ss} + m_{ss}^{(\nu)}
& m_{st} + m_{st}^{(\nu)} 
& m_{sR} + m_{sR}^{(\nu)} \\  
 m_{st} + m_{st}^{(\nu)}
& m_{tt} + m_{tt}^{(\nu)}
& m_{tR} + m_{tR}^{(\nu)} \\ 
m_{sR} + m_{sR}^{(\nu)}
& m_{tR} + m_{tR}^{(\nu)}
& m_{RR} + m_{RR}^{(\nu)} \end{array} \right),
\end{align}
where
\begin{align}
 m_{ss} + m_{ss}^{(\nu)} &= \kappa^{-1} \left(\frac{\alpha}{b}\right)^2 \frac{ a e^{\frac{1}{2} \left(t^2+\alpha \right)} \left(-2+4 \alpha +\alpha ^2\right) (1+t \gamma )}{3 \left(\frac{\alpha }{b}\right)^{5/2}},  \notag \\
 m_{st} + m_{st}^{(\nu)} &= \kappa^{-1} \left(\frac{\alpha}{b}\right) \frac{a e^{\frac{1}{2} \left(t^2+\alpha \right)} (-1+\alpha ) \left(t+\gamma +t^2 \gamma \right)}{3 \left(\frac{\alpha }{b}\right)^{3/2}}, \notag \\
 m_{sR} + m_{sR}^{(\nu)} &= - \kappa^{-1} \left(\frac{\alpha}{b}\right) \frac{i \sqrt{2} b^2 c \left(-2-2 \alpha +\alpha ^2\right)}{3 \alpha ^2}, \notag \\
 m_{tt} + m_{tt}^{(\nu)} &= \kappa^{-1} \frac{a e^{\frac{1}{2} \left(t^2+\alpha \right)} \left(2 t \gamma +2 t^3 \gamma -2 \gamma ^2+t^4 \gamma ^2+t^2 \left(1+2 \gamma ^2\right)\right)}{3 \sqrt{\frac{\alpha }{b}} (1+t \gamma )},  \notag \\
 m_{tR} + m_{tR}^{(\nu)} &= -\kappa^{-1} \frac{i \sqrt{2} b c (-1+\alpha ) \left(t+\gamma +t^2 \gamma \right)}{3\alpha ( 1+t  \gamma )}, \notag \\
 m_{RR} + m_{RR}^{(\nu)} &= \kappa^{-1} \frac{b^2 c^2 \sqrt{\frac{\alpha}{b}} e^{\frac{1}{2} \left(-t^2-\alpha \right)} \left(1-\frac{1}{\alpha }\right)^2}{3 a(1+ t \gamma )}  .
 \end{align}
It has been checked that the determinant of this matrix vanishes for $\alpha$ and $t$ satisfying eqs.~(\ref{ca}) and (\ref{model:polonyi_tunability}). 
The masses of the physical fermions are the two non-zero eigenvalues of this matrix. The result however is quite tedious and we only state the numerical values for $\gamma=1$:
\begin{align}   m_{\chi_1} & \approx  2.57 \ m_{3/2}, \notag \\  m_{\chi_2} & \approx  0.12 \ m_{3/2}. \end{align}

\section{Mass of the $U(1)_R$ gauge boson} \label{sec:appendix_gauge}

The kinetic term $\mathcal L_s $ of the scalar field $s$ can be written as
\begin{align} \mathcal L_s /e = - g_{s \bar s} \hat \partial_\mu s \hat \partial^\mu \bar s, \label{theory:scalar_kin} \end{align}
where the covariant derivative is defined as \begin{align} \hat \partial_\mu s &= \partial_\mu s - k^s A_\mu  \notag \\ &= \partial_\mu s + ic A_\mu, \end{align}
where $k^s$ is the Killing vector associated with the shift symmetry given by $k^s = -ic$ (see eq.~(\ref{shift})). Equation~(\ref{theory:scalar_kin}) can then be written as 
\begin{align} \mathcal L_s/e &=  \frac{-p \kappa^{-2} }{(s + \bar s)^2} \left( \partial^\mu s \partial_\mu \bar s - ic A^\mu \partial_\mu(s -\bar s)+ c^2 A^\mu A_\mu \right) .  \end{align}
 The local shift symmetry allows us to gauge fix $s-\bar s = 0$, which results in a mass term $\mathcal L_m /e = -\frac{1}{2} m_{A_\mu}^2 A_\mu A^\mu$ for the gauge boson.
This mass term however, needs a rescaling due to a non-canonical kinetic term\footnote{
This rescaling was absent in an earlier version of this thesis. We thank A.~Chatrabhuti for pointing this out.}
\begin{align}
 \mathcal L_{\text{kin}} / e = - \frac{\text{Re}f(s)}{4} F_{\mu \nu} F^{\mu \nu}.
\end{align}
The result is 
\begin{align}  m_{A_\mu} = \sqrt{\frac{2p}{\text{Re}f(s)}} \frac{\kappa^{-1} c }{\langle s + \bar s \rangle} , \label{appendix:gaugebosonmass} \end{align}
where $\text{Re}f(s) = 1$ for $p=1$, and $\text{Re}f(s)= \frac{s + \bar s}{2} $ for $p=2$.
  
\section{Scalar masses}  \label{sec:appendix_scalars}

In this section we demonstrate how the mass of the scalar fields is calculated for $p=2$. The masses of the scalar fields for the other models is calculated analogously. 

We isolate the quadratic contribution of the scalar potential eq.~(\ref{scalarpotp2}), which is repeated here for convenience
\begin{equation} \mathcal V =  \kappa^{-4} a^2 e^{b(s+\bar s)} \left(  \frac{b^2}{2}  - \frac{2b}{s+\bar s}  - \frac{1}{(s+ \bar s)^2 }    \right)  + \kappa^{-4} \frac{c^2}{s + \bar s + 2d} \left( \frac{2}{s+\bar s} - b \right)^2. \notag \end{equation}
The quadratic term $\mathcal V_{quad} = \frac{1}{2} \mathcal V''(\alpha/b) \left( s + \bar s - \frac{\alpha}{b} \right)^2$ in the series expansion of $\mathcal V$ 
around its minimum at $\langle s + \bar s \rangle = \frac{\alpha}{b}$ is given by
(simplified by using equation (\ref{model:tunability_p2}))
\begin{align} \mathcal V_{quad} = 
 \kappa^{-4} \frac{\left(48+192 \alpha-128 \alpha^2-8 \alpha^3+24 \alpha^4-8 \alpha^5+\alpha^6\right) b^5 c^2 (s + \bar s - \frac{\alpha}{b})^2}{2 \alpha^5 \left(2+4 \alpha-\alpha^2\right)} . \notag \end{align}
Taking into account the non-canonical form of the kinetic terms\footnote{ 
The kinetic term of $\text{Re}(s)$ is given by 
\begin{align} \mathcal L_{\text{kin}}  = \frac{-\kappa^{-2}}{2\text{Re}(s)^2} \partial_\mu \text{Re}(s) \partial^\mu \text{Re}(s).  \end{align} 
Expanding around the minimum $\phi = \text{Re}(s) - S_0$, with $S_0 = \frac{\alpha \kappa}{2b}$, gives 
\begin{align} \mathcal L = \frac{-1}{2S_0^2} \left( 1 - 2 \frac{\phi}{S_0} +\dots \right) \partial^\mu \phi \partial_\mu \phi . \end{align} 
We should thus rescale the mass with a factor $S_0$. }, the mass of the dilaton $\text{Re}(s) = \frac{s + \bar s}{2}$ is then given by
\begin{align} m_s^2 =    \frac{   -\kappa^{-2} b^4 c^2 \left(48+192 \alpha-128 \alpha^2-8 \alpha^3+24 \alpha^4-8 \alpha^5+\alpha^6\right) }{ \alpha^4 \left(-2-4 \alpha+\alpha^2\right)}.   \end{align}
For the earlier calculated numerical value of $\alpha$ (see eq.~(\ref{model:tunability_p2})) this expression is indeed positive: $m_s^2 > 0$.

\cleardoublepage

\chapter{Linear-Chiral duality} \label{ch:appendix_linear_chiral}

In this appendix we present the details of the linear-chiral (or tensor-scalar) duality which lead to the result presented in section~\ref{sec:linearchiral}.
This duality is presented in global supersymmetry in section~\ref{appendix:linearchiral_SUSY}.
Since a superspace formalism is more convenient in this context, the conventions of Wess and Bagger~\cite{wessbagger} were used in this section.
Following, a dual version of the theory in section~\ref{sec:p=2} in the K\"ahler frame with an exponential superpotential is presented in section~\ref{sugraduality}.
An important identity is proven in section~\ref{Appendix:Proof}.
This appendix is based on work with I.~Antoniadis~\cite{RK1}.

\section{Tensor-scalar duality in global supersymmetry} \label{appendix:linearchiral_SUSY}
In this section we will use the superfield notation and conventions of~\cite{wessbagger}.
It is important to note that in this notation, a linear multiplet $L$ is defined from a general superfield by imposing the relation
\begin{align} D^2 L = \bar D^2 L = 0, \label{linear:constraintL} \end{align}  
where the differential operators $\bar D^2$ and $D^2$ are given by
\begin{align}
 D_\alpha &= \frac{ \partial}{\partial \theta^\alpha} + i \sigma^\mu_{\alpha \dot \alpha} \bar \theta^{\dot \alpha} \partial_\mu, \notag \\
 \bar D_{\dot \alpha} &= - \frac{\partial}{\partial \bar \theta^{\dot \alpha}} - i \sigma^\mu_{\alpha \dot \alpha} \partial_\mu. \label{chiralprojection}
\end{align}
Here, $\theta^\alpha \ (\bar \theta^{\dot \alpha} )$ are the fermionic coordinates of the (anti-)chiral superspace. 

\subsubsection{Easy warm-up: No superpotential} 

As a warm-up, we will show how one can describe a (globally) supersymmetric theory, 
given by a single chiral multiplet $S$ with vanishing superpotential and a K\"ahler potential $\mathcal K(S + \bar S)$ given by eq.~(\ref{model:kahlerpotential}) in terms of a linear multiplet.
The Lagrangian of this theory is given by
\begin{align}  \mathcal L = \int d^2 \theta d^2 \bar \theta \mathcal K (S + \bar S). \end{align}
To obtain the dual version of this theory, we consider the Lagrangian
\begin{align} \mathcal L = \int d^2 \theta d^2 \bar \theta \left( \mathcal F(L) - p L(S + \bar S) \right), \label{L_FO_1} \end{align}
where $L$ is a priori an unconstrained real vector superfield. For purposes that become clear in an instant, we take $\mathcal F(L)$ to be given by $\mathcal F(L) =p \log(L)$. 
The equations of motion for the (anti-)chiral superfields $S$($\bar S$) give $D^2 L = \bar D^2 L = 0$, which is exactly the condition (\ref{linear:constraintL}) for $L$ to be a linear multiplet.  The second term in (\ref{L_FO_1}) is thus a generalised Lagrange multiplier that ensures the linearity of $L$. The resulting theory in terms of the linear multiplet is given by
\begin{align} \int d^2 \theta d^2 \bar \theta \mathcal F(L). \end{align}
On the other hand, the equations of motion of (the now unconstrained real superfield) $L$ give
\begin{align} \mathcal F'(L) = p \left( S + \bar S \right), \end{align} 
such that  
\begin{align} S + \bar S = \frac{1}{L} \label{SdualL}. \end{align}
Some basic algebra then shows that
\begin{align} \mathcal F(L) =p \log \left( \frac{1}{S + \bar S} \right) = - p \log \left(S + \bar S \right). \end{align}
The chiral theory is then given by
\begin{align}\int d^2 \theta d^2 \bar \theta \mathcal K(S + \bar S),\end{align}
with $\mathcal K(S + \bar S)$ given in (\ref{model:kahlerpotential}).

Since the natural partner of the graviton in the massless sector of superstrings is the scalar dilaton and an antisymmetric tensor, the above exercise should be repeated in supergravity (local supersymmetry). 
Below, we first perform the duality in the presence of a superpotential given by eq.~(\ref{model:superpotential}) in global supersymmetry, and consequently in supergravity.

\subsubsection{With a superpotential} \label{rigidsusy}

We are now ready to perform the above duality in the presence of a non-trivial (exponential) superpotential, given by eq.~(\ref{model:superpotential}).
This superpotential has the following property: 
under the shift symmetry~(\ref{shift}) the superpotential transforms as $W(S) \longrightarrow W(S) e^{-i\alpha b}$,
where the gauge parameter is now denoted as $\alpha$ to distinguish it from the superspace coordinates $\theta$.
This transformation can be compensated by the R-transformation
$\theta \longrightarrow \theta e^{-\frac{i}{2}\alpha b} $ such that $\int d^2 \theta W(S)$ is invariant under the combined shift and R-transformations. 
Note that one can not gauge the shift symmetry in this model, since it is entangled with an R-symmetry, which can not be gauged in rigid supersymmetry, in contrast with supergravity.

We are now interested in the dual description of the theory
\begin{align} \mathcal L= \int d^4 \theta \mathcal K(S +\bar S) + \int d^2 \theta W(S) + \int d^2 \bar \theta \bar W(\bar S), \label{L_chiral} \end{align}
where the  K\"ahler potential is given by eq.~(\ref{model:kahlerpotential}) and the superpotential given by (\ref{model:exp_superpot}).
 To find the dual theory, we start from the Lagrangian
\begin{align} \mathcal L & = \int d^4 \theta \mathcal F(L)  -p \int d^4 \theta L (S + \bar S) + \int d^2 \theta W(S) + \int d^2 \bar \theta \bar W(\bar S) \label{FO_shift}, \end{align}
where as before, $S$ and $\bar S$ are chiral and anti-chiral superfields, respectively, and $L$ is an a priori unconstrained real superfield. 
As we will see below, it turns out that in the dual theory, $L$ will still be a real superfield off-shell. 
However, it will contain the same number of propagating degrees of freedom as those of a linear multiplet. 
As in the previous example, we take $\mathcal F(L) = p \log(L)$. The equations of motion for $L$ give, as before
\begin{align} S + \bar S = \frac{1}{L} \label{SdualL2}. \end{align}
By substituting this relation into the Lagrangian (\ref{FO_shift}) one retrieves (\ref{L_chiral}).

We now derive the equations of motion for the chiral superfield. This results in modified linearity conditions
\begin{align} p \bar D^2 L &= -4 W'(S), \label{LW} \\ p D^2 L &= -4 \bar W'( \bar S), \label{LbarW} \end{align}
and thus $L$ is still an unconstrained real multiplet. One can use these equations to write $S$ and $\bar S$ in terms of $L$
\begin{align} S = \frac{1}{b} \log \left( \frac{-p}{4ab} \bar D^2 L  \right), \label{SL}  \\ \bar S = \frac{1}{b} \log \left( \frac{-p}{4ab} D^2 L \right) \label{SbarL}. \end{align}
Note that one can rewrite the Lagrangian (\ref{FO_shift}) as
\begin{align} \mathcal L =&  \int d^4 \theta  \mathcal F(L) + \int d^2 \theta \left( \frac{p}{4} \bar D^2 L S + W(S)  \right) 
\notag \\ &+ \int d^2 \bar \theta \left( \frac{p}{4} D^2 L \bar S + \bar W (\bar S )  \right) . \label{FO_rewritten} \end{align}
By substituting (\ref{SL}) and (\ref{SbarL}) into (\ref{FO_rewritten}) one obtains the dual theory in terms of the superfield $L$
\begin{align} \mathcal L_{L} =&  \int d^4 \theta  \mathcal F(L)   \notag \\ &+ \int d^2 \theta \left( \left( \frac{p}{4b} \bar D^2 L\right)  \log \left( \frac{-p}{4ab} \bar D^2 L \right) - \frac{p}{4b} \bar D^2 L   \right) \notag \\ &+ \int d^2 \bar \theta \left( \left( \frac{p}{4b} D^2 L\right)  \log\left( \frac{-p}{4ab} D^2 L \right) - \frac{p}{4b}  D^2  L   \right) . \label{L_L1} \end{align}
Note that $L$ is a real vector superfield instead of linear superfield. 
However, as we show below, the vector multiplet $L$ contains auxiliary fields that can be eliminated by their equations of motion, 
and the physical content of the theory can still be described in terms of only a real scalar $l$, a Majorana fermion $\chi$ and a field strength vector field $v_\mu$. 
In fact, a more natural description of this theory will be given by a real scalar $l$, a Majorana fermion $\chi$ and a  pseudoscalar $w$, 
which is the phase of the $\theta^2$-component of $L$. 

In the conventions of \cite{wessbagger}, the general expression of the real superfield $L$ is defined as follows
\begin{align} L =& l +i \theta \chi - i \bar \theta \bar \chi + \theta^2 Z + \bar \theta^2 \bar Z - \theta \sigma^\mu \bar \theta v_\mu \notag \\ &+ i \theta^2 \bar \theta (\bar \lambda - \frac{i}{2}\bar \sigma^\mu \partial_\mu \chi) -  i \bar \theta^2 \theta ( \lambda - \frac{i}{2} \sigma^\mu \partial_\mu \bar \chi) + \frac{1}{2} \theta^2 \bar \theta^2 (D - \frac{1}{2} \Box l) . \end{align}
Rewriting $Z = \rho e^{iw}$ and $\bar Z = \rho e^{-iw}$, it follows that the Lagrangian (\ref{L_L1}) is given in components by
\begin{align} \mathcal L =& \frac{\mathcal F'(l)}{2} \left( D - \frac{1}{2} \Box l \right) \notag \\  & + \frac{\mathcal F''(l)}{2} \left( -\chi  \left( \lambda - \frac{i}{2} \sigma^\mu \partial_\mu \bar \chi \right) - \bar \chi \left( \bar \lambda - \frac{i}{2} \bar \sigma^\mu \partial_\mu \chi \right)  + 2 \rho^2 - \frac{1}{2} v^\mu v_\mu \right) \notag \\  & +\frac{1}{4} \mathcal F'''(l) \left( \chi^2 \rho e^{-iw} + \bar \chi^2 \rho e^{iw} + \chi \sigma^\mu \bar \chi v_\mu \right) + \frac{\mathcal F''''(l)}{16} \chi^2 \bar \chi^2  \notag \\ &  + \frac{p}{b} D \log \left( \frac{p}{ab} \rho \right) + \frac{p}{b} w \partial^\mu v_\mu  -  \frac{p}{4b} \frac{\lambda^2}{\rho}e^{iw} -  \frac{p}{4b} \frac{\bar \lambda^2}{\rho}e^{-iw}.  \label{LF_L} \end{align}
Note that the fields $D,\lambda, v_\mu$ and $\rho$ appear without kinetic terms in the Lagrangian. They are therefore auxiliary fields that can be eliminated by their equations of motion. These are
\begin{align} \rho &= \frac{ab}{p} e^{-\frac{b}{2p}(\mathcal F '(l))}, \notag \\ p \frac{\lambda}{b \rho} e^{iw} 
&= - \mathcal F''(l) \chi, \notag \\ \frac{\mathcal F''}{2} v_\mu &=  \frac{\mathcal F'''}{4} \chi \sigma^\mu \bar \chi - \frac{p}{b}\partial^\mu w,  \notag \\ 
-\frac{p}{b} D &=2 \mathcal F'' \rho^2 + \frac{\rho}{4} \mathcal F''' \left( \chi^2 e^{-iw} + \bar \chi^2 e^{iw}\right) + \frac{p}{4b} (\frac{\lambda^2}{\rho}e^{iw} + \frac{\bar \lambda^2}{\rho}e^{-iw}). \end{align}
Indeed, these equations of motion can be used to eliminate these fields from the Lagrangian by substituting them back into eq.~(\ref{LF_L}) 
\begin{align} \mathcal L =&- \frac{1}{4} \mathcal F'(l) \Box l +  \frac{i}{4} \mathcal F''(l) \left( \chi \sigma^\mu \partial \bar \chi + \bar \chi \bar \sigma^\mu \partial_\mu \chi \right) + \frac{a^2 b^2}{p^2} \mathcal F''(l) e^{\frac{b}{p} \mathcal F'(l)} \notag \\ &+ \frac{1}{\mathcal F''(l)} \left( \frac{- \mathcal F'''(l)}{2} \chi \sigma^\mu \bar \chi \partial_\mu w +\frac{p^2}{b^2} \partial^\mu w\partial_\mu w \right) + \left( \frac{\mathcal F''''(l)}{16} - \frac{\mathcal F'''(l)^2}{32 \mathcal F''(l)}\right) \chi^2 \bar \chi^2 \notag \\ &+ \frac{ab}{4p} e^{\frac{b}{p} \mathcal F'(l)/2} \left( \chi^2 e^{-iw} + \bar \chi^2 e^{iw}\right) \left( \mathcal F'''(l) - \frac{b}{p} \mathcal F''(l)^2 \right).  \end{align}
It follows that the Lagrangian can be written only in terms of the physical propagating fields $l, \chi$ and $w$. Note that it is also possible to write the Lagrangian in terms of $l, \chi$ and $v_\mu$. The above calculation proves that the field strength $v_\mu$ contains only one degree of freedom.

Note that the resulting scalar potential for the field $l$ (last term of the first line of the above expression) is:
\begin{align} \mathcal V &=  -\frac{a^2 b^2}{p^2} \mathcal F''(l) e^{\frac{b}{p} \mathcal F'(l)} = \frac{a^2 b^2}{p} l^{-2} e^{\frac{b}{l} }. \end{align}
As a check, we can substitute ({\ref{SdualL2}}) into the above equation to find
\begin{align} \mathcal V&=\frac{a^2 b^2}{p} (s + \bar s)^2 e^{b (s + \bar s)}, \label{Vchiral} \end{align}
which corresponds to the scalar potential that can be directly obtained in the chiral formulation from equation (\ref{L_chiral}).

\section{Tensor-scalar duality in supergravity} \label{sugraduality}

We now calculate the same duality in supergravity by using the chiral compensator formalism \cite{TheBook}. 
In order to simplify the notation, we put $\kappa=1$ in this section.
We start with the Lagrangian
\begin{align} \mathcal L =& \left[ - \frac{1}{2} \left( S_0 e^{-gV_R} \bar S_0 \right)^3 L^{-2} \right]_D + \left[S_0^3 W(S)\right]_F + \left[f \mathcal W^2 \right]_F + \text{h.c.} 
\notag \\ &-   \left[L (S + \bar S + c V_R) \right]_D, \label{SUGRALFO} \end{align}
where, as before, $S$($\bar S$) is an (anti-)chiral multiplet with zero Weyl weight, 
$L$ is a (unconstrained) real multiplet with Weyl weight 2 and $V_R$ is a vector multiplet of zero Weyl weight associated with a gauged R-symmetry
with coupling constant $g$ given by $g = \frac{bc}{3}$. $S_0$ ($\bar S_0$) is the (anti-)chiral compensator superfield with Weyl weight $+1$ ($-1$). 

An F-term $[\ \ ]_F$ is defined on a chiral multiplet $X=(s, P_L \zeta, F)$ with weight 3 
\begin{align} \left[ X \right]_F =  e \  \text{Re} \left( F + \frac{1}{\sqrt 2} \bar \psi_\mu \gamma^\mu P_L \zeta + \frac{1}{2} s \bar \psi_\nu \gamma^{\mu \nu} P_R \psi_\nu \right), \label{F-term} \end{align}
where $e$ is the determinant of the vierbein. From now on, we write $P_{L,R} \zeta = \zeta_{L,R}$. 
A D-term $[\ \ ]_D$ is defined on a real multiplet $C = (C, \chi, Z, v_a, \lambda, D)$ with weight 2 as
\begin{align}  \left[ C \right]_D = & \ e  \left( D - \frac{1}{2} \bar \psi \cdot \gamma i \gamma_* \lambda - \frac{1}{3} C R(\omega) 
+ \frac{1}{6} \left( C \bar \psi_\mu \gamma^{\mu \rho \sigma } - i \bar \chi \gamma^{\rho \sigma} \gamma_* \right) R'_{\rho \sigma}(Q) \right. \notag \\ &\left. + \frac{1}{4} \epsilon^{abcd} \bar \psi_a \gamma_b \psi_c \left( v_d - \frac{1}{2} \bar \psi_d \chi \right) \right), \label{D-term} \end{align}
where $R(\omega)$ and $R'(Q)$ are the graviton and gravitino curvatures.
The superpotential is given by eq.~(\ref{model:exp_superpot}) and $\mathcal W^2 = \mathcal W^\alpha \mathcal W_\alpha$ contains the gauge kinetic terms for $V_R$ 
with gauge kinetic function $f$, which is assumed here to be constant $f = \gamma$.
A linear gauge kinetic function, as is necessary to obtain a tunable cosmological constant (see section~\ref{sec:p=2}), will be discussed below.
 
Note that since the R-symmetry generator $T_R$ does not commute with the supersymmetry generators $Q_\alpha$ by eq.~(\ref{algebra:R-symmetry}), 
repeated here for convenience of the reader
\begin{align}  \left[ T_R , Q_\alpha \right] = -i (\gamma_*)_\alpha^{\ \ \beta} Q_\beta, \notag  \end{align}
the global R-symmetry in the previous section necessarily becomes gauged when one couples this theory to gravity. 
The above Lagrangian thus has a gauged R-symmetry (coupled to the gauged shift symmetry), under which the various fields transform as
\begin{align} S &\longrightarrow S - ic \Lambda , &  \bar S &\longrightarrow \bar S + ic \bar \Lambda, \notag \\ S_0 & \longrightarrow e^{\frac{ibc}{3} \Lambda} S_0,   &    \bar  S_0  &\longrightarrow  S_0 e^{\frac{-ibc}{3} \bar \Lambda}, \notag \\ V_R &\longrightarrow V_R + i \left( \Lambda - \bar \Lambda \right), \notag \\ L & \longrightarrow L, \end{align}
 where the gauge parameter $\Lambda$($\bar \Lambda$) is a (anti-)chiral multiplet.

The equations of motion for $L$ are given by
\begin{align} L = \frac{S_0 e^{-gV_R} \bar S_0  }{(S + \bar S)^{1/3} } \label{SUGRALS}. \end{align}
Below, we fix the conformal gauge by choosing the lowest components $s_0$ and $\bar s_0$ of $S_0$ and $\bar S_0$ as $s_0 \bar s_0 = l^{\frac{2}{3}}$ (see equation (\ref{conformalgauge})), so that the lowest components of equation (\ref{SUGRALS}) reads
\begin{align} l = \frac{1}{s + \bar s}. \label{SUGRAls} \end{align}
If one substitutes equation (\ref{SUGRALS}) back into the Lagrangian (\ref{SUGRALFO}), one retrieves
\begin{align} \mathcal L =& \left[ - \frac{3}{2} S_0  e^{-gV_R} \bar S_0  (S + \bar S + cV_R )^{2/3}\right]_D \notag \\
&+ \left[S_0^3 W(S)\right]_F + \left[f \mathcal W^2 \right]_F +  \text{h.c.}, \end{align}
which corresponds to a theory with K\"ahler potential $\mathcal K = -2 \log \left( S + \bar S \right) $, a superpotential $W(S)$ and a gauged (shift) R-symmetry in the old minimal formalism.

We are now ready to derive the analogue of the Lagrangian (\ref{L_L1}) in supergravity and confirm that the propagating degrees of freedom are still two real scalars and a Majorana fermion, as was the case in rigid supersymmetry.

To obtain the theory in terms of the superfield $L$, we use the following identity which is proven in section~\ref{Appendix:Proof}
\begin{align} \left[ L(S + \bar S)\right]_D = \left[ T(L) S \right]_F + \text{h.c.}, \label{D-F} \end{align}
where $T$ is the chiral projection operator.
In global supersymmetry, the chiral projection is in superspace the operator $\bar D^2$, where $\bar D_{\dot \alpha}$ is defined in equation (\ref{chiralprojection}). 
In supergravity one can only define such an operator $T$ on a real multiplet with Weyl weight $w$ and chiral weight $c$ given by $(w,c)=(2,0)$. 
Consider now a real multiplet $L$ with components $L = (l, \chi, Z, v_a, \lambda, D)$, where $a$ from now on is used to indicate a flat frame index, 
related to a curved Lorentz index $\mu$ by $v_a = e_a^{\ \mu} v_\mu$. 
Since a chiral multiplet $\Sigma$ can be defined by its lowest component, we define the chiral projection of $L$ via the operator $T$ as a chiral multiplet with lowest component $-\bar Z$
\begin{align} T(L) \ : \ L \ (2,0) \ \longrightarrow \ \Sigma(L) = T(L), \ \ \ (w,c) = (3,3). \label{T(L)} \end{align}
In global supersymmetry, this definition of the chiral projection operator $T(L)$ would correspond to $T(L) =- \frac{1}{4} \bar D^2 L$.

By using the relation (\ref{D-F}), one can now calculate the equation of motion for $S$
\begin{align} S = \frac{1}{b} \log \left( \frac{T(L)}{ab S^3_0} \right), \end{align}
and insert this back into the original Lagrangian (\ref{SUGRALFO}) to obtain
\begin{align} \mathcal L =  & \left[ - \frac{1}{2} \left( S_0 e^{-gV_R} \bar S_0 \right)^3 L^{-2} \right]_D - \left[ c L V_R \right]_D \notag \\& + \frac{1}{b} \left[ T(L)- T(L) \log \left( \frac{ T(L)}{ab S^3_0} \right) \right]_F + \left[ f \mathcal W^2 \right]_F+ \text{h.c.} \label{SUGRAL} \end{align}
Below, it is shown that apart from the obvious extra fields compared to global supersymmetry, namely the graviton, the gravitino, a gauge boson and a gaugino\footnote{
Actually, a linear combination of the two 2-component fermions present in this model is the Goldstino 
which will be eaten by the gravitino according to the super-BEH mechanism, see section \ref{sec:p=2}. }, 
this theory does not contain any additional degrees of freedom and the (non-gravitational) spectrum can be described in terms of two real scalars $l$ and $w$, 
where $w$ is defined as $Z = \rho e^{iw}$, and a Majorana fermion $\chi$. 

We now look for a component expression of the Lagrangian (\ref{SUGRAL}).
Since we are interested in comparing the F-term scalar potential of the above theory, we will from now on neglect the vector multiplet $V_R$ putting $g=c=f=0$. 
The components of the real multiplet $L$ are given by $L = (l, \chi, Z, v_a, \lambda, D)$, 
while the components of the (auxiliary) chiral multiplet are given by $S_0 = (s_0, P_L \zeta, F)$, where $P_{L(R)}$ is the left-handed (right-handed) projection operator. 
The Weyl and chiral weights of the $Z$-component of a real linear multiplet are given by $(w,c) = (w+1, -3)$. 
Since the lowest component of a chiral multiplet must satisfy $w=c$ for the superconformal algebra to close, it is clear that the operation $T(L)$, given by equation (\ref{T(L)}), 
can only be defined on a real multiplet with $w=2$, so that $\bar Z$ has weights $(w,c)=(3,3)$ and can thus be used as lowest component of a chiral multiplet.  
This agrees with the assumption that $L$ has Weyl weight 2. The auxiliary field $S_0$ is chosen to have weight 1. 

The chiral multiplet obtained by the projection operator $T$ has components 
$T(L) = (- \bar Z, -\sqrt 2 i P_L \left(\lambda + \cancel{ \mathcal D} \chi \right), D + \mathcal D^a \mathcal D_a l + i \mathcal D^a v_a)$, 
where the covariant derivatives are given by
\begin{align} \mathcal D_a l &= \partial_a l - 2 \mathcal{B}_a l - \frac{1}{2} i \bar \psi_a \gamma_* \chi, \notag \\ 
\mathcal D_a v_b &= \partial_a v_b - 3 \mathcal B_a v_b + \frac{1}{2} \bar \psi_a \left( \gamma_b \lambda + \mathcal D_b \chi \right) - \frac{3}{2} \bar \varphi_a \gamma_b \chi, \notag \\ 
\mathcal D_a P_L \chi &= \left( \partial_a + \frac{1}{4} \omega_a^{\ bc} \gamma_{bc} - \frac{5}{2} \mathcal B_a + \frac{3}{2} i \mathcal A_a \right) \chi \notag \\
& \ \ - \frac{1}{2} P_L \left( i Z - \cancel v - i \cancel {\mathcal D} l \right) \psi_a + 2i l P_L \varphi_a, \end{align}
where $\psi_\mu$ is the gravitino, $\varphi_\mu$ is the gauge field corresponding to the conformal supercharge $\mathcal S$, $\mathcal B_\mu$ is the gauge field of dilatations 
and $\mathcal A_\mu$ is the gauge field of the T-symmetry (which is the $U(1)$ R-symmetry of the superconformal algebra). 

To obtain a real multiplet from a real function $\phi (s, \bar s)$ of chiral superfields $(s, \zeta_L, F)$ and their anti-chiral counterparts, we use \cite{VanProeyen:1983wk}
\begin{align} &\left(  \phi (s, \bar s) ; - \sqrt 2 i \phi_s \zeta_{L}  ; - 2 \phi_s F + \phi_{ss} \bar \zeta_L \zeta_L ; i \phi_s \partial_a s - i \phi_{\bar s} \partial_a \bar s + i \phi_{s \bar s} \bar \zeta_L \gamma_a \zeta_R ; \right.  \notag \\ & -\sqrt 2 i \phi_{s \bar s} F \zeta_R + \sqrt 2 i \phi_{s \bar s} \cancel \partial \bar s \zeta_L + \frac{i}{\sqrt 2} \phi_{s s \bar s} \zeta_R \cdot \bar \zeta_L \zeta_L ; \notag \\ &\phi_{s \bar s} \left[ 2 F \bar F - 2 \partial_a s \partial^a \bar s - \partial_a \bar \zeta_L \gamma^a \zeta_R + \bar \zeta_L \cancel \partial \zeta_R \right] - \phi_{s s \bar s} \left[ \bar \zeta_L \zeta_L \bar F + \bar \zeta_L \cancel \partial s \zeta_R \right] \notag \\ &\left.  - \phi_{s \bar s \bar s} \left[ \bar \zeta_R \zeta_R F + \bar \zeta_R \cancel  \partial \bar s \zeta_L \right] + \frac{ \phi_{s s \bar s \bar s}}{2} \bar \zeta_L \zeta_L \cdot \bar \zeta_R \zeta_R \right), \label{chiralcomponents} \end{align}
where $\phi_s = \frac{\partial \phi}{\partial s}$, $\phi_{\bar s} = \frac{\partial \phi}{\partial \bar s}$, etc. To obtain a real superfield from a function $f(C)$ of real multiplets with components $C_i, \chi_{Li}, Z_i, v_{ai}, \lambda_{Ri}, D_i$ we use
\begin{align} &\left( f(C); f^i \chi_{Li} ;  f^i Z_i - \frac{1}{2} f^{ij} \bar \chi_{Li} \chi_{Lj} ; f^i v_{ai} + \frac{i}{2} f^{ij} \bar \chi_{Li} \gamma_a \chi_{Rj}; \right. \notag \\ & f^i \lambda_{Ri} + \frac{1}{2} P_L \left[ Z_i -  i \cancel v_i - \cancel \partial C_i \right] \chi_j - \frac{1}{4} f^{ijk} \chi_{Ri} \cdot \bar \chi_{Lj} \chi_{Lk} ; \notag \\ & f^i D_i - \frac{1}{2} f^{ij} \left[ 2 \bar \chi_i \lambda_j - Z_i \bar Z_j + v_{ai} v^a_j + \partial_a C_i \partial^a C_j - \left( \partial_a \bar \chi_{Li} \right) \gamma^a \chi_{Rj} + \bar \chi_{Li } \cancel \partial \chi_{Rj} \right] \notag \\ & \left. - \frac{1}{4} f^{ijk} \left[ \bar \chi_{Li} \bar Z_j \chi_{Lk} + \bar \chi_{Ri} Z_j \chi_{Rk} + 2i \bar \chi_{i} \cancel v_j \chi_k  \right] + \frac{1}{8} f^{ijkl} \bar \chi_{Li} \chi_{Lj} \cdot \bar \chi_{Rk} \chi_{Rl} \right). \notag \end{align}

We are now ready to write down the (non-gauge) bosonic part of the Lagrangian (\ref{SUGRAL}) as
\begin{align} \mathcal L/e &=  \frac{1}{2} (s_0 \bar s_0)^3 l^{-2} R +  (s_0 \bar s_0)^{3} l^{-3} D  + \frac{3}{2}  (s_0 \bar s_0)^{3} l^{-4}  v^\mu v_\mu  \notag \\ & - \frac{1}{b} \log \left( \frac{\rho^2}{a^2 b^2 (s_0 \bar s_0)^3} \right) D + \frac{1}{b} 2 v^\mu \partial_\mu w   + \mathcal L_{kin} - \mathcal V_F, \label{VkinVF} \end{align}
where we used that 
\begin{align} \Box l = \mathcal D^a \mathcal D_a l = \partial^a \partial_a l - \frac{1}{3} l R \label{boxl} \end{align}
 to rewrite a term involving $\Box l$ in the third term in (\ref{SUGRAL}). The scalar field $w$ is defined as 
\begin{align} Z = \rho e^{iw}, \label{Zw} \end{align}
 and it will turn out that this field inherits the gauged shift symmetry. $\mathcal L_{kin}$ and $\mathcal V_F$ in equation (\ref{VkinVF}) are given by
\begin{align} \mathcal L_{kin} =&  \ 3  (s_0 \bar s_0)^{2} l^{-2}   \partial^\mu s_0 \partial_\mu \bar s_0    + \frac{3}{2}  (s_0 \bar s_0) l^{-2} \partial^\mu(s_0 \bar s_0) \partial_\mu (s_0 \bar s_0)  
\notag \\ &-3   (s_0 \bar s_0)^{2} l^{-3}  \partial^\mu l \partial_\mu (s_0 \bar s_0) 
 + \frac{3}{2}  (s_0 \bar s_0)^{3} l^{-4}  \partial^\mu l \partial_\mu l  -\left( \frac{s_0 s_0}{l} \right)^3 \partial^\mu \partial_\mu l,   \notag \\ 
\mathcal V_F =& \  9   (s_0 \bar s_0)^{2} l^{-2}  F \bar F    + 3   (s_0 \bar s_0)^{2} l^{-3} \left( \bar s_0 Z \bar F + s_0 \bar Z F \right)  \notag \\
&+ 3 \frac{Z \bar F}{b \bar s_0} + 3 \frac{\bar Z F}{b s_0}  +  \frac{3}{2}  (s_0 \bar s_0)^{3} l^{-4} Z \bar Z \notag \\ 
=& - \frac{\rho^2}{(s_0 \bar s_0)^3} l^2 \left( \frac{1}{b} +  (s_0 \bar s_0)^2  l^{-3} \right)^2  +  \frac{3}{2} \rho^2 (s_0 \bar s_0)^{3} l^{-4} , \end{align}
where we solved the equations of motion of the auxiliary field $F$ and used equation (\ref{Zw}) to go from the first to the second line of $\mathcal V_F$. On the other hand, the equations of motion for $D$ and $v_\mu$ are given by
\begin{align} \rho^2= a^2 b^2 (s_0 \bar s_0)^3 e^{ b \left( \frac{s_0 \bar s_0}{l}\right)^3 }, \notag \\  (s_0 \bar s_0)^3 l^{-4} v_\mu =  - \frac{2}{3b} \partial_\mu w.\end{align}
We now fix the conformal gauge as
\begin{align} (s_0 \bar s_0)^3 = l^2 \label{conformalgauge} \end{align}
to obtain a correctly normalized Einstein-Hilbert term. Upon this choice $\mathcal L_{kin}$ reduces to $\mathcal L_{kin} = -\frac{1}{2} \frac{1}{l^2} \partial^\mu l \partial_\mu l$ and we obtain the final result
\begin{align} \mathcal L = \frac{1}{2} R - \frac{1}{2} \frac{1}{l^2} \partial^\mu l \partial_\mu l - \frac{2}{3 b^2} \partial^\mu w \partial_\mu w - \mathcal V_F,\end{align}
 with 
 \begin{align} \mathcal V_F = a^2 e^{\frac{b}{l}} \left( \frac{b^2}{2} -2bl - l^2 \right) . \end{align}
Upon substituting the relation (\ref{SUGRAls}), 
one indeed finds the correct F-term scalar potential computed with the chiral formulation (see equation (\ref{scalarpotp2})), namely
\begin{align} \mathcal V_F =   a^2 e^{b(s+\bar s)} \left(  \frac{b^2}{2}  - \frac{2b}{s+\bar s}  - \frac{1}{(s+ \bar s)^2 }    \right) .\end{align}

In order to obtain a tunable dS vacuum (see section \ref{sec:p=2}), 
one has to extend the above formalism to include a linear gauge kinetic function $f(s)$. By following the same method as above, but by including an extra term
\begin{align} \left[ \beta S \mathcal W^2 \right]_F + \text{h.c.}, \end{align}
which contains the gauge kinetic terms of $V_R$ with gauge kinetic function $f(S) = \beta S + \gamma$ to the Lagrangian (\ref{SUGRALFO}), one finds that the result (\ref{SUGRAL}) is still valid upon the substitution
 \begin{align}  T(L) \longrightarrow T(L) - \beta \mathcal W^2.  \end{align}
The theory dual to the one described in section \ref{sec:p=2} is given by
\begin{align} \mathcal L =  & \left[ - \frac{1}{2} \left( S_0 e^{-gV_R} \bar S_0 \right)^3 L^{-2} \right]_D - \left[ c L V_R \right]_D \notag \\ 
& + \frac{1}{b} \left[ \left( T(L) - \beta \mathcal W^2 \right) - \left( T(L) - \beta \mathcal W^2 \right) \log \left( \frac{ T(L) - \beta \mathcal W^2 }{ab S^3_0} \right) \right]_F + \text{h.c.} \notag \\
& + \left[ \gamma \mathcal W^2 \right]_F+ \text{h.c.}. \label{SUGRAL2} \end{align}
 
  \section{Proof of an important identity} \label{Appendix:Proof}
 
 In this Appendix we prove equation (\ref{D-F}), repeated here for convenience
 \begin{align} \left[ L(S + \bar S)\right]_D = \left[ T(L) S \right]_F + \text{h.c.}  . \label{D-Fappendix} \end{align}
To do this, we calculate both sides in components and see that they coincide. It is sufficient to show this only for the bosonic terms. If the (bosonic) components of the chiral multiplet $S$ are given by $\left( s, 0, F \right)$ and the (bosonic) components of the real multiplet $L$ are given  by $\left( l \ ; 0 \ ; Z \ ; v_\mu \ ; 0 \ ; D \  \right)$, then the components of $S + \bar S$ can be calculated using (\ref{chiralcomponents}) 
\begin{align} S + \bar S &= \left(s + \bar s \ ; 0 \ ; -2(F + \bar F) \ ; i \partial_\mu (s - \bar s) \ ; 0 \ ; 0 \  \right). \end{align}
The chiral multiplet $T(L)$ is defined as a chiral multiplet with lowest component $-\bar Z$ and has components
\begin{align} T(L) = (- \bar Z \ ; 0 \ ; D + \Box l + i \mathcal D^\mu v_\mu \ ). \end{align}
The left-hand side of equation (\ref{D-Fappendix}) can be written in components by using equations (\ref{chiralcomponents}) and (\ref{D-term}) (neglecting all terms involving fermions)
\begin{align} \left[L (S + \bar S) \right]_D = & -\frac{1}{3} l (s + \bar s) R + (s + \bar s) D - \bar Z F - Z \bar F \notag \\
&- i v^\mu \partial_\mu (s - \bar s) - \partial^\mu \partial_\mu (s + \bar s). \end{align}
Similarly, by using equation (\ref{chiralcomponents}), $\left[ T(L) S \right]_F$ can be written in components as
\begin{align} \left[ T(L) S \right]_F = - \bar Z F + s D + s \Box l + i s \partial^\mu v_\mu. \end{align}
The right-hand side is thus given by
\begin{align} \left[ T(L) S \right]_F + \text{h.c.}  = - \bar Z F - Z \bar F + (s + \bar s) D + (s + \bar s) \Box l + i (s + \bar s) \partial^\mu v_\mu .\notag \end{align}
The equality (\ref{D-Fappendix}) is then proven by using equation (\ref{boxl}) for $\Box l$ and then partially integrating the last two terms.

\cleardoublepage

\chapter{R-symmetries and anomalies} \label{ch:Appendix_anomalies}

In rigid supersymmetry, the Lagrangian of the model in eqs.~(\ref{model:exp_superpot}) is given by
\begin{align}
 \mathcal L = \left[ \mathcal K(S + \bar S) \right]_D + \left( \left[ W(S) \right]_F  + \left[ f(S) \mathcal W^2 \right]_F + \text{h.c.} \right) . \label{appendix:SUSYlagrangian_superfields}
\end{align}
Here, $S = (s, \chi, F)$ is the dilaton multiplet and the operations $\left[ \ \ \right]_F$ and $\left[ \ \ \right]_D$ are defined in eqs.~(\ref{F-term}) and (\ref{D-term}) respectively.
The K\"ahler potential $\mathcal K(S+ \bar S)$ is given by eq.~(\ref{model:kahlerpotential}), 
the superpotential is given by eq.~(\ref{model:superpotential}), and the gauge kinetic function is given by eq.~(\ref{model:gauge kinetic function}), 
summarized here for convenience of the reader
\begin{align}
 \mathcal K &= - \kappa^{-2} p \log (s + \bar s), \notag \\
 W &= \kappa^{-3} a e^{bs}, \notag \\
 f_R (s) &= \gamma + \beta_R s.
\end{align}
The subscript $R$ is explicitly written to distinguish the gauge kinetic function corresponding to $U(1)_R$ from the gauge kinetic functions of the Standard Model gauge groups.

In eq.~(\ref{appendix:SUSYlagrangian_superfields}), $\mathcal W^2$ is the multiplet whose scalar component is $\bar \lambda P_L \lambda$, where $\lambda$ is the R-gaugino.
For example, the last term in eq.~(\ref{appendix:SUSYlagrangian_superfields}) contains the $U(1)_R$ kinetic terms, given by
\begin{align}
\mathcal L \supset  - \frac{1}{4} \text{Re}(f(s))   F_{\mu \nu} F^{\mu \nu}  + \frac{1}{8} \text{Im}(f(s) ) \epsilon^{\mu \nu \rho \sigma} F_{\mu \nu} F_{\mu \nu}.
\end{align}
Since the superpotential transforms under the shift symmetry eq.~(\ref{shift}), this cannot be a gauge symmetry in a rigid supersymmetric theory (in contrast with supergravity).
However, it can be considered as a global R-symmetry. When $s \rightarrow s - ic$, the superpotential transforms as
\begin{align} W(s) \rightarrow W(s)  e^{-ibc}. \label{appendix:superpot_transform}\end{align}
It follows that under this global symmetry, the chiral fermions and gauginos transform as\footnote{
This can be more intuitively seen in the superfield formalism, where the Lagrangian contains a term
\begin{align} \mathcal L \supset \int d^2 \theta W(S). \end{align}
Since the superpotential transforms as (\ref{appendix:superpot_transform}), $d \theta$ should transform as $d \theta \rightarrow d\theta e^{ibc/2}$ such that the Lagrangian remains invariant. 
The Grassmann variable then transforms as $\theta \rightarrow \theta e^{-ibc/2}$.
In the conventions of~\cite{wessbagger}, a chiral superfield is given by $\Phi = C + \sqrt 2 \theta \chi + \theta^2 F $. 
To compensate for the transformation of $\theta$, the component fields should transform as
$C \rightarrow C$, $\chi \rightarrow \chi e^{ibc/2}$ and $F \rightarrow F e^{ibc}$.
In a gauge multiplet however, 
the gauge boson appears as the term proportional to $\theta \bar \theta$ and the D-component field appears proportional to $\theta^2 \bar \theta^2$. 
They are therefore invariant under the global R-symmetry.
Finally the gaugino, appearing in a vector superfield as the field proportional to $\bar \theta^2 \theta$, 
has a transformation with opposite sign to the chiral fermions $\lambda \rightarrow \lambda^{-ibc}$.
}
\begin{align}
P_L \chi &\rightarrow P_L \chi e^{ibc/2}, \notag \\
 P_L \lambda &\rightarrow P_L \lambda e^{-ibc/2},
\end{align}
Respectively.
With $\xi = bc $, the R-charges of the chiral fermions, the gauginos and the gravitino\footnote{
Although the gravitino does not appear in the spectrum of a rigid supersymmetric theory, 
its contribution to the anomalies should be taken into account in section~\ref{sec:Anomalies-FI}.
It can be seen from eqs.~(\ref{cov_derivatives})
that the R-charge of the gravitino is the same as that of a gaugino.
The contributions of the gravitino to the cubic anomaly and the mixed gravitational anomaly are discussed in~\cite{grisaru}.
}
in the theory are then given by
\begin{align}
R_\chi &= Q + \xi/2, \notag \\
 R_\lambda &= -\xi/2, \notag \\
 R_{3/2} &= -\xi/2, \label{R-charge:allocations}
\end{align}
where we included a superfield charge $Q$ for the chiral multiplets (see eq.~(\ref{Q-killing})). 
This implies that the components of a matter multiplet $(z, \chi, F)$ carry a charge $R_z = Q$, $R_\chi = Q + \xi/2$ and $R_F = Q + \xi$ respectively.
The scalar component of the dilaton multiplet transforms as a shift under $U(1)_R$, and has therefore a vanishing charge $Q$.

Even though an R-symmetry can not be gauged in a rigid supersymmetric theory, since the R-symmetry generator does not commute with supersymmetry (see eq.~\ref{algebra:R-symmetry}),
we continue by analyzing the anomalous contributions to the Lagrangian originating from a gauge symmetry under which the fermions carry a charge given by eqs.~(\ref{R-charge:allocations}).

We now define 
\begin{align}
\mathcal  C_A \delta^{ab} &= \text{Tr}_{A} \left[ R_\chi (\tau^a \tau^b )_A \right] + T_G \delta^{ab} R_\lambda, \label{Appendix:CA} \\
\mathcal C_R &= \text{Tr} \left[ R_\chi^3 \right] +  n_\lambda R_\lambda^3  + 3 R_{3/2}^3,
\end{align}
where $\mathcal C_A$ denotes the mixed anomaly of $U(1)_R$ with the Standard Model gauge groups $A = U(1)_Y, SU(2)_L, SU(3)$.
In eq.~(\ref{Appendix:CA}), one has $T_G \delta^{ab} = f^{acd} f^{bcd}$, with $T_G = N$ for $SU(N)$ and $0$ for $U(1)$.
$\text{Tr}_R (\tau^a \tau^b ) = \delta^{ab} T_R $ is the trace over irreducible representations. 
The contribution of the gravitino to the cubic anomaly is three times that of a gaugino~\cite{grisaru}.

The anomalous $U(1)_R$ generates a cubic anomaly. 
As a result, the variation of the Lagrangian does not vanish. Instead, it is given by
\begin{align}
 \delta \mathcal L_{\text{1-loop}} =   - \frac{(1/3)\mathcal C_R \theta }{32 \pi^2} \epsilon^{\mu \nu \rho \sigma} F_{\mu \nu} F_{\rho \sigma}.
\end{align}
This anomaly can however be canceled by a Green-Schwarz mechanism (see eqs.~(\ref{model:Green-Schwarz term}) and (\ref{model:Green-Schwarz contribution})).
Cancellation of $\delta \mathcal L_{GS}  + \delta \mathcal L_{\text{1-loop} } = 0 $ 
\glossary{name=GCS,description=Generalized Chern-Simons (term)} gives\footnote{
The presence of Generalized Chern-Simons (GCS) terms~\cite{Anastasopoulos:2006cz,Anastasopoulos:2007qm,DeRydt:2007vg} is assumed in eqs.~(\ref{an:cubic}) and (\ref{an:mixed}).
Note that there is an additional symmetry factor $1/3$ for the cubic anomaly.
} 
\glossary{name=GCS,description=Generalized Chern-Simons (term)} 
\begin{align}
 \beta_R c =  - \frac{\mathcal C_R }{12 \pi^2}. \label{an:cubic}
\end{align}
This anomaly cancellation condition is to be compared with the result in eq.~(\ref{fieldtheory}).
Similarly, we have for the mixed anomalies ($U(1)_R$ times Standard Model gauge groups)
\begin{align}
 \beta_A c = - \frac{\mathcal C_A }{4 \pi^2}. \label{an:mixed}
\end{align}
Regarding the mixed gravitational anomaly, one has 
\begin{align}
 \delta \mathcal L_{\text{1-loop}} = - i \theta \frac{\mathcal C_{\text{grav}}}{384 \pi^2} R_{\mu \nu} \tilde R^{\mu \nu},
\end{align}
where 
\begin{align}
\mathcal C_{\text{grav}}=  \text{Tr}[R_\chi] + n_\lambda R_\lambda + (-21) R_{3/2}.
\end{align}
Here it was used that the contribution of the gravitino is $(-21)$ times the contribution of a gaugino~\cite{grisaru}.
We assume that an (unknown) UV-completion of the theory cancels this anomaly due to a term in the Lagrangian
\begin{align}
 \mathcal L_{\text{grav}} =  i \beta_{\text{grav}} \frac{ \text{Im}(s)}{4} R_{\mu \nu} \tilde R^{\mu \nu} .
\end{align}
The anomaly is canceled $ \delta \mathcal L_{\text{1-loop}} + \delta \mathcal L_{\text{grav}} = 0$ if
\begin{align}
 \mathcal C_{\text{grav}} = 96 \pi^2 \beta_{\text{grav}} c . \label{an:grav}
\end{align}
This indeed agrees with eqs.~(\ref{ee}) for $\beta_{\text{grav}} = b_{\text{grav}} / 12 \pi^2$.

\chapter{Comments the case with $p=1$ and non-zero $\beta$} \label{ch:Appendix_p1}

\section{On the small value of $\beta$} \label{Appendix:nonzerobeta}

In this Appendix we first demonstrate how tunable dS vacua can be found for the model~(\ref{model:other}) for a general $\beta$.
In section~\ref{sec:other} it was assumed that $\beta \ll 1$. Here, we show that the anomaly cancellation conditions indeed force $\beta$ to be very small.
 
\begin{figure} \centering \medskip \includegraphics[width=0.5\textwidth]{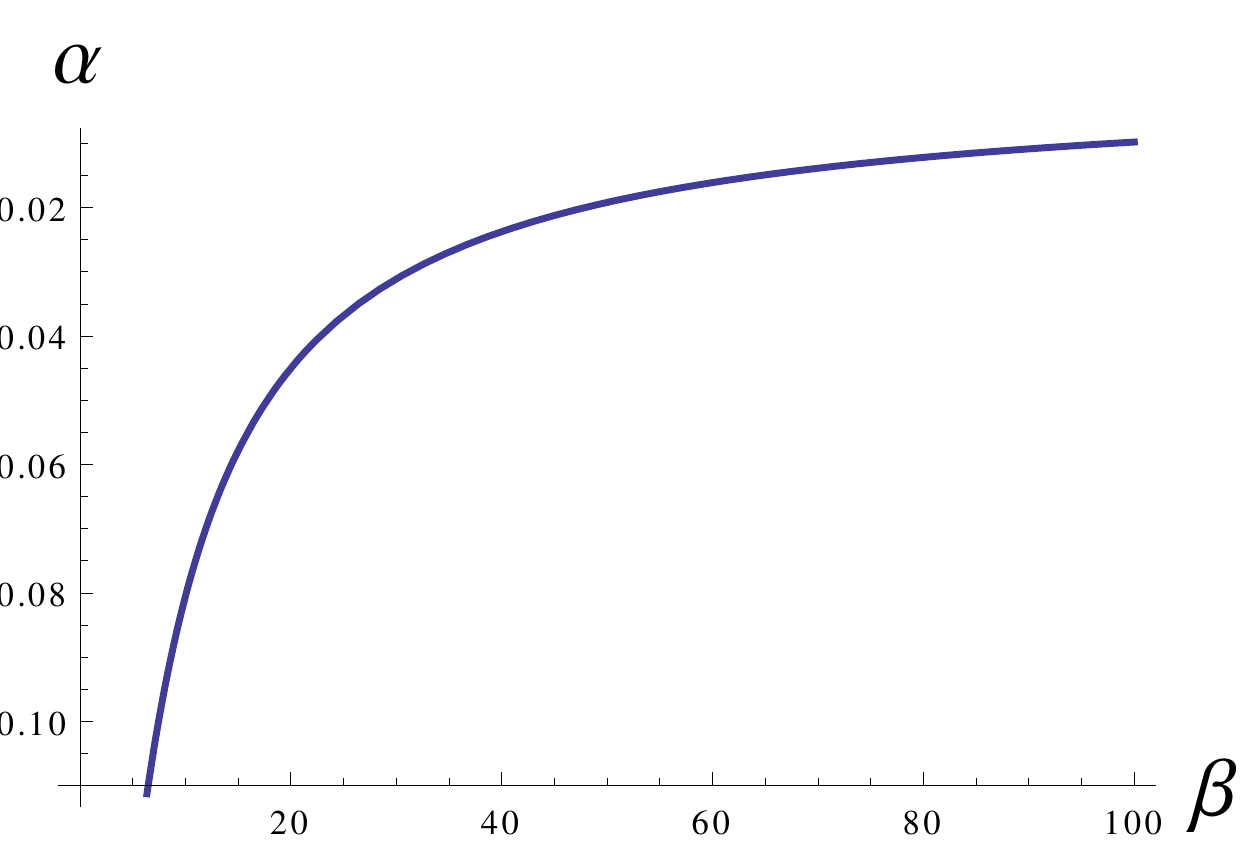} 
 \caption{The solutions to eq.~(\ref{betaprime}) are plotted with $\alpha$ as a function of $\beta$. Although up to four (complex) solutions can exist for every $\beta$, only a single solution satisfies $\alpha < 0$ (to ensure a negative $b$) and $A(\alpha,\beta) < 0$ to satisfy eq.~(\ref{Aalphabeta}). Only positive $\beta$ is plotted since the positivity is required to satisfy the anomaly cancellation conditions (see below). }  \label{plot_alphabeta} \end{figure}

If $\beta$ is not neglected, the D-term contribution to the scalar potential is given by
\begin{align} \mathcal V_D =  \frac{1}{2 + \beta b (s + \bar s)} \left(\kappa^{-2}bc - \frac{\kappa^{-2} c}{s+ \bar s} - q_\alpha \varphi_\alpha \bar \varphi_\alpha \right)^2.\end{align}
The solution to $\partial \mathcal V = \mathcal V = 0$ gives 
\begin{align} (\alpha - 1)^2 \left( \alpha^2 -2 \right) + \left( \frac{\beta \alpha (\alpha-1)}{2+\beta \alpha} -2 \right) \left(-3 + (\alpha-1)^2 \right)=0. \label{betaprime}\end{align}
For $\beta = 0$, this gives indeed $\alpha \approx -0.233$, in agreement with equation~(\ref{model:p1_tunability}). For other values of $\beta$ the result is plotted in figure~\ref{plot_alphabeta}.
The relation between the parameters to obtain a vanishing cosmological constant becomes
\begin{align} \frac{bc^2}{a^2} = - \frac{\alpha e^{\alpha} (2+ \beta \alpha) (-3 + (\alpha-1)^2 )}{(\alpha-1)^2} =A(\alpha,\beta). \label{Aalphabeta}\end{align}
One concludes that for every (finite) $\beta$ the vacuum is tunable. The gravitino mass eq.~(\ref{p1:gravitino_mass}) is repeated here for convenience of the reader
\begin{align} m_{3/2} &=  \kappa^{-1} a e^{\alpha/2} \sqrt{\frac{b}{\alpha}}. \notag \end{align}
The scalar soft mass squared and the trilinear couplings are given by
\begin{align}  m_0^2 &= m_{3/2}^2 (\sigma_s + 1) +\kappa^{-2} \frac{q_\alpha bc}{1+ \beta \alpha/2} \frac{ 1 - \alpha}{\alpha}, \notag \\ 
 A_0 &= m_{3/2} \ \rho_s,  \notag \\ 
 \rho_s &= -1 + (\alpha-1)\left(\alpha - 1 + P(\alpha,\beta) \xi \right), \notag \\
 P(\alpha,\beta) &= \frac{2 \alpha e^\alpha (\sigma_s +1) }{(\alpha-1) A(\alpha,\beta)}  \left(1+\frac{ \beta \alpha}{2}\right). \label{Palphabeta} \end{align}
The relations~(\ref{Aalphabeta}) can be written as
\begin{align} bc = -\kappa m_{3/2} e^{-\alpha/2} \sqrt{\alpha A(\alpha, \beta)}, \label{bc=Aalphabeta}\end{align}
which are used to write the scalar soft mass squared as 
\begin{align}  m_0^2 &= m_{3/2}^2 (\sigma_s + 1) + m_{3/2} \kappa^{-1} q_\alpha \ Q(\alpha,\beta), \notag \\  
Q(\alpha, \beta) &= \frac{\alpha-1}{ \alpha} e^{-\alpha/2} \frac{\sqrt{\alpha A(\alpha,\beta)} }{1+\beta \alpha/2}.  \end{align}
To avoid tachyonic masses, the charges of the MSSM fields should satisfy
\begin{align} q_\alpha > q_0 = - \kappa  m_{3/2} \frac{\sigma_s +1}{Q(\alpha,\beta )}. \label{qlimitbeta} \end{align}

\begin{figure}[H] \centering \medskip \includegraphics[width=0.5\textwidth]{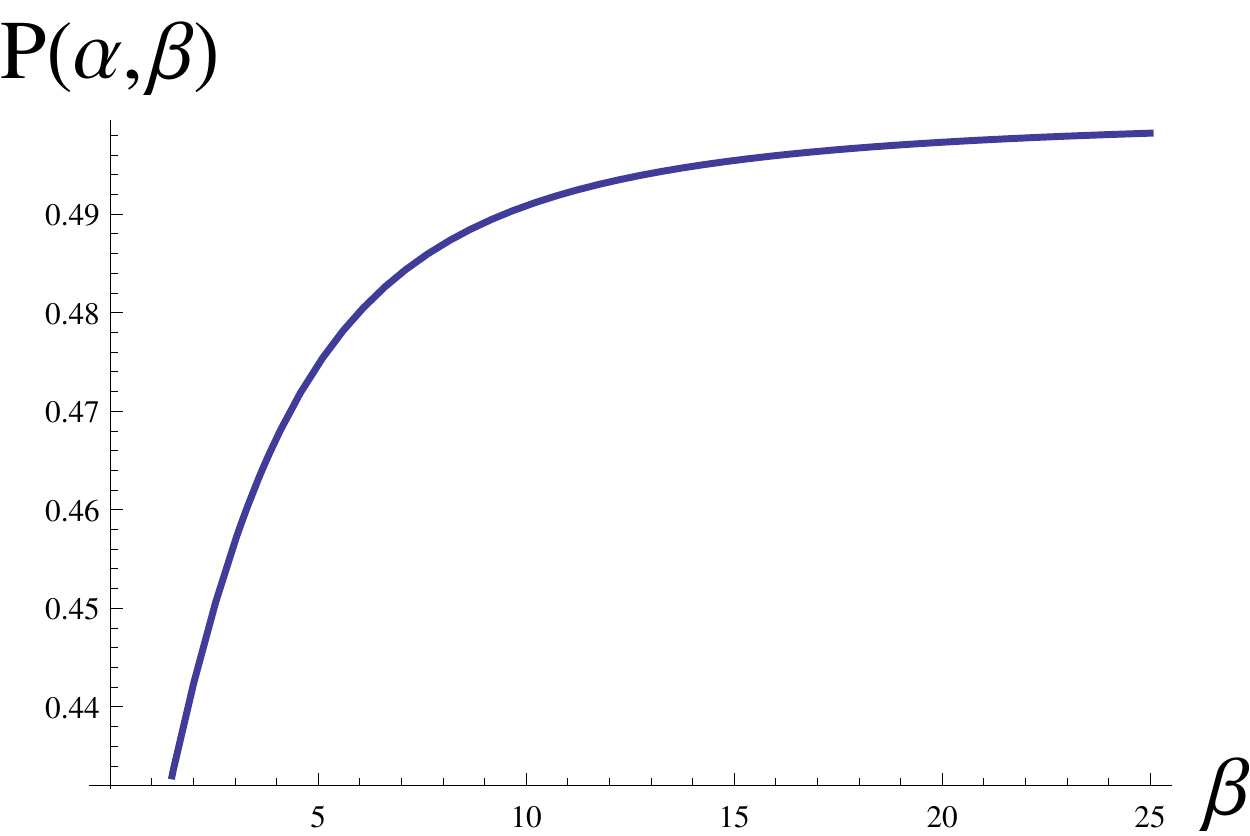} 
\caption{A plot of $P(\alpha,\beta)$ given by eq.~(\ref{Palphabeta}) as a function of $\beta$, where $\alpha$ and $\beta$ are solutions of eq.~(\ref{betaprime}). It follows that the trilinear coupling $A_0$ (as a function of $\xi$) has only a very small dependence on $\beta$. One should therefore focus on $\xi \lesssim 10$ (as in section \ref{sec:other}) since otherwise $|A_0|$ becomes too large to allow for a viable electroweak vacuum.}  \label{plot_Pbeta} \end{figure}

Next, the anomaly cancellation conditions (see eqs.~(\ref{other:anomalycancellation})) for the cubic anomaly give
\begin{align} \beta &= - \frac{f(\theta,\theta_L,\theta_Q) q^3}{12 \pi^2 bc} \notag \\
 &= f(\theta,\theta_L,\theta_Q) \xi^3 (\kappa m_{3/2})^2 g(\alpha,\beta), \label{betaprimeanomaly}\end{align}
where eqs.~(\ref{bc=Aalphabeta}) and (\ref{qlimitbeta}) were used, and
\begin{align} g(\alpha,\beta ) = - \frac{e^{\alpha/2} (\sigma_s +1)^3}{12 \pi^2 Q(\alpha,\beta )^3 \sqrt{\alpha A(\alpha,\beta )}}. \label{galphabeta}\end{align}
From eq.~(\ref{ftheta}) it follows that $ 9.47 \leq f(\theta,\theta_L,\theta_Q) \leq 91$, for all $\theta,  \theta_L,  \theta_Q  \in \  ]0,1[$, such that $\beta >0$  from eq.~(\ref{betaprimeanomaly}).
From figure~\ref{plot_Pbeta}, where $P(\alpha,\beta)$ is plotted as a function of $\beta$, it follows that the trilinear coupling $A_0$ has only a very slight dependence on $\beta $. 
We can therefore assume $\xi < \mathcal O(10)$ (as in section~\ref{sec:other})  since otherwise the trilinear coupling $A_0$ would be too large to allow for a realistic electroweak vacuum.
\begin{figure}[H] \centering \medskip \includegraphics[width=0.5\textwidth]{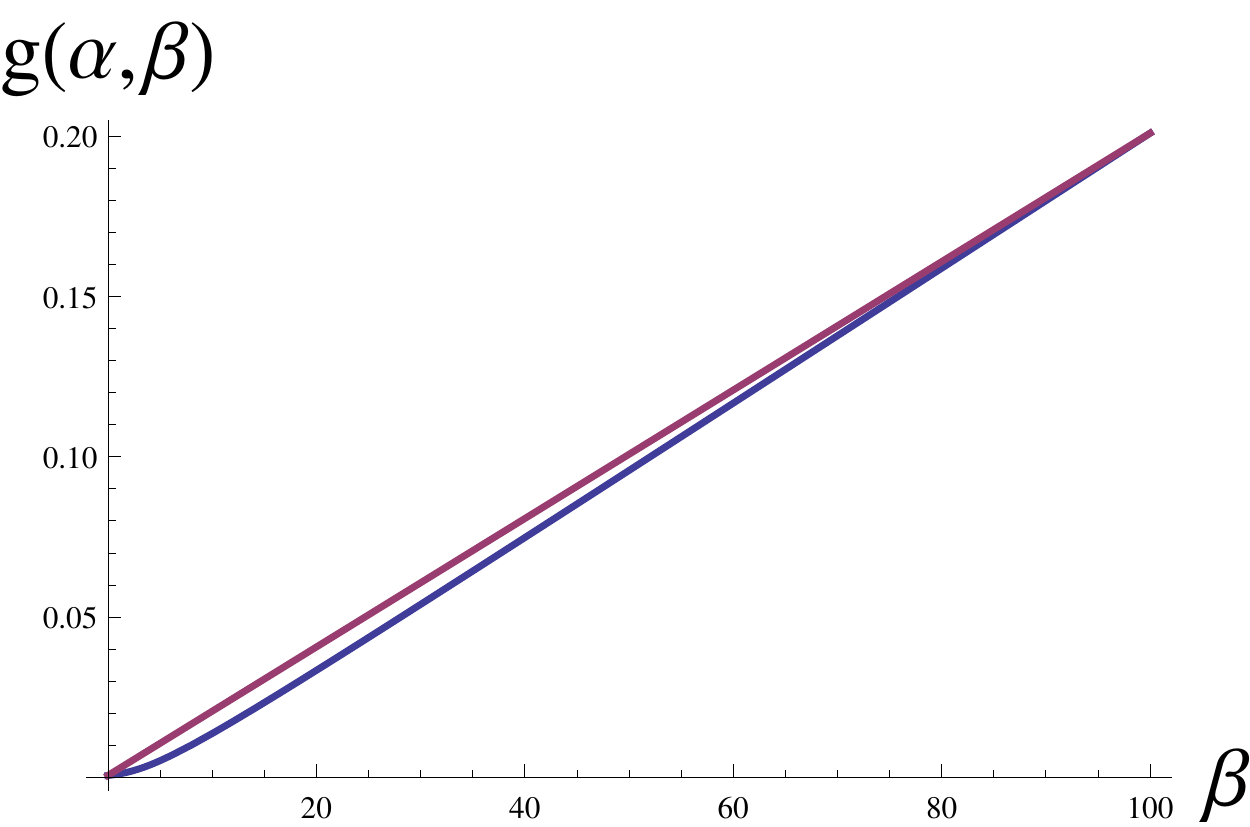} 
\caption{In blue: A plot of $g(\alpha,\beta)$ given by eq.~(\ref{galphabeta}) as a function of $\beta$, where $\alpha$ and $\beta$ are solutions of eq.~(\ref{betaprime}). The approximation $g_0(\beta)$ given by eq.~(\ref{g0}) is plotted in purple.}
  \label{plot_gbeta} \end{figure} 

In figure~\ref{plot_gbeta},  $g(\alpha,\beta)$ is plotted as a function of $\beta $. As is shown, we can approximate $g(\alpha,\beta )$ by a linear function of $\beta$ (In fact, $g_0(\beta)$ is also plotted in figure~\ref{plot_gbeta} and completely overlaps with the actual function $g(\alpha,\beta)$)
\begin{align} g(\alpha,\beta) \approx g_0(\beta) &= g(0) + \omega \ \beta,\label{g0} \end{align} 
where \begin{align}g(0) &= 0.0005618, \notag \\ \omega &= 0.002003. \end{align}
For a given $m_{3/2}$, $\xi$ and $f(\theta,\theta_L,\theta_Q)$, one can then solve 
\begin{align} \beta = f(\theta,\theta_L,\theta_Q) \xi^3 (\kappa m_{3/2})^2 g_0(\beta), \notag \end{align} to find
\begin{align} \beta = \frac{g(0)}{1/\mathcal A - \omega}, \label{betaapprox} \end{align}
where \begin{align}\mathcal A =  f(\theta,\theta_L,\theta_Q) \xi^3 (\kappa m_{3/2})^2.   \end{align}
For $\xi < 10$, $m_{3/2} <40 \text{ TeV}$ and $f(\theta,\theta_L,\theta_Q) <91$, this corresponds to $\beta \lesssim  \mathcal O(10^{-26}) $. 
Note that for small $\beta$ one can approximate eq.~(\ref{betaapprox}) by $\beta \lesssim g(0) \mathcal A$ which leads to the same result.

We conclude that the anomaly cancellation condition can only be satisfied for very small $\beta$.

\section{On the inconsistency of $\gamma=0$} \label{Appendix:p1_gamma0}

In this Appendix we address the tunability of the scalar potential~(\ref{scalarpot_generalp}) for where $p=1$, $\gamma=0$ and non-zero $\beta$. In this case, the constraints $\partial \mathcal V = \mathcal V = 0$ give
\begin{align} (\alpha - 1)^2 \left( \alpha^2 -2 \right) + \left( \alpha -3 \right) \left(-3 + (\alpha-1)^2 \right)=0.\end{align}
The real and negative solutions to this equation are $\alpha \approx -2.47$ and $\alpha \approx -0.39$. 
The first solution however is inconsistent since $\langle \sigma_s \rangle >0$, 
while the F-term contribution to the scalar potential,
which is at the minimum of the potential given by $\mathcal V_F = \kappa^{-4} |a|^2 \frac{e^\alpha}{\langle s + \bar s \rangle} \sigma_s$, should be negative.

Indeed, a vanishing cosmological constant can be found when $b \langle s + \bar s \rangle = \alpha \approx -0.39$, while the other parameters are related by
\begin{align} \frac{b^2 c^2}{\beta a^2} = - \frac{\alpha^2 e^{\alpha}  (-3 + (\alpha-1)^2 )}{(\alpha-1)^2} \approx 0.057. \end{align}

However, due to the presence of a Green-Schwarz contribution~(\ref{model:Green-Schwarz term}) because of the gauge kinetic function which is linear in $s$,
the model is not gauge invariant. One could argue that this can be solved by introducing an extra field $z$ which is charged under the $U(1)$,
where its charge is carefully chosen such that its anomalous contribution to the variation of the Lagrangian cancels the Green-Schwarz contribution~(\ref{model:Green-Schwarz contribution}).
However, an argument similar to the one presented in section~\ref{sec:Anomalies:p=2}  for $p=2$ shows that this is inconsistent.

\cleardoublepage


\backmatter
\includebibliography
\makebackcoverXII
\end{document}